\documentclass[sn-mathphys-num, iicol]{sn-jnl}

\usepackage{multirow}%
\usepackage{amsmath,amssymb,amsfonts}%
\usepackage{amsthm}%
\usepackage{mathrsfs}%
\usepackage[title]{appendix}%
\usepackage{xcolor}%
\usepackage{manyfoot}%
\usepackage{booktabs}%
\usepackage{algorithm}%
\usepackage{listings}%
\usepackage[caption=false,font=normalsize,labelfont=sf,textfont=sf]{subfig}
\usepackage{stfloats}
\usepackage{url}
\usepackage{graphicx}
\usepackage{algpseudocode} % [noend]

\newcommand{\R}{\mathbb R}

\theoremstyle{thmstyleone}%
\theoremstyle{thmstyletwo}%

\theoremstyle{thmstylethree}%
\begin{document}

\title[Quantifying the Resolution of a Template after Image Registration]{Quantifying the Resolution of a Template after Image Registration}

\author*[1]{\fnm{Matthias} \sur{Glock}}\email{matthias.glock@tu-ilmenau.de}

\author[1]{\fnm{Thomas} \sur{Hotz}}\email{thomas.hotz@tu-ilmenau.de}

\affil*[1]{\orgdiv{Group for Probability Theory and Mathematical Statistics}, \orgname{Technische Universität Ilmenau}, \postcode{98693}, \orgaddress{\city{Ilmenau}, \country{Germany}}}


\abstract{In many image processing applications (e.g. computational anatomy) a groupwise registration is performed on a sample of images and a template image is simultaneously generated.
From the template alone it is in general unclear to which extent the registered images are still spatially misaligned.
The local spatial variation left after registration may be seen as the resolution of the resulting template.
In this sense we develop the first template resolution measure (TRM) quantifying the misalignment at each location of the template.
The TRM is based on the key insight that the size of such misalignments can be determined as the amount of smoothing required to bring the registered images in agreement. This relationship is mathematically derived from characteristic examples in one dimension.
This way, the TRM quantifies the remaining spatial variation in the template's units of length. Furthermore we propose to enhance the template by an effective visualization of its resolution measure.
Finally we demonstrate the TRM's applicability and validate its interpretability for example datasets in two and three dimensions. The corresponding code is publicly available on GitHub.}




\maketitle

\section{Introduction and state of the art}
\label{sec:intro}

Given a sample of images, obtained for example by structural magnetic resonance imaging (MRI) from different individuals, how can one make comparisons between the samples despite variations in shape and size? A common step to approach such problems is a groupwise registration of all the images with a simultaneously generated template image, also called an atlas in the anatomical context.
The registration, resulting in a pointwise correspondence between the images and the template, may be performed with various classes of transformations such as rigid, affine or more general non-linear deformations \citep{atlas_joshi}. In all these cases there may be still misalignments left even after the registration procedure, which is the starting situation we will assume here. This is either due to the rigidity of the transformations, which are unable to accurately match the equivalent image features, as in the rigid and affine case, or because there simply is no global pointwise correspondence between the images -- their features are just different at some locations.
In the first case some features may be easier to match than others, when they have less variation across the samples, which is usually reflected by the template being sharper around such locations. Accordingly the template may be blurry wherever there is less consensus on a feature between the images.
In general, a naive way of evaluating the accuracy of a groupwise registration is to consider the sharpness (or entropy) of the generated template \citep{aladdin, sharp_tmp_1, sharp_tmp_2, sharp_tmp_3}.
But this is not always reliable and raises the questions of how to determine which regions in a template are a faithful representation of the structures found in the sample images, and how to quantify the regions where this ceases to be the case.

\begin{figure}[t!]
   \centering
   \includegraphics[trim={1.9cm 0.8cm 2.2cm 1.3cm}, clip, width=\linewidth]{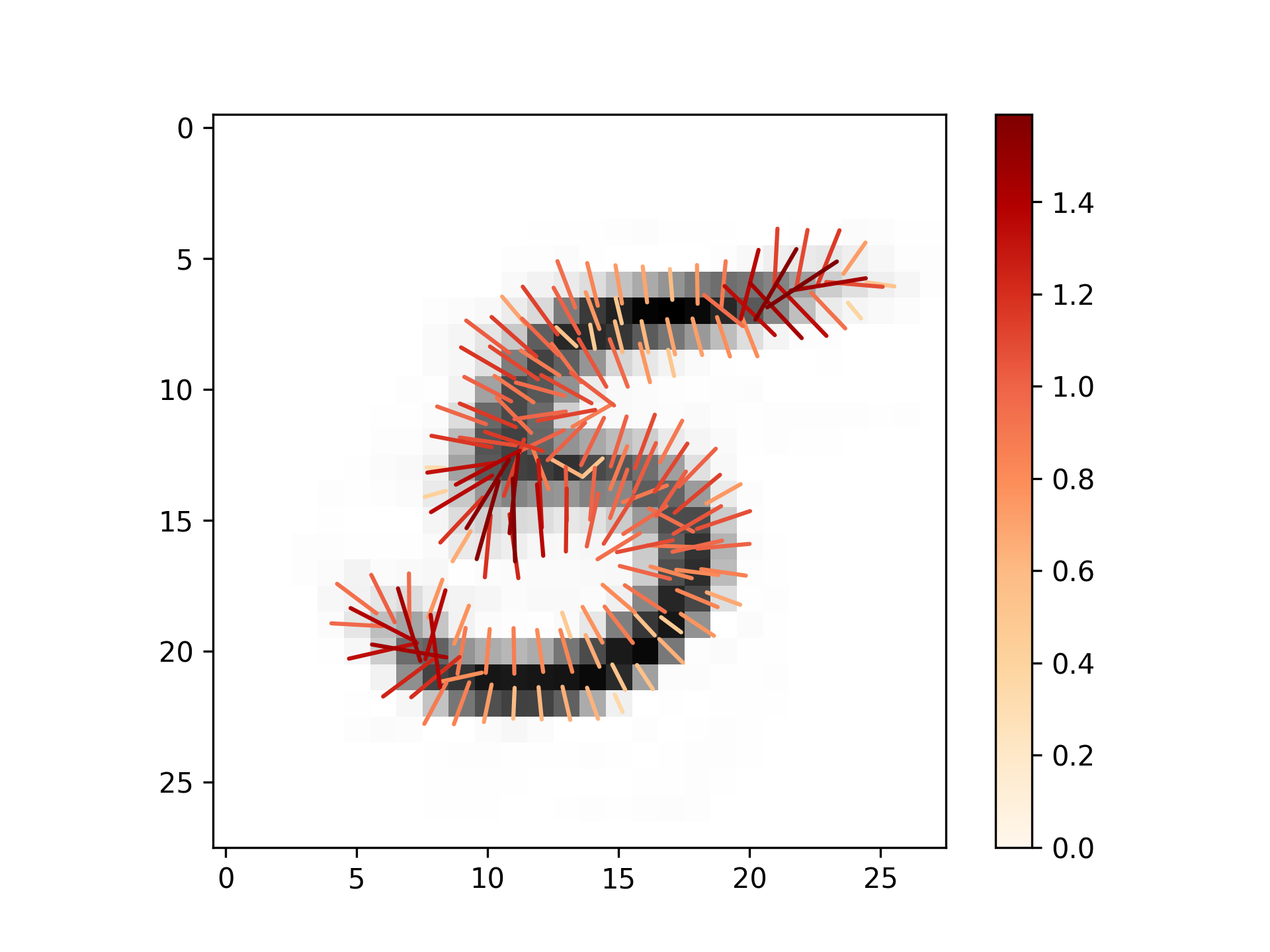}
   \caption{The final resolution measure visualized by coloured bars on top of an affine template for the sample images of handwritten digits ``5'' from the MNIST dataset \citep{mnist_data}. The length and colour of the bars indicate the resolution at the corresponding locations.}
   \label{fig:mnist_tmp_bars_0}
\end{figure}

Our goal is the quantification of the misalignment remaining after image registration, leading to the development of a pointwise measure of the resolution of the template generated in a groupwise registration.
More specifically the misalignment of the images can be thought of to be composed of a ``vertical'' part corresponding to image intensity variations at the same location of a common feature and a ``horizontal'' part due to spatial misalignments of the features themself (see Section \ref{sec:vert_hor}). While the vertical variation may be quantified by something as simple as the pixelwise standard deviation of the registered images, the horizontal variation is much harder to capture because at every pixel one has to incorporate the information of a neighbourhood of that location. The measure developed below is interpretable as a horizontal resolution of the generated template measured in pixels, see Figure \ref{fig:mnist_tmp_bars_0}; we hence term it template resolution measure (TRM).

Commonly used tools that quantify the alignment accuracy of a registration between images, be it groupwise or just pairwise registration, are volume overlap measures such as the Dice score/coefficient or Jaccard index \citep{dice_1, aladdin, atlas_joshi, lombaert2014spectral, klein2009evaluation}.
These measures require the availability of an additional segmentation and labelling of the images and produce a single scalar per segment. This additional information is not required by our method, which even provides a pixelwise measure.
In fact, our novel approach based on Gaussian smoothing of the registered images offers a simple visual way of assessing the local reliability of the template image, identifying the regions where the samples agree on a feature and quantifying those where they are horizontally misaligned.

%-------------------------------------------------------------------------
\section{Background: registration and templates}\label{sec:reg}

We are concerned with the groupwise registration of a sample of images, which is often based on pairwise registration of the images to a common template, that is created simultaneously during the registration.

The pairwise registration of two images consists of finding a transformation from a given class of transformations (e.g. affine) that warps one of the images such that it matches the other (fixed) image as closely as possible with respect to a suitable similarity metric.

In the following, images are given by (square integrable) functions $I:\Omega \to \R$ assigning to each point of the image domain $\Omega \subseteq \R^d$ a real valued intensity (grey value).
The dimensions of the image domain that are considered here are $d = 1, 2, 3$. 

Note that in practice, the continuum $\Omega$ is, of course, approximated by a discrete grid of pixels and even though the registration framework below is formalized in the continuous setting, it is solved approximately for suitably discretized versions of the corresponding mathematical concepts.

The transformations acting on the images are now given by diffeomorphic (i.e. bijective and in both directions differentiable) functions $\Psi:\Omega \to \Omega$ on the image domain, which create a warped image through the expression $I \circ \Psi^{-1}$. This means that the grey value of the warped image at a point $y \in \Omega$ is given by $I(x)$ for $x = \Psi^{-1}(y)$, which is just the point that gets mapped to $y$ under $\Psi$.

A common approach for the creation of a template of $n$ images $I_1, \dots, I_n$ is the minimization of a functional based on pairwise image registrations \citep{aladdin,atlas_joshi}.
Such functionals often are of the form
\[\mathcal E(T, \theta_1, \dots, \theta_n) = \sum_{i=1}^n \mathcal L_{\rm sim}(I_i \circ \Psi^{-1}_{\theta_i}, T) + \lambda\,\mathcal L_{\rm reg}(\Psi^{-1}_{\theta_i})\,,\]
where an image similarity term $\mathcal L_{\rm sim}$ measures the discrepancy between the warped images and the template $T$ and a regularization term $\mathcal L_{\rm reg}$ penalizes the complexity of the transformation, thus avoiding deformations that are unnecessarily large. The terms are coupled through a trade-off parameter $\lambda > 0$. Here every image $I_i$ has its own associated transformation $\Psi_{\theta_i}$ parametrized by a suitable parameter $\theta_i$.

\begin{figure}[h!]
   \centering
   \includegraphics[trim={4.1cm 0.7cm 2.1cm 1.2cm}, clip, width=0.62\linewidth]{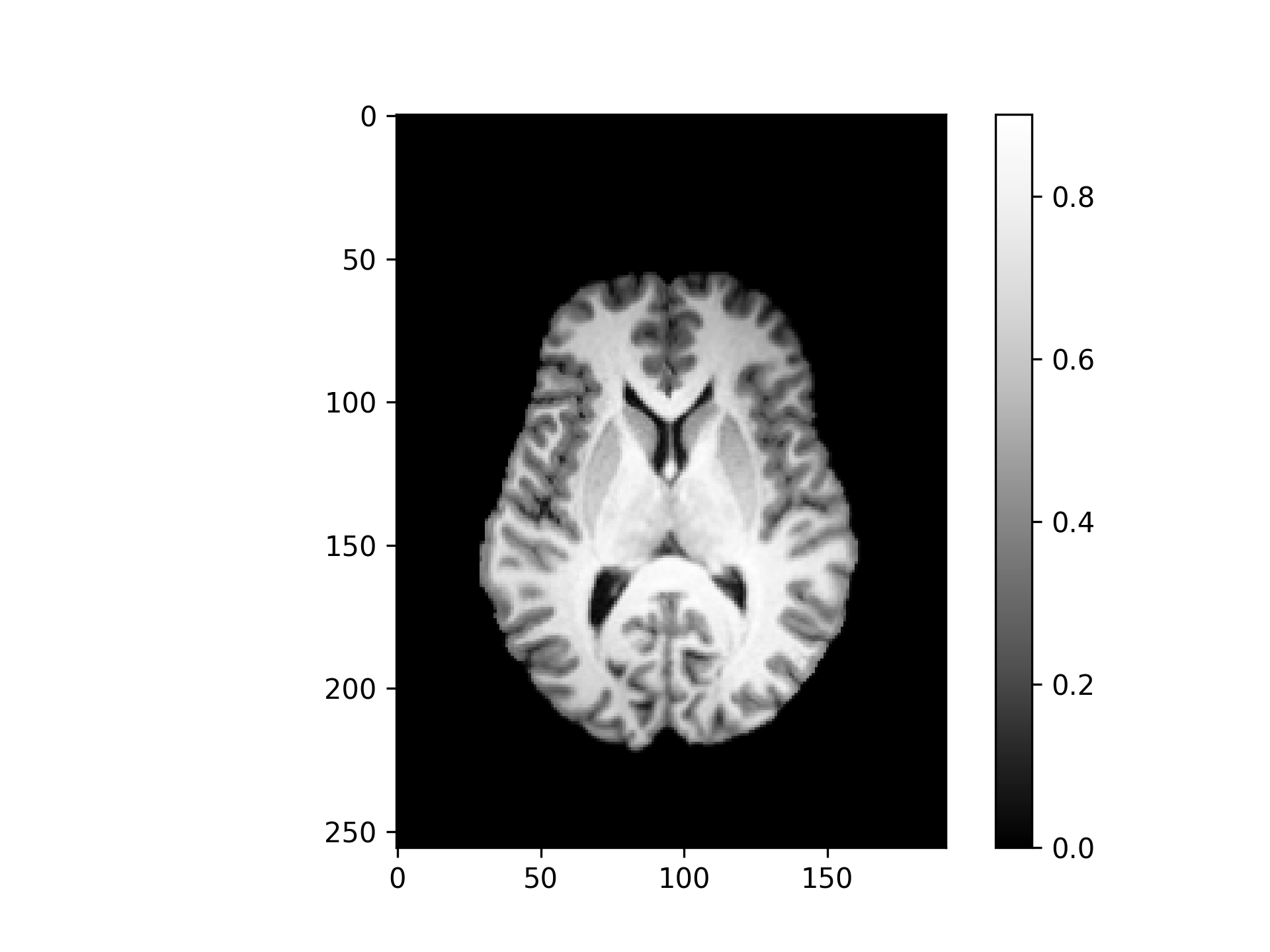}
   \\[2em]
   \includegraphics[trim={4.1cm 0.7cm 2.1cm 1.2cm}, clip, width=0.62\linewidth]{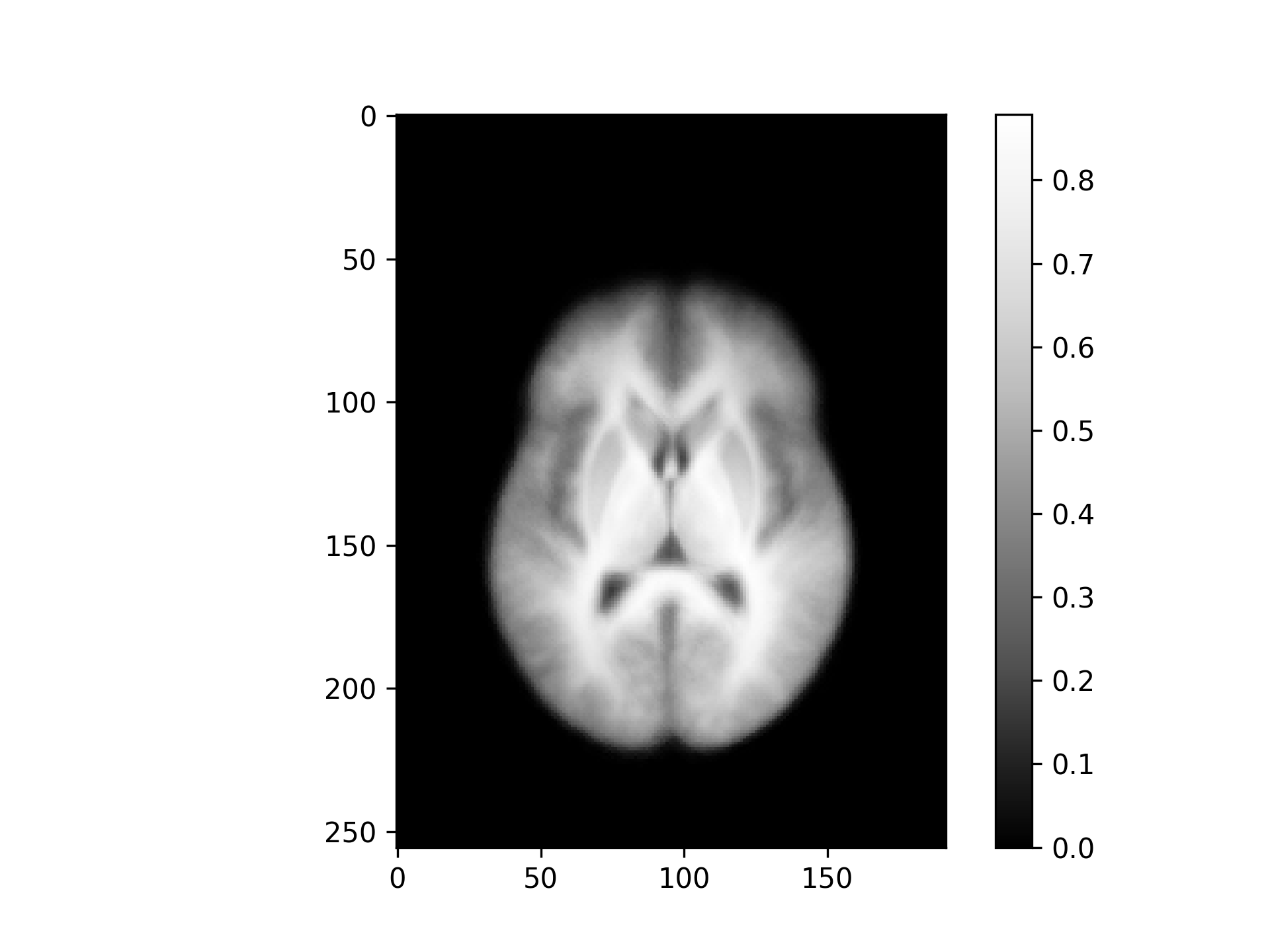}
   \\[2em]
   \includegraphics[trim={4.0cm 0.7cm 2.2cm 1.2cm}, clip, width=0.62\linewidth]{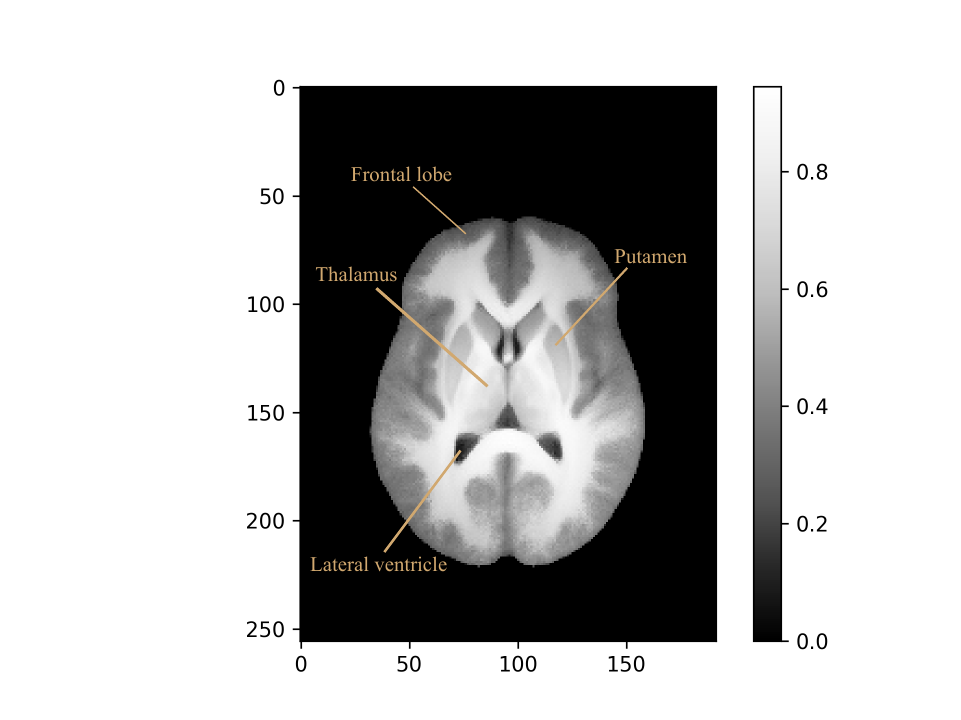} % with labels
   \caption{Horizontal slice for a sample image from the NFBS dataset \citep{nfbs_data} after registration (top) and the corresponding template (middle) based on $L^2$-norm similarity and affine transformations, as well as an annotated template (bottom) based on the $L^1$-norm similarity and affine transformations.}
   \label{fig:brain_tmp_sample}
\end{figure}

A simple choice for the similarity term is the $L^2$-norm:
\[\mathcal L_{\rm sim}(I, J) = \|I - J\|_2^2 = \int_\Omega (I(x) - J(x))^2 \,\mathrm{d}x\,.\]
Other common choices are metrics based on (normalized) cross correlation or (normalized) mutual information \citep{sim1}. Here, the $L^2$-norm is used due to its simplicity while the less common $L^1$-norm $\|I - J\|_1 = \int_\Omega |I(x) - J(x)| \,\mathrm{d}x$ is also used for demonstration purposes below, as it leads to (artificially) sharper templates, see Figures \ref{fig:brain_tmp_sample}, \ref{fig:edges_sample_sig}, and \ref{fig:mnist_tmp}.

The regularization term $\mathcal L_{\rm reg}$ depends on the chosen class of transformations. We will consider the following two classes.

%\subsubsection*{Affine transformations}
\textbf{Affine transformations:}
If $\Omega = \R^d$ the transformations are called \textit{affine} if they are of the from
\[\Psi_\theta(x) = Ax + b \,,\]
for $x \in \Omega$ and $\theta = (A, b)$ with an invertible matrix $A \in \text{GL}(d)$ and a translation vector $b \in \R^d$.
Since affine maps are always defined on all of $\R^d$, it is often necessary to extend the images, e.g. by defining them to have the constant value zero outside the image domain.

%\subsubsection*{Rigid transformations}
\textbf{Rigid transformations:}
If the matrix $A$ of an affine transformation is a rotation matrix $A \in \mathrm{SO}(d)$, i.e. it satisfies $A^\mathsf{T} A = I_d$ and $\det A = 1$, the transformation is called \textit{rigid} or an Euclidean transformation.

In both cases these affine transformations can be regularized by
\[\mathcal L_{\rm reg}(\Psi_\theta) = \|A - I_d\|_2^2 + \|b\|_2^2\,,\]
which penalizes the transformation from deviating too much from the identity transformation $\Psi_\text{id}(x) = x$.

More complicated non-linear transformations are commonly used to achieve state of the art registration accuracy \citep{zhao2019recursive, ma2023hierarchical, meng2023non, ma2023deformable}; these are often created by integrating suitable vector fields \citep{aladdin, atlas_joshi, balakrishnan2019voxelmorph, hernandez2009registration, beg2005computing}.
%Non-linear registration has its own issues separate from misalignments such as controlling the complexity of the non-linear transformations.
Even though non-linear transformations are frequently used for registration, simpler methods like the rigid or affine case still serve their purpose as they are often used as a preprocessing step \citep{mok2022affine, zhou2014novel}.
Typically there are still misalignments left after such a rough registration step.
In the following, the focus will be on the simpler affine and rigid type transformations as non-linear transformations have additional difficulties such as choosing the right regularization to control the complexity of these transformations.
Quantifying the horizontal misaligned after affine or rigid registration is still useful, since, on the one hand, one gains insight into the possible improvement that can be gained by a following non-linear registration step, and on the other hand the quantification of the horizontal variation is useful in its own right as it visualizes the variation of the underlying dataset. In the non-linear case this variation would be absorbed by the transformations themselves, rendering it much harder to visualize.

As already mentioned, for real images given on a discrete domain that approximates the underlying continuum $\Omega$, the minimization is performed numerically with a discretized version of the functional $\mathcal E$.

More specifically we are using the PyTorch framework \citep{pytorch} to perform the optimization of the registration functional $\mathcal E$ using automatic differentiation and
%Here we use
the Adam optimizer.
%with a learning rate of $0.01$ and a maximum of $200$ iterations (sufficient for convergence in our settings). The trade-off parameter $\lambda$ is set to $1.0$.
For additional details, in particular the values for the hyperparameters used, see the source code available on GitHub (\url{https://github.com/Stochastik-TU-Ilmenau/image-template-resolution}).

After minimizing $\mathcal E$ the optimal parameters $\hat\theta_1, \dots, \hat\theta_n$ and the template $\hat T$ are obtained. The registered images $R_i = I_i \circ \Psi_{\hat\theta_i}^{-1}$ are now aligned with the template.
Figure \ref{fig:brain_tmp_sample} shows an example of a registered sample image for the 3d example dataset from Subsection \ref{sec:app_3d} together with template images for the $L^2$- and $L^1$-norm for affine transformations.

\begin{figure}[t!]
   \centering
   \includegraphics[trim={4.1cm 0.7cm 2cm 1.2cm}, clip, width=0.7\linewidth]{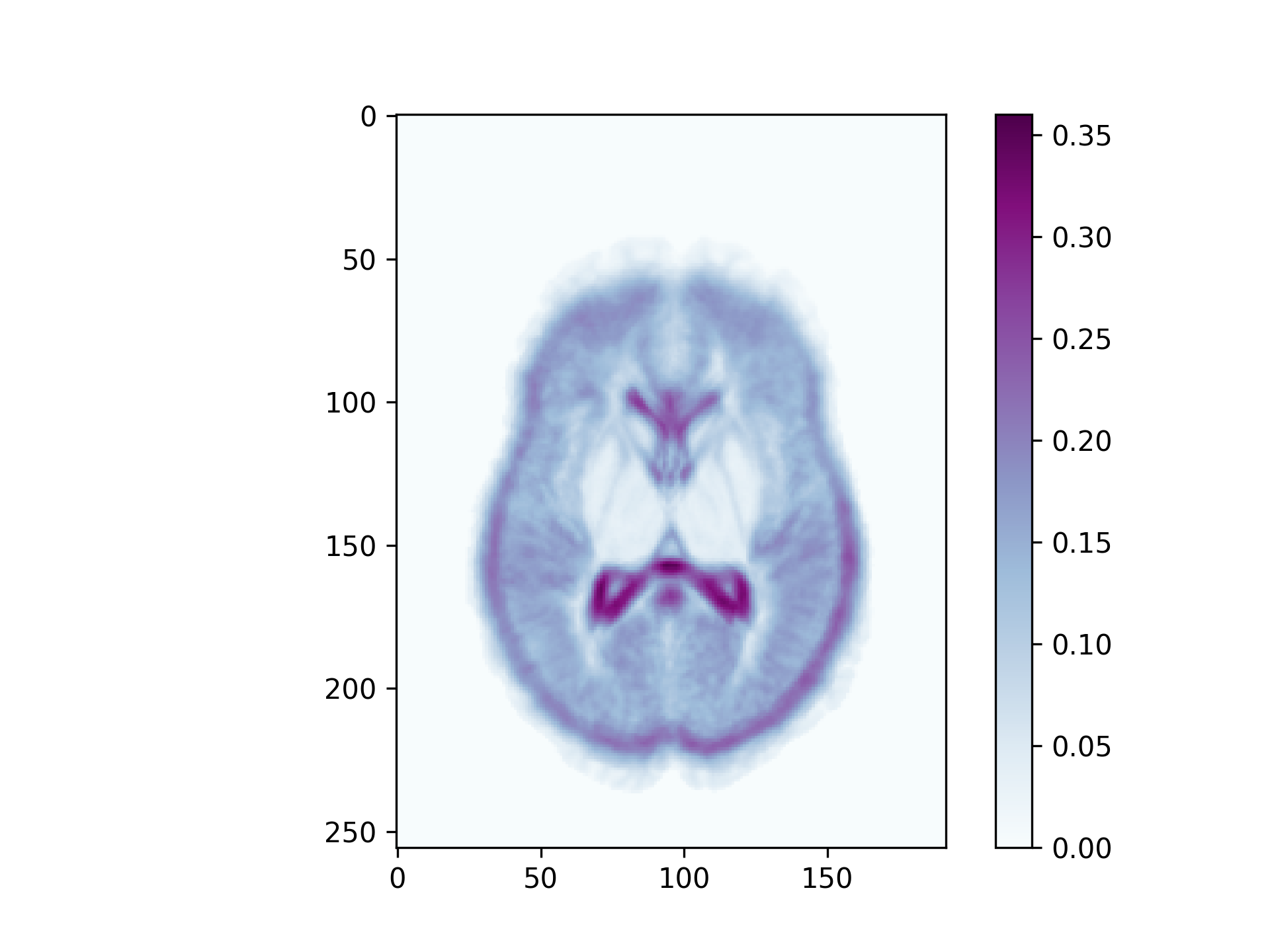}
   \includegraphics[trim={4.1cm 0.7cm 2cm 1.2cm}, clip, width=0.7\linewidth]{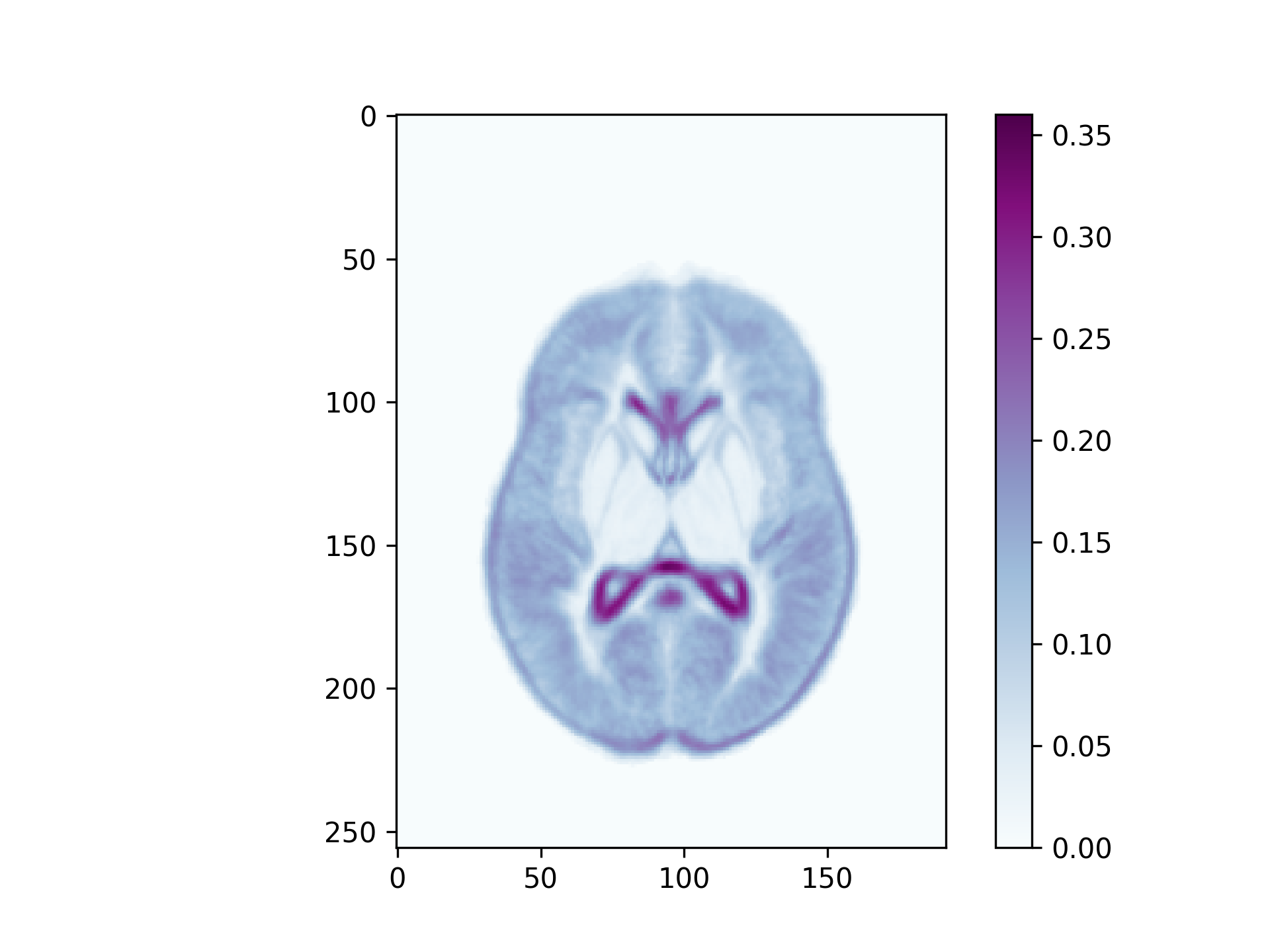}
   \caption{Pixelwise standard deviations of the entire registered NFBS dataset based on the $L^1$-norm for rigid (top) and affine (bottom) transformations.}
   \label{fig:nfbs_l1_vert_var}
\end{figure}

\begin{figure}[t!]
   \centering
   \includegraphics[trim={3.8cm 0.7cm 4.3cm 1.2cm}, clip, width=0.56\linewidth]{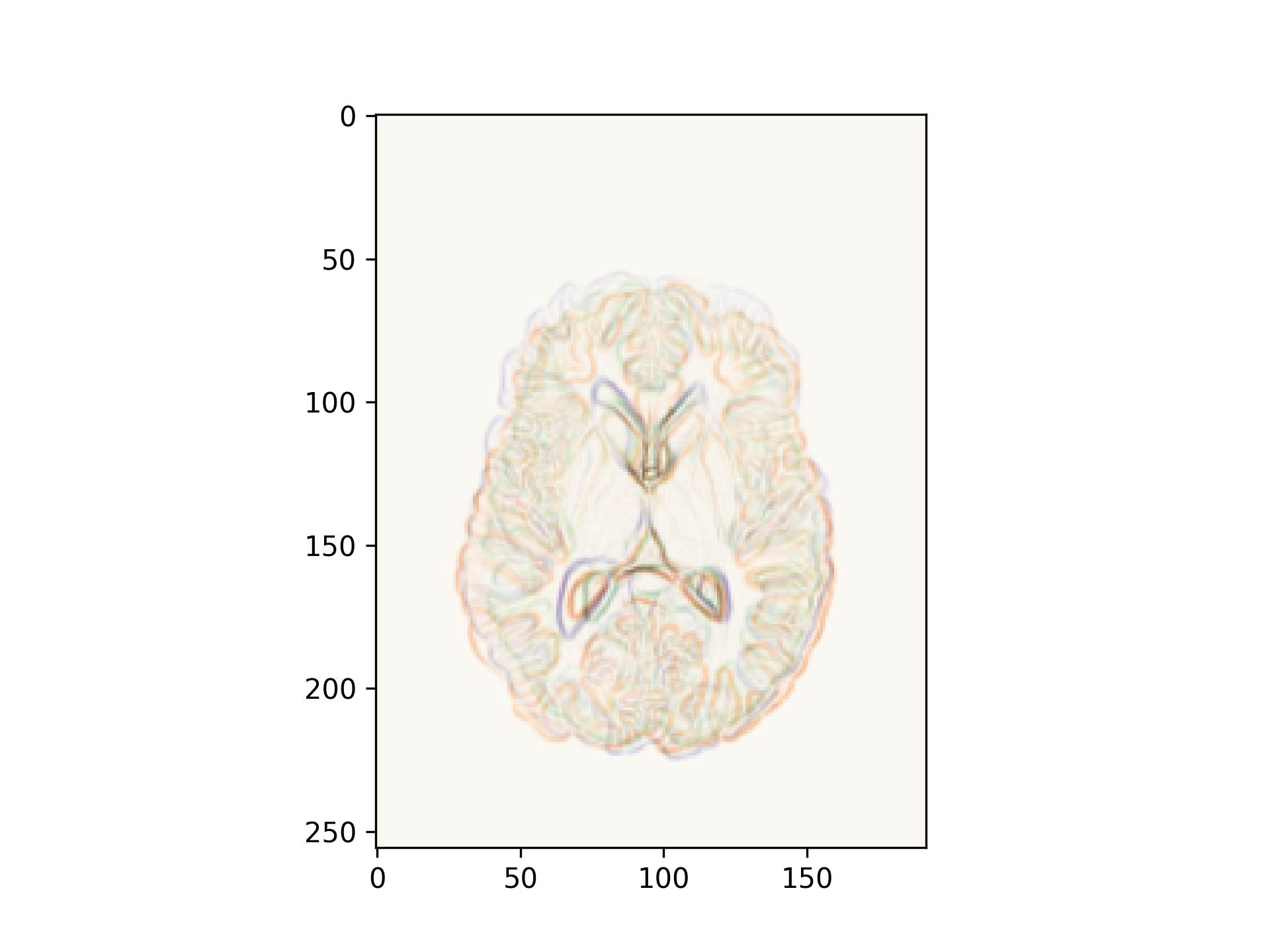}
   \includegraphics[trim={3.8cm 0.7cm 4.3cm 1.2cm}, clip, width=0.56\linewidth]{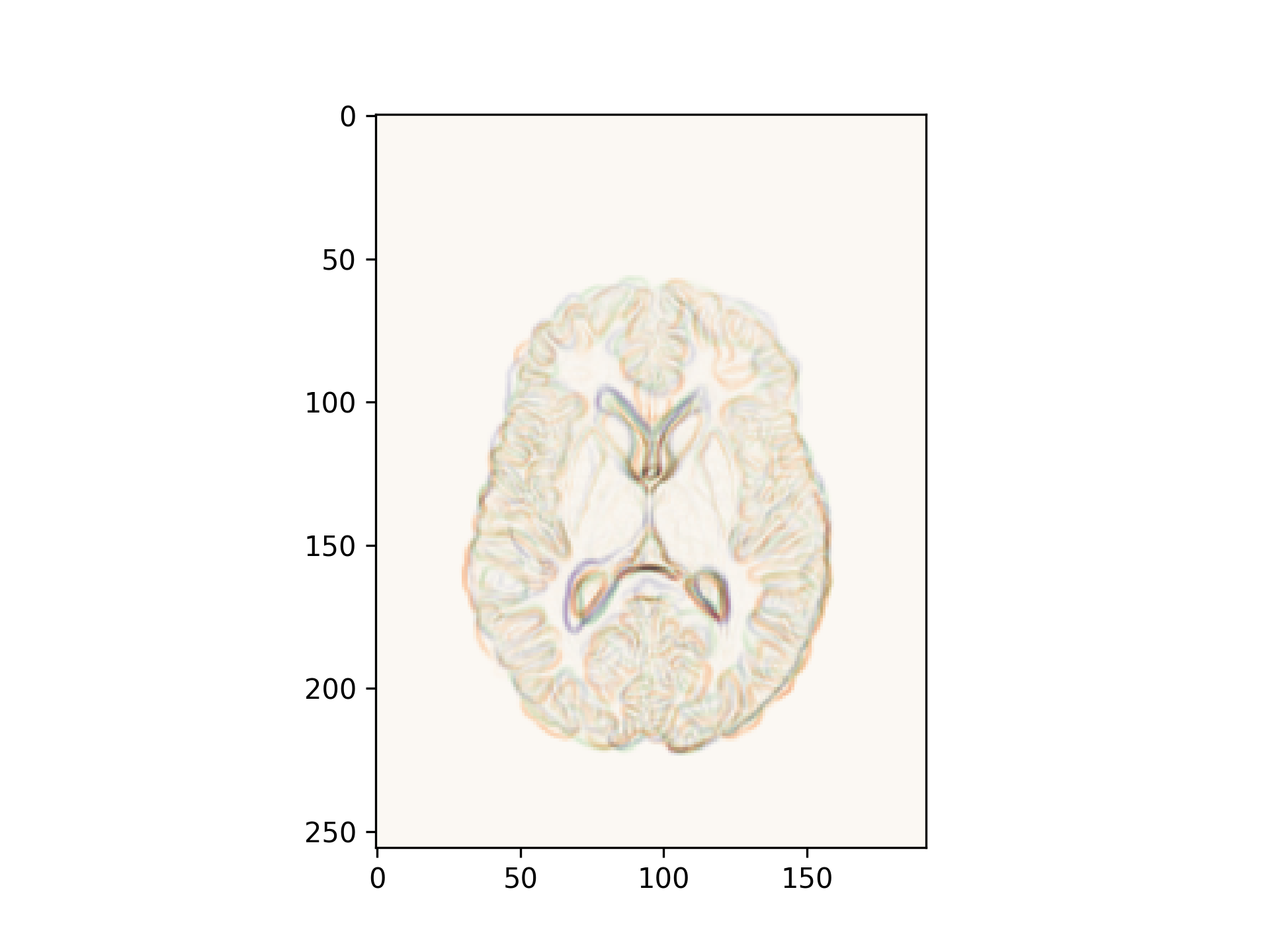}
   \caption{Edges (computed with a Sobel filter) of three (purple, orange, green) randomly selected registered images of the NFBS dataset based on the $L^1$-norm for rigid (top) and affine (bottom) transformations.}
   \label{fig:nfbs_l1_hor_var_edge}
\end{figure}

%-------------------------------------------------------------------------
\section{Vertical and horizontal variation}\label{sec:vert_hor}

Given the registered images, after the optimization in the last section has been carried out, we can try to visualize the variation that is still left between the images.
As mentioned in the introduction, the variation can be either interpreted as being vertical, corresponding to variations in the image intensities at each pixel location, or as horizontal, corresponding to the misalignment of common features like edges.

\begin{figure*}[t!]
   \centering
   \subfloat{\includegraphics[width=0.5\linewidth]{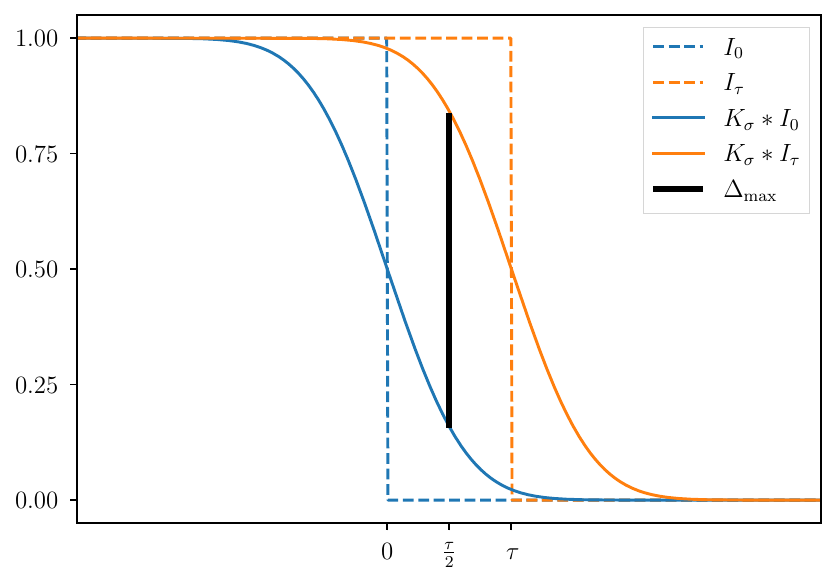}}
   \hfill
   \subfloat{\includegraphics[width=0.5\linewidth]{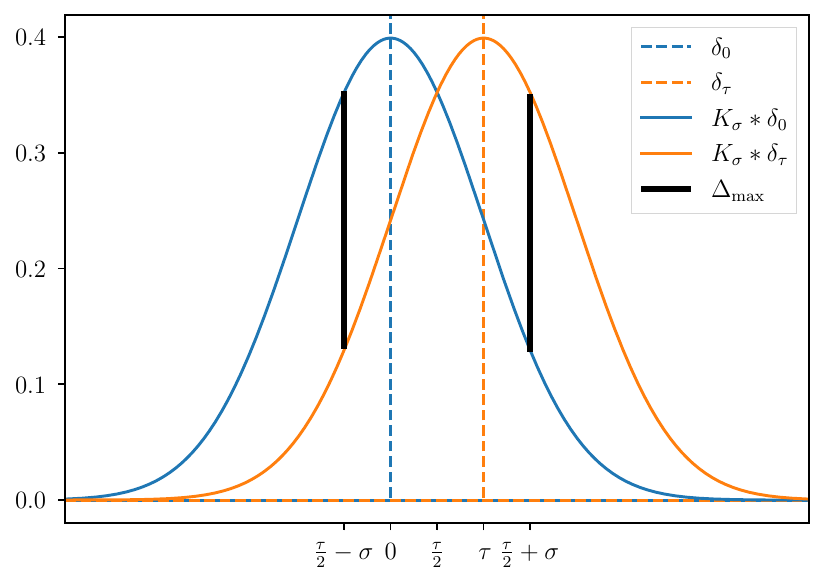}}
   \caption{Behaviour of example images under Gaussian smoothing, on the left for two shifted edges and on the right for two shifted point masses. The maximal difference $\Delta_{\max}$ decreases in both cases for increasing bandwidth $\sigma$.}
   \label{fig:edges_points}
\end{figure*}

Here the terms vertical and horizontal are based on viewing images as the graph of a function mapping from the (one, two or three dimensional) spatial domain to scalar intensity values. This terminology is more generally used in the differential geometric context of fibre bundles, where vertical corresponds to the direction of the fibres (here the space $\R$ of intensities) and horizontal to the base space (here the spatial domain $\Omega$).
The Figures \ref{fig:edges_points}, \ref{fig:edges_sample}, \ref{fig:edges_sample_sig} and \ref{fig:mnist_slice} are based on this point of view for the one dimensional case, where the horizontal axis corresponds to the spatial domain and the vertical axis to the intensity values. We use the terms analogously in the case of two and three dimensional spatial domains.

In Figure \ref{fig:nfbs_l1_vert_var} the registered images from the Neurofeedback Skull-stripped (NFBS) dataset \cite{nfbs_data} (using the $L^1$-norm) are used to compute the pixelwise standard deviation for rigid and affine transformations quantifying (vertical) intensity variations.
Our goal is now to find a suitable quantification of the horizontal variation, which is not easily visualizable. In Figure \ref{fig:nfbs_l1_hor_var_edge} we plot the edges of three randomly selected registered images to give an impression of the horizontal variation left in the registered samples.

The improvement when switching from rigid to affine transformations is clearly visible when looking at the edges in Figure \ref{fig:nfbs_l1_hor_var_edge}. One can see that the outer boundaries of the brains and the lateral ventricles are better matched in the affine case, as expected. This effect is only indirectly visible through a sharpening in the standard deviation plots (Fig. \ref{fig:nfbs_l1_vert_var}).

In the following our goal is to develop a better measure to quantify the registered samples' horizontal variation at every pixel location of the template image.
The resulting measure (Figures \ref{fig:brain_tmp_bars} and \ref{fig:brain_tmp_res}) is able to reveal variation not visible in the standard deviation plots (Fig. \ref{fig:nfbs_l1_vert_var}) as discussed in Section \ref{sec:app_3d}.

%-------------------------------------------------------------------------
\section{The template resolution measure}

The basic idea underlying our method is that smoothing, i.e. local averaging, reduces those differences between registered images which are due to horizontal variation. Therefore, the amount of smoothing necessary to make these differences small allows to quantify the horizontal misalignment.

\subsection{Horizontal translations under Gaussian smoothing}\label{sec:motivation}

Given registered images with local features like edges or peaks, how do small horizontal misalignments between such features still left after registration behave under Gaussian smoothing?
Gaussian smoothing can be achieved by the convolution of an image with a Gaussian kernel of a certain bandwidth $\sigma>0$, which is given in the one dimensional case by
\[K_\sigma(x) = \tfrac1{\sigma\sqrt{2\pi}} e^{-\frac12 \frac{x^2}{\sigma^2}}\,.\]
The following two stylized examples of one dimensional ``images'', i.e. with spatial domain $\Omega = \mathbb R$, are used to study the influence of smoothing on horizontal misalignment.

\subsubsection*{Shifted edges}
%\bmhead{Shifted edges}
For a sharp edge $I_s(x) = \mathbf1_{(-\infty, s]}(x)$ (indicator function of the interval $(-\infty, s]$) at a location $s\in\mathbb R$ we have for the smoothed image $(K_\sigma * I_s)(x) = 1 - \Phi(\tfrac{x-s}{\sigma})$, where $\Phi$ is the distribution function of the standard normal distribution. For two misaligned images $I_0$ and $I_\tau$ the difference of the smoothed images
\[\left|(K_\sigma * I_0)(x) - (K_\sigma * I_\tau)(x)\right| = \left|\Phi(\tfrac{x}{\sigma}) - \Phi(\tfrac{x-\tau}{\sigma})\right|\]
attains its maximum, due to symmetry, at the location $x = \tfrac\tau2$ with the value
\[\Delta_{\max} = \left|2\Phi(\tfrac12 \tfrac\tau\sigma) - 1\right|\,,\]
which depends only on the ratio $\tfrac\tau\sigma$ and decreases monotonically for growing $\sigma$, see the left panel in Figure \ref{fig:edges_points}.

\begin{figure*}[t!]
   \centering
   \subfloat{\includegraphics[width=0.5\linewidth]{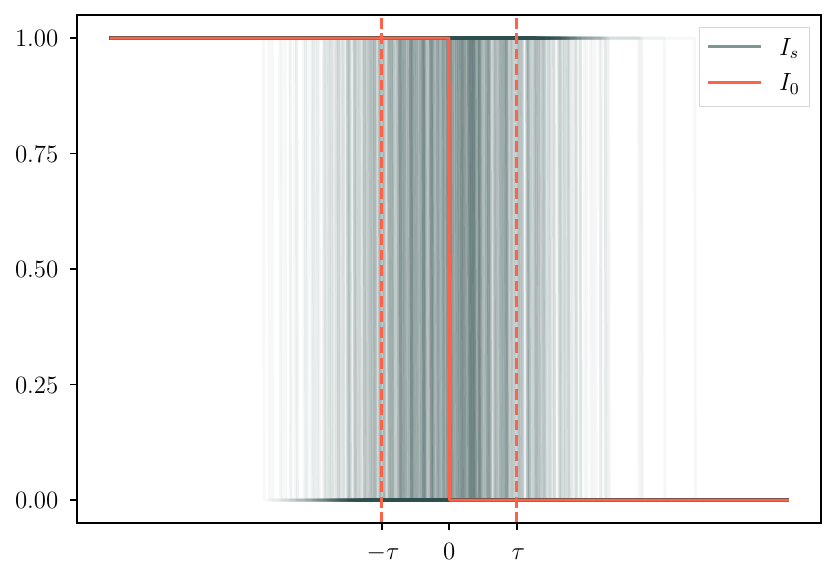}}
   \hfill
   \subfloat{\includegraphics[width=0.5\linewidth]{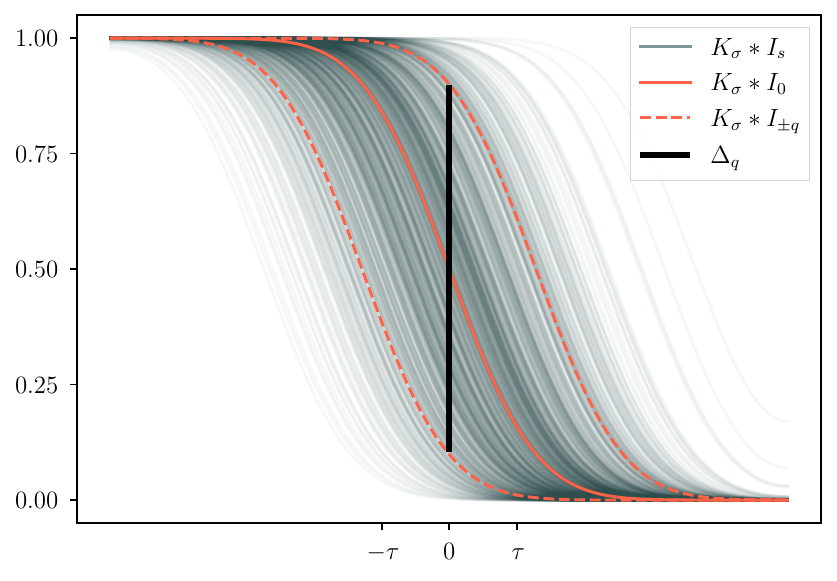}}
   \caption{On the left $n=1000$ sharp edges whose shifts $s$ are randomly sampled from $\mathcal N(0,\tau^2)$ are shown. On the right the samples are smoothed with bandwidth $\sigma = \tau$ and the corresponding empirical quantile range (here for the $0.1$- and $0.9$-quantile) at the location $x=0$ is shown as a bar of length $\Delta_q$. The smoothed versions of the edges $I_{\pm q}$ on the right belong to the $0.1$- and $0.9$-quantiles $\pm q$ of the $\mathcal N(0,\tau^2)$ distribution and approximately match the endpoints of the empirical quantile range.}
   \label{fig:edges_sample}
\end{figure*}

\subsubsection*{Shifted point masses}
%\bmhead{Shifted point masses}
For a point mass (Dirac delta) $\delta_s$ at a location $s\in\R$ we have $(K_\sigma * \delta_s)(x) = K_\sigma(x - s)$.
Since the Gaussian kernel satisfies $K_\sigma(x) = \tfrac1\sigma K_1(\tfrac x\sigma)$, the difference
\[|(K_\sigma * \delta_0)(x) - (K_\sigma * \delta_\tau)(x)| = |K_\sigma(x) - K_\sigma(x - \tau)|\]
between two misaligned, smoothed point masses located at $0$ and $\tau \in \mathbb R$ converges to zero for increasing $\sigma$ simply due to the effect of the smoothing itself. But even after rescaling by a factor of $\sigma$, which prevents the smoothed images from vanishing, the following approximation
\begin{align*}
   \sigma \left|K_\sigma(x) -  K_\sigma(x - \tau)\right| &= \left|K_1(\tfrac x\sigma) - K_1(\tfrac{x - \tau}\sigma)\right| \\
   %&= |\tfrac\tau{\sigma} \tfrac\sigma\tau \left( K_1(\tfrac1\sigma(x - \tfrac\tau2) + \tfrac\tau{2\sigma}) - K_1(\tfrac1\sigma(x - \tfrac\tau2) - \tfrac\tau{2\sigma}) \right)| \\
   &= \left|\tfrac\tau\sigma \left( K_1'(\tfrac1\sigma(x - \tfrac\tau2)) + o(\tfrac\tau\sigma) \right)\right| \\
   &\approx \left|\tfrac\tau\sigma K_1'(\tfrac1\sigma(x - \tfrac\tau2))\right|\,,
   %\\
   %&= |\tfrac\tau\sigma \left( K_1'(\tfrac x\sigma) + o(\tfrac\tau\sigma) \right)| \\
   %&\approx |\tfrac\tau\sigma K_1'(\tfrac x\sigma)|
\end{align*}
obtained by a linearization at the midpoint, holds for $\sigma$ large relative to $\tau$, where the derivative of the Gaussian kernel is given by $K_1'(x) = -xK_1(x)$. This approximation attains its maximum at the two locations $x = \tfrac\tau2 \pm \sigma$ with a value of
\[\Delta_{\max} = \tfrac1{\sqrt{2\pi e}} \left|\tfrac\tau\sigma\right|\,,\] 
which also depends only on the ratio $\tfrac\tau\sigma$; see the right panel in Figure \ref{fig:edges_points}.
\\

In both cases the maximum of the difference goes to zero for increasing $\sigma$. For the shifted edges this decrease is monotone in $\sigma$ and in the case of the smoothed point masses this behaviour also occurs for sufficiently large $\sigma$, when the above approximation becomes accurate enough. For a larger misalignment between the images, i.e. for a larger $\tau$, a larger $\sigma$ is needed in both cases to decrease the distance of the images to a given value.

More generally the misalignments that are still left between registered images can be reduced by smoothing with larger horizontal misalignments usually requiring more smoothing for the same reduction in the difference of the image intensities.

In the case of shifted edges one can now devise a strategy to quantify horizontal misalignment: since $\Delta_{\max}$ depends only on the ratio $\tfrac\tau\sigma$ one could match the strength of the smoothing with the size of the horizontal shift by choosing $\sigma = \tau$, which gives $\Delta_{\max}^* = 2\Phi(\tfrac12) - 1 \approx 0.38$ independent from $\sigma$ and $\tau$.
Thus, when $\tau$ is not known, two misaligned edges can be smoothed with increasing $\sigma$ until the maximum difference $\Delta_{\max}$ fulfils $\Delta_{\max} \leq \Delta_{\max}^*$ for the first time and the corresponding smoothing factor $\sigma^*$ gives an estimate of $\tau$.

For a sample of $n$ shifted edges, one can replace $\Delta_{\max}$ by a difference of two suitably chosen quantiles:
if the edges are samples of a random edge $I_S$ with $S \sim \mathcal N(0, \tau^2)$
and $p_0, p_1 \in (0, 1)$ with $p_0 < p_1$ (e.g. $p_0 = 0.1$ and $p_1 = 0.9$) are two given probabilities we want to know the resulting range of the $p_0$- and $p_1$-quantiles of the smoothed images at the location with the highest variance in the image intensities, which is at $x=0$ due to symmetry.
As above, we consider the case $\sigma = \tau$, where the strength of the smoothing now matches the standard deviation of the horizontal shifts.
If
\[X = (K_\tau * I_S)(0) = 1 - \Phi\!\left(-\tfrac S\tau\right) = \Phi\!\left(\tfrac S\tau\right)\]
is the (random) intensity value of such a smoothed image at $x=0$ we can calculate the $p_0$- and $p_1$-quantiles of the distribution of $X$ as follows:
since $\tfrac S\tau \sim \mathcal N(0, 1)$ we have
\[\mathbf{P}\!\left(\tfrac S\tau \leq q_k\right) = p_k\]
where $q_k = \Phi^{-1}(p_k)$ for $k=0,1$ by the definition of quantiles for a continuous distribution, which can be equivalently transformed into
\begin{equation}
   \mathbf{P}(X \leq p_k) = p_k \label{eq:xd}
\end{equation}
since $\Phi$ is a monotonically increasing function. Thus the $p_0$- and $p_1$-quantiles of $X$ are just the values $p_0$ and $p_1$ themselves with the range $\Delta_q^* = p_1 - p_0$. Another way of viewing \eqref{eq:xd} is that the distribution of $X$ is just the uniform distribution on the interval $[0, 1]$.

Thus, if a sample $I_{s_1}, \dots, I_{s_n}$ of shifted edges with unknown $\tau$ is smoothed until the range $\Delta_q$ of the empirical $p_0$- and $p_1$-quantiles of $(K_\sigma * I_{s_1})(0), \dots, (K_\sigma * I_{s_n})(0)$ satisfies
\begin{equation}
\Delta_q \leq \Delta_q^* = p_1 - p_0 \label{eq:qc}
\end{equation}
the corresponding smoothing factor $\sigma^*$ gives an estimate of the standard deviation $\tau$ and therefore quantifies the horizontal misalignment. This setting is visualized in Figure \ref{fig:edges_sample}.

Instead of the quantile range $\Delta_q^*$ one could also use the standard deviation $\sqrt{\mathbf{Var}\,X} = \tfrac1{\sqrt{12}}$ (variance of uniform distribution) as a criterion for when to stop the smoothing, but for real images quantiles are preferred due to a higher robustness against outliers.

\subsection{Application to real images}

\begin{figure}[ht!]
   \centering
   \includegraphics[width=1\linewidth]{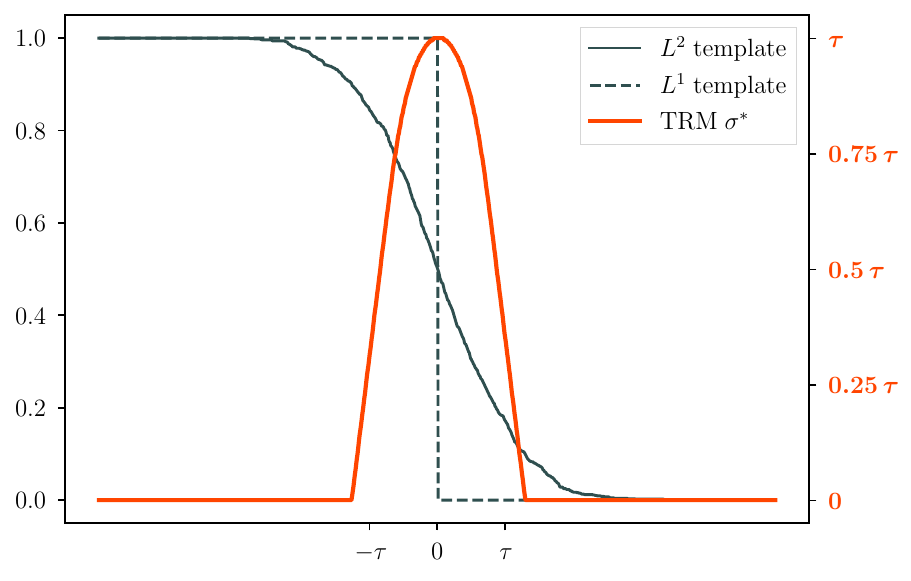}
   \caption{Based on the (sharp) edge samples from Figure \ref{fig:edges_sample} the smoothing bandwidth $\sigma^*(x)$ satisfying condition \eqref{eq:qc} is shown at each location $x$ together with the templates for the $L^2$- and $L^1$-norm (here given by the pixelwise mean and median of the sample images).}
   \label{fig:edges_sample_sig}
\end{figure}

In the last section the sample of edges varied around the location $x=0$ and the quantile condition \eqref{eq:qc} was only evaluated at this single location.
For real images one can now determine the smallest bandwidth $\sigma^*$ at every location $x \in \Omega$ satisfying condition \eqref{eq:qc} at $x$. In Figure \ref{fig:edges_sample_sig} this is visualized for the sample of shifted edges, where the orange line is the corresponding $\sigma^*$ as a function of the location $x$.
This is useful as a visual way of quantifying the horizontal variability of a sample of images in a localised manner, especially when the dimension of $\Omega$ is larger than one.
Here, the computed $\sigma^*$ values are zero wherever most samples agree on having a plateau of height zero or one and the maximum at $x=0$ corresponds to the resolution of the edge, all other values are to be understood to continuously interpolate between these two cases.
For the simultaneously generated template
the values $\sigma^*(x)$ can be interpreted as the resolution of the template at the location $x$. Therefore $\sigma^*$ constitutes the desired template resolution measure (TRM).

Natural images, especially in more than one dimension, deviate from the model situation above, where idealized one-dimensional edges were shifted based on a Gaussian distribution, in a number of ways: edges might not be perfectly sharp; their horizontal misalignment can be caused by more complicated transformations like rotations, shearings or even non-linear deformations; and even in the translation case the Gaussian distribution assumption might be inappropriate. Additionally, edges in real images usually belong to plateaus of different heights.

Despite these challenges we use the model situation as a rule of thumb and augment the quantile range $\Delta_q^*$ by multiplying it with an additional parameter $\eta$, that can be interpreted as an effective height of the edge for which one wants to quantify the horizontal misalignment. The effective height needs to be hand-tuned to the specific sample of images under consideration, e.g. for the examples in Section \ref{sec:app} the value of $\eta$ is chosen to be about half the intensity range.
The goal range for the reduction in the quantile range is now $\eta \Delta_q^*$, since the inequality in \eqref{eq:xd} is simply scaled with $\eta$.

\begin{figure*}[bt!]
   \begin{minipage}{\textwidth}
   \begin{algorithm}[H]
      \caption{Template resolution measure}\label{alg}
      \begin{algorithmic}
      \renewcommand{\algorithmicrequire}{\textbf{Input:}}
      \renewcommand{\algorithmicensure}{\textbf{Output:}}
      \Require registered images $R_1, \dots, R_n$, effective height $\eta$, threshold probabilities $(p_0, p_1)$, step size $s$
      \Ensure smoothing sigmas $\sigma^*(x)$ per pixel $x\in\Omega$
   
      \State initialize $\sigma \gets 0$, $\sigma^*(x) \gets 0$, $\mathbf M(x) \gets \texttt{false}$ for all $x\in\Omega$
   
      \While {$\mathbf M(x)$ is $\texttt{false}$ for any $x\in\Omega$} \Comment{run until mask $\mathbf M$ is \texttt{true} for all pixels}
         \For{$i = 1$ to $n$}
         \State $S_i \gets \textsc{GaussianFilter}(R_i, \sigma)$ \Comment{create smoothed images with bandwidth $\sigma$}
         \EndFor
         \State $(Q_0, Q_1) \gets$ \textsc{Quantile}$(\{S_1, \dots, S_n\}, \{p_0, p_1\})$ \Comment{pixelwise empirical $p_0$- and $p_1$-quantiles of smoothed images}
         \State $R \gets (Q_1 - Q_0)$ \Comment{pixelwise quantile range}
         \State $\mathbf M_\text{all} \gets (R \leq \eta (p_1 - p_0))$ \Comment{Boolean mask of all pixels with quantile range $\leq \eta (p_1 - p_0)$}
         \State $\mathbf M_\text{new} \gets \mathbf M_\text{all}$ and not $\mathbf M$ \Comment{mask of pixels that fullfill the quantile range condition for the first time}
         \State $\mathbf M \gets \mathbf M$ or $\mathbf M_\text{all}$ \Comment{update mask of all pixels that fullfilled the condition at least once}
         \For{$x \in \Omega$ where $\mathbf M_\text{new}(x)$ is \texttt{true}}
         \State $\sigma^*(x) \gets \sigma$ \Comment{save current $\sigma$ for all pixels fullfilling the quantile range condition for the first time}
         \EndFor
         \State $\sigma \gets \sigma + s$ \Comment{increase $\sigma$ by stepsize $s$}
      \EndWhile
   
      \end{algorithmic}
   \end{algorithm}
   \end{minipage}
\end{figure*}

%------------------------------------------------------------------------
\subsection{Template resolution measure}

Given the registered images $R_i = I_i\circ\Psi_{\hat\theta_i}^{-1}$ from Section \ref{sec:reg} the remarks of the last subsection now turn into the following procedure:
all registered images are progressively smoothed with increasing bandwidth $\sigma$, starting with $\sigma = 0$ (original non-smoothed images).
At every location $x \in \Omega$ in the image domain one now searches for the smallest bandwidth $\sigma^*(x)$, which decreases the difference between the empirical $p_0$- and $p_1$-quantiles of the smoothed images for given $p_0, p_1 \in (0, 1)$ at $x$ below the $\eta \Delta_q^* = \eta (p_1 - p_0)$ threshold.
This is also described with more detail in Algorithm~\ref{alg}.
The resulting TRM image showing $\sigma^*$ (see Figure \ref{fig:mnist_tmp_res} and \ref{fig:brain_tmp_res}) now quantifies the horizontal variation of the images registered with the template at every location of the image domain.

In the examples we assume that all images are naturally extended beyond $\Omega$ by defining the intensity to be zero there. This choice also guarantees the convergence of the above procedure, since such images vanish under Gaussian smoothing for sufficiently large $\sigma$ (similar to the model situation with the point masses) and will eventually fulfil the quantile condition everywhere.

\subsection{Visualization of the template resolution}\label{sec:vis}

\begin{figure}[ht!]
   \centering
   \subfloat{\includegraphics[trim={1.8cm 0.7cm 2cm 1.2cm}, clip, width=0.5\linewidth]{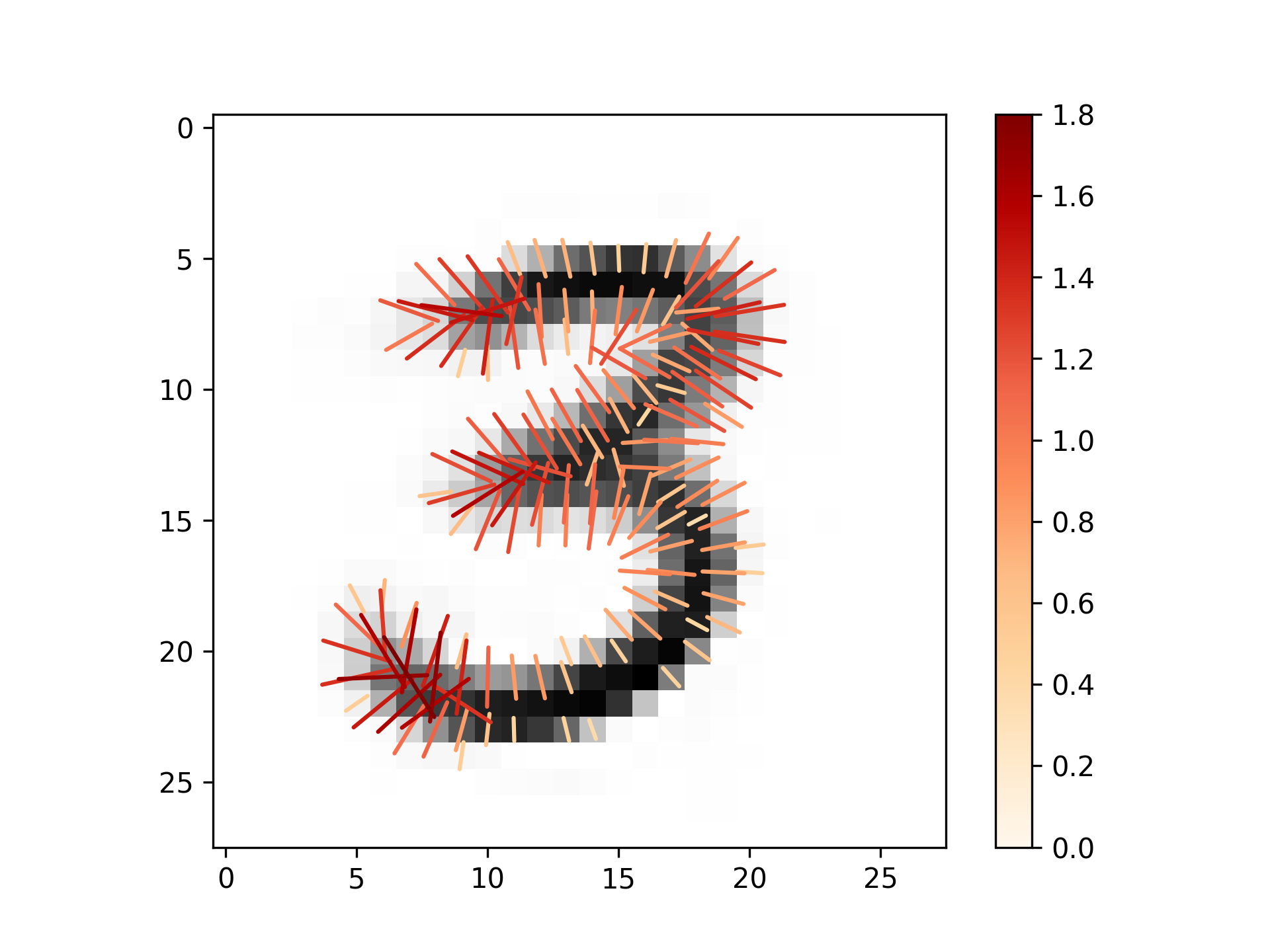}}
   \hfill
   \subfloat{\includegraphics[trim={1.8cm 0.7cm 2cm 1.2cm}, clip, width=0.5\linewidth]{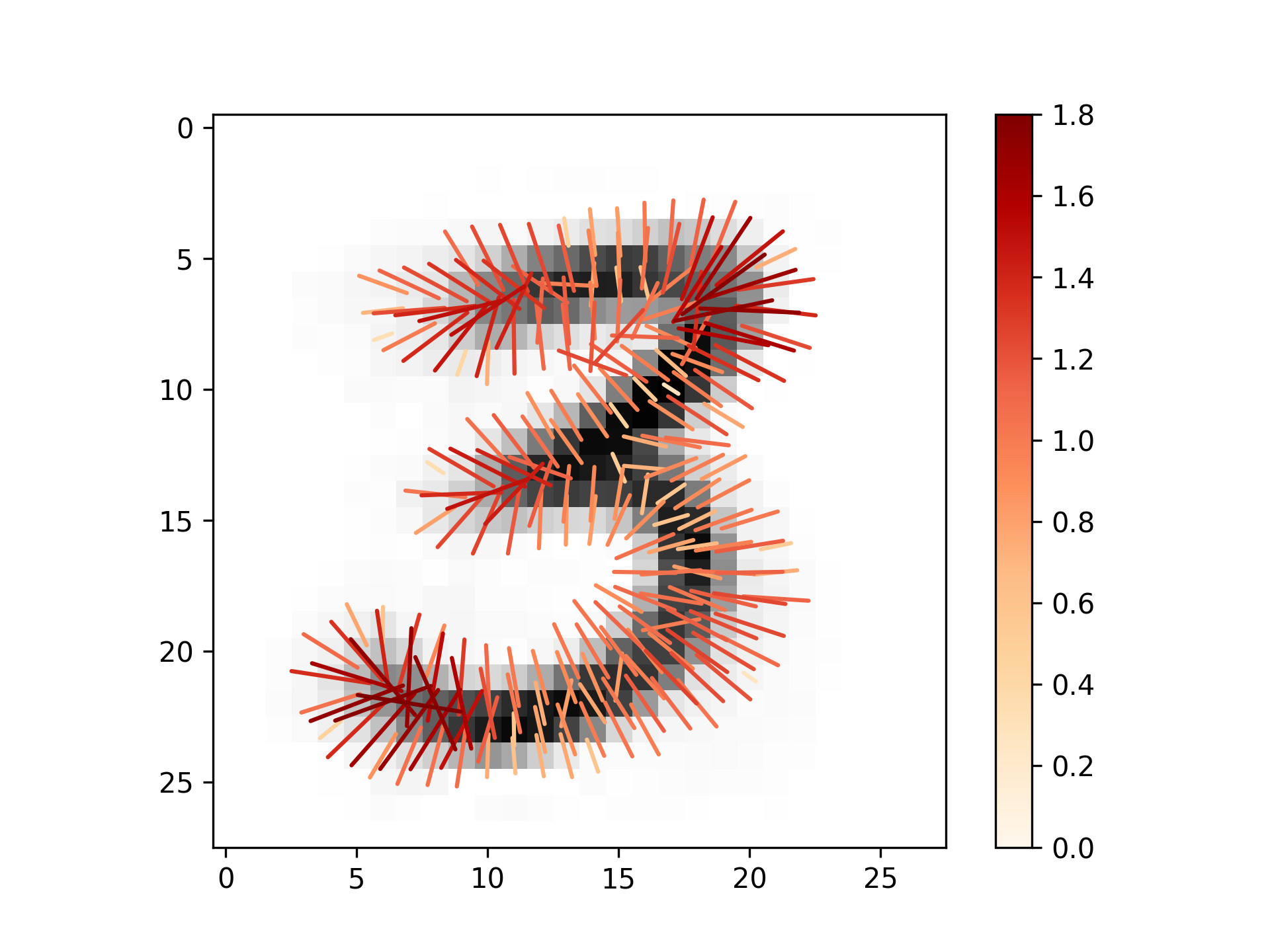}}
   \\
   \subfloat{\includegraphics[trim={1.8cm 0.7cm 2cm 1.2cm}, clip, width=0.5\linewidth]{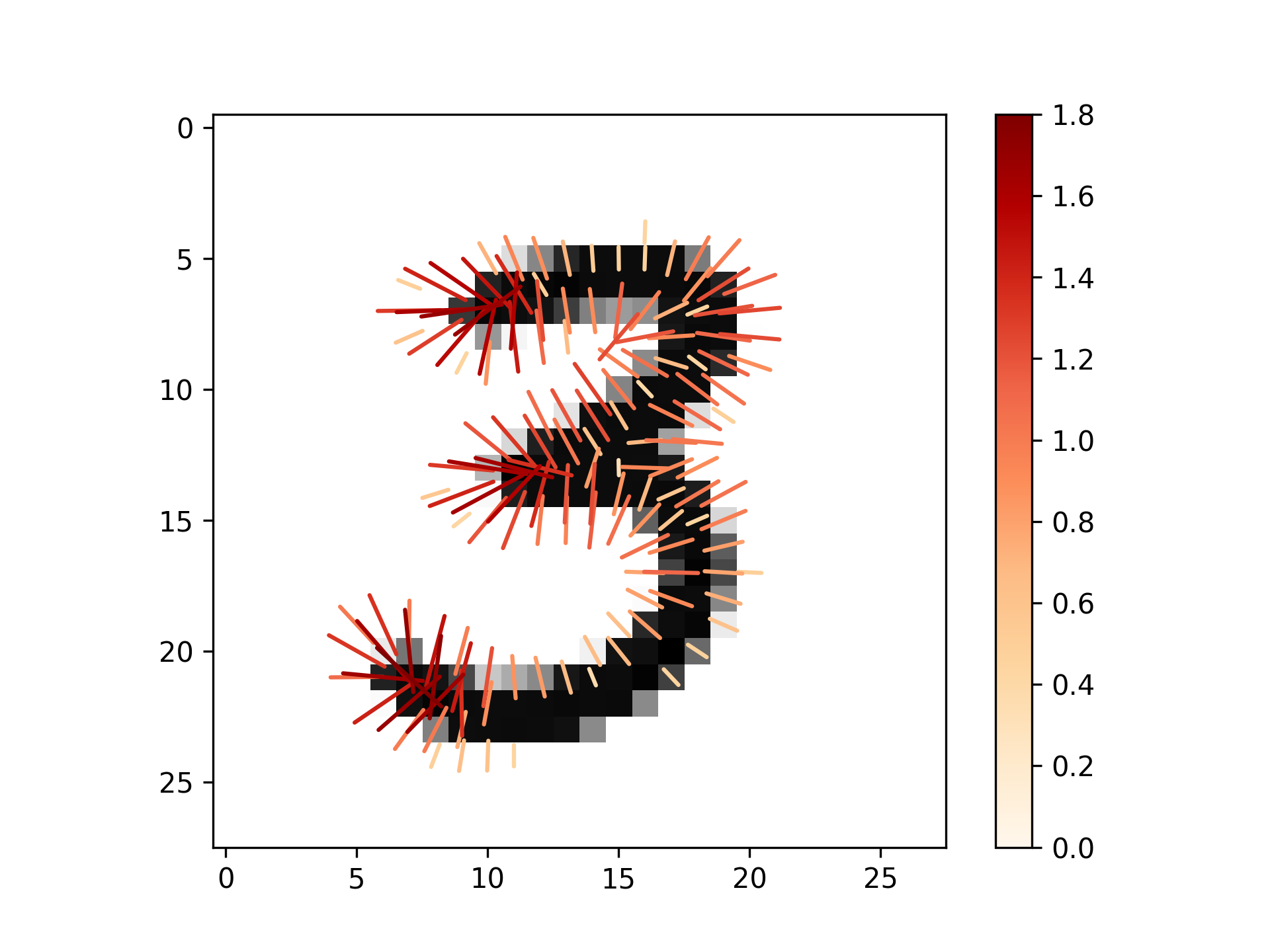}}
   \hfill
   \subfloat{\includegraphics[trim={1.8cm 0.7cm 2cm 1.2cm}, clip, width=0.5\linewidth]{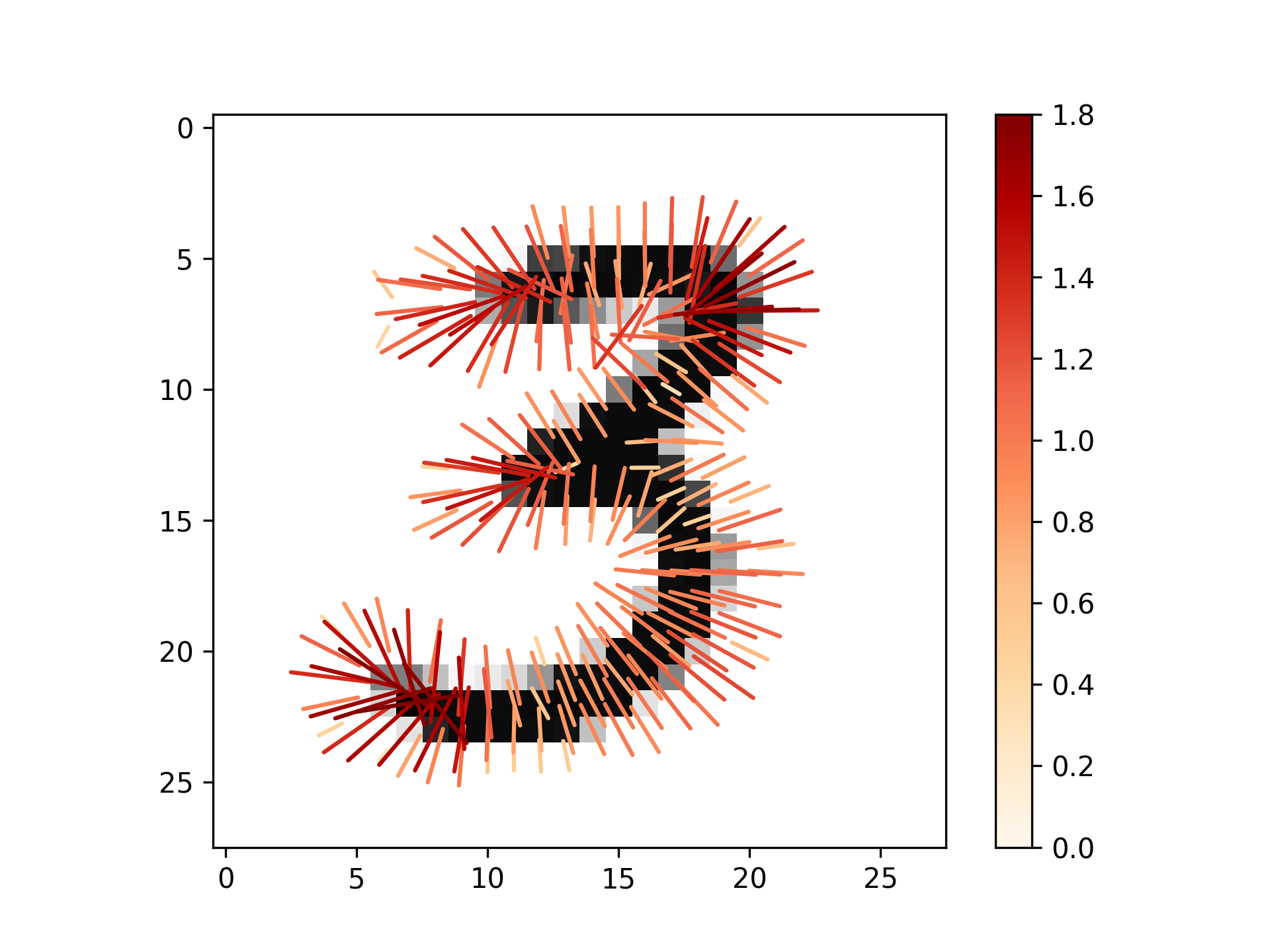}}
   \caption{Visualization of $\sigma^*$ on top of the templates based on the $L^2$-norm (top row) and the $L^1$-norm (bottom row) for affine (left column) and rigid (right column) transformations.}
   \label{fig:mnist_tmp_bars}
\end{figure}

Since the values of the TRM $\sigma^*$ quantify the horizontal resolution of the template $T$, a visualization of the $\sigma^*(x)$ values on top of the template at the corresponding locations is desired. For this, at every pixel location $x \in \Omega$ of the template the gradient $\nabla T(x)$ is estimated (e.g. via a Gaussian derivative filter). The value $\sigma^*(x)$ ist then visualized as a bar of length $2\,\sigma^*(x)$ centred at $x$ along the direction given by $\nabla T(x)$, as shown in Figure \ref{fig:mnist_tmp_bars_0}, \ref{fig:mnist_tmp_bars} and \ref{fig:brain_tmp_bars}. 
When the gradient $\nabla T(x)$ is zero at $x$ no bar is plottet. In the 3d case the bar is projected orthogonally onto the plane of the 2d image slice being visualized.
The gradient direction is a natural choice for displaying the horizontal variation in so far as it is orthogonal to the level sets of the template image and thus the horizontal variation at a point on an edge of the template is also visualized orthogonal to this edge.

%------------------------------------------------------------------------
\section{Experiments}\label{sec:app}

Algorithm \ref{alg} is now used to analyse example datasets in two and three dimensions.
The threshold probabilities $p_0$ and $p_1$ are always chosen as $0.1$ and $0.9$, respectively, i.e., 80\% of the smoothed images are required to deviate from each other by at most 80\% of the effective height $\eta$.

\def\digit{3}

\begin{figure}[ht!]
   \centering
   \subfloat{\includegraphics[trim={1.8cm 0.7cm 2cm 1.2cm}, clip, width=0.5\linewidth]{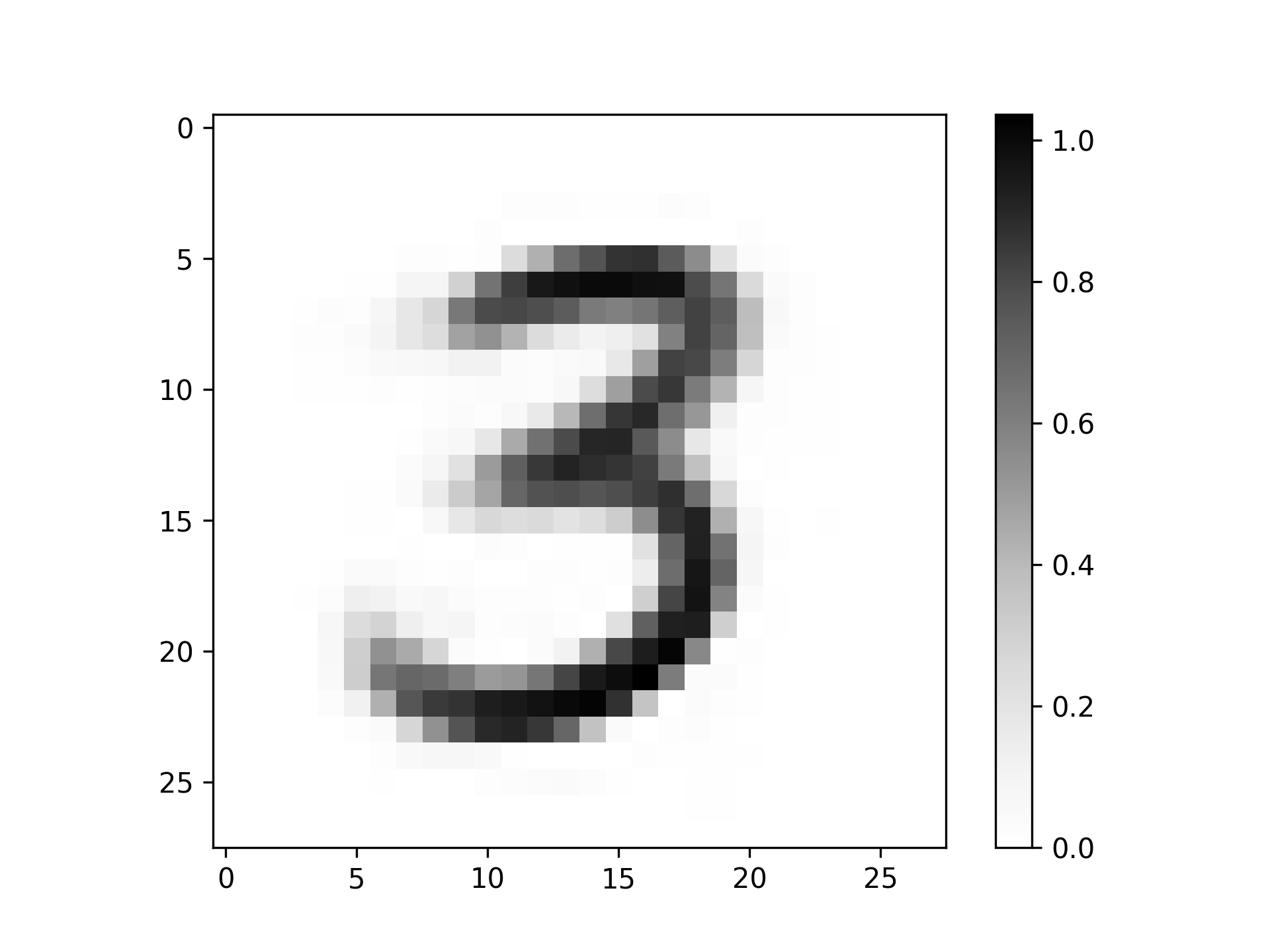}}
   \hfill
   \subfloat{\includegraphics[trim={1.8cm 0.7cm 2cm 1.2cm}, clip, width=0.5\linewidth]{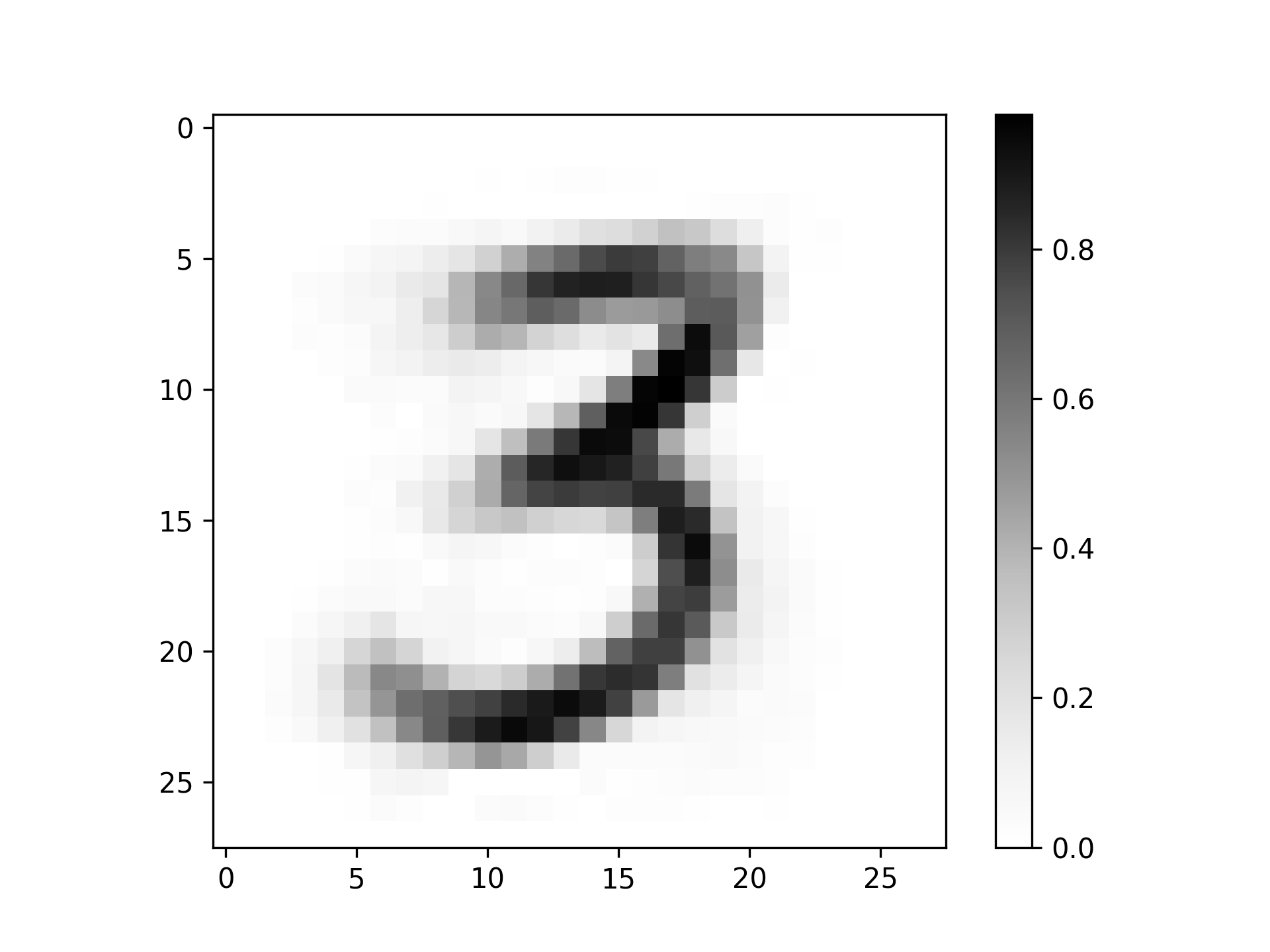}}
   \\
   \subfloat{\includegraphics[trim={1.8cm 0.7cm 2cm 1.2cm}, clip, width=0.5\linewidth]{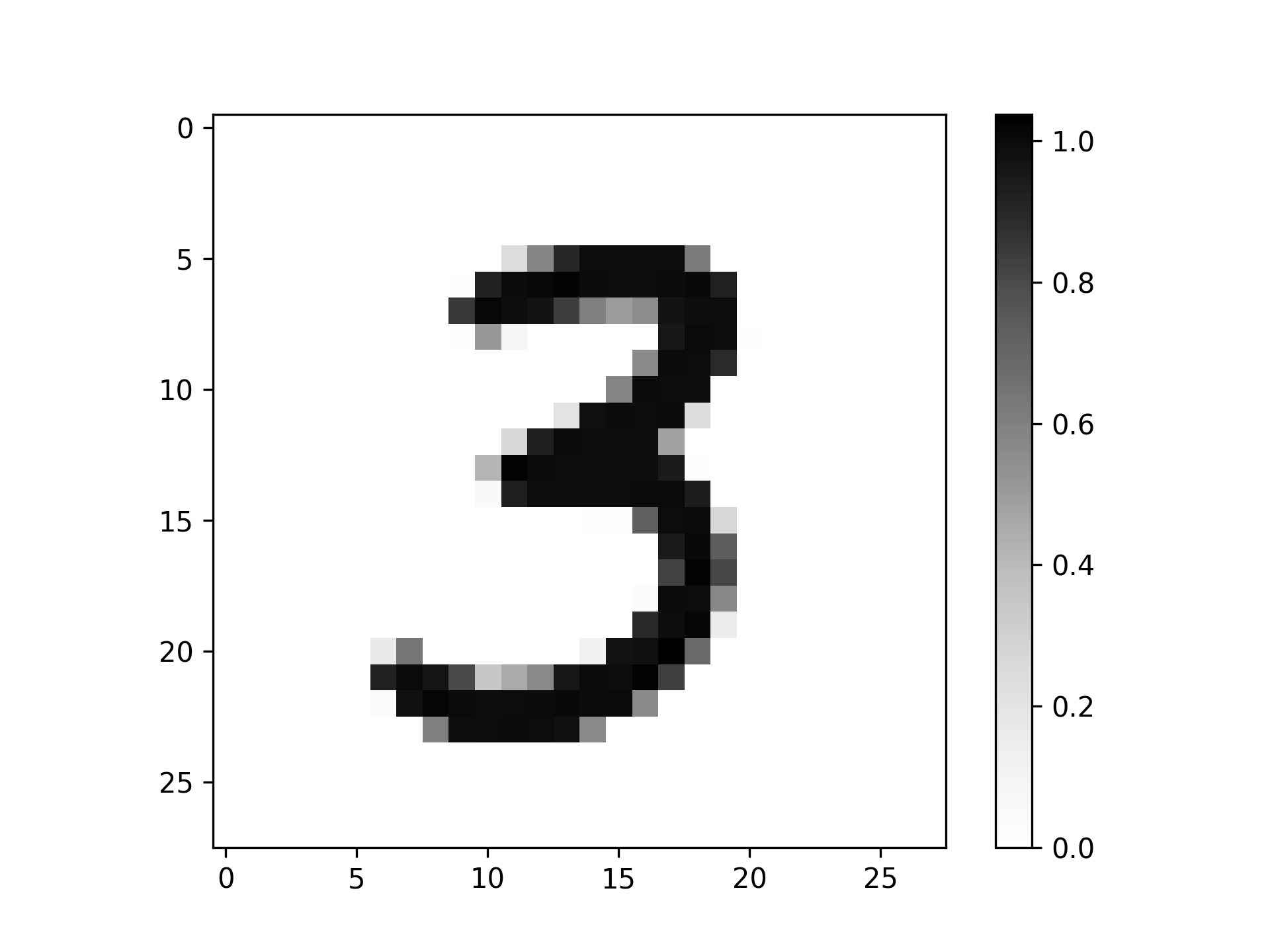}}
   \hfill
   \subfloat{\includegraphics[trim={1.8cm 0.7cm 2cm 1.2cm}, clip, width=0.5\linewidth]{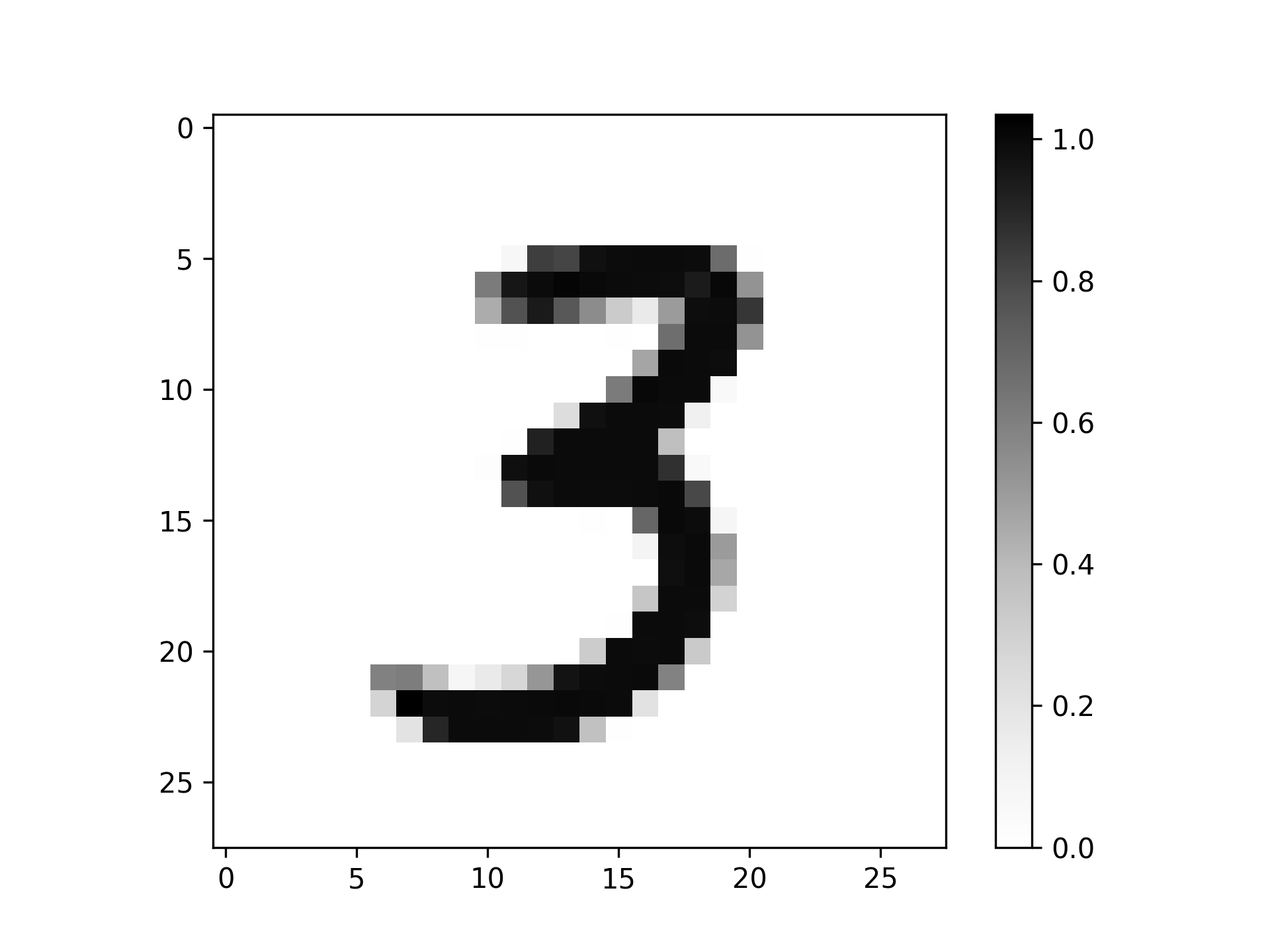}}
   \caption{Templates for the digit \digit{} from the MNIST dataset \citep{mnist_data} based on the $L^2$-norm (top row) and the $L^1$-norm (bottom row) for affine (left column) and rigid (right column) transformations.}
   \label{fig:mnist_tmp}
\end{figure}

\begin{figure}[ht!]
   \centering
   \subfloat{\includegraphics[trim={1.8cm 0.7cm 2cm 1.2cm}, clip, width=0.5\linewidth]{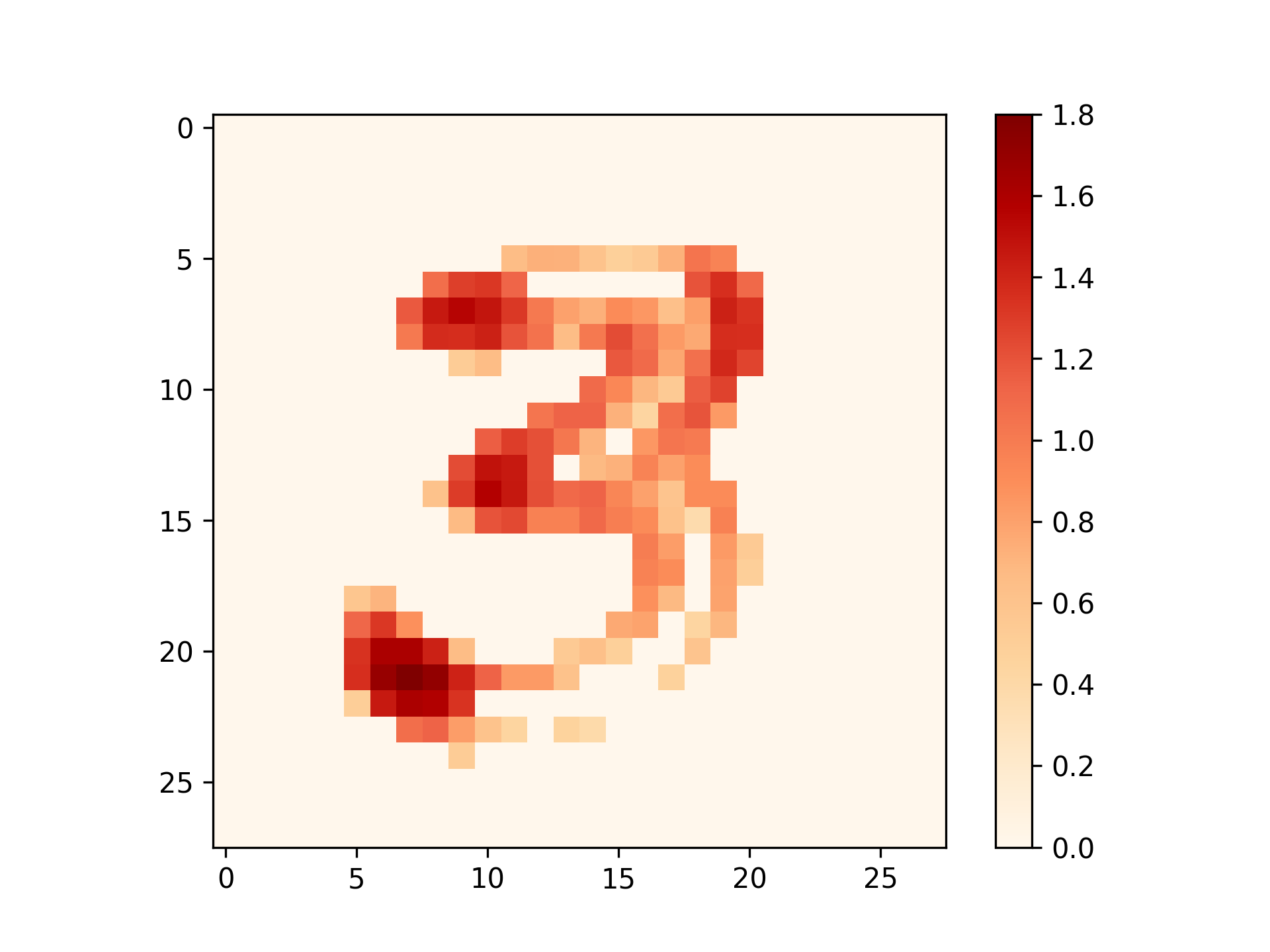}}
   \hfill
   \subfloat{\includegraphics[trim={1.8cm 0.7cm 2cm 1.2cm}, clip, width=0.5\linewidth]{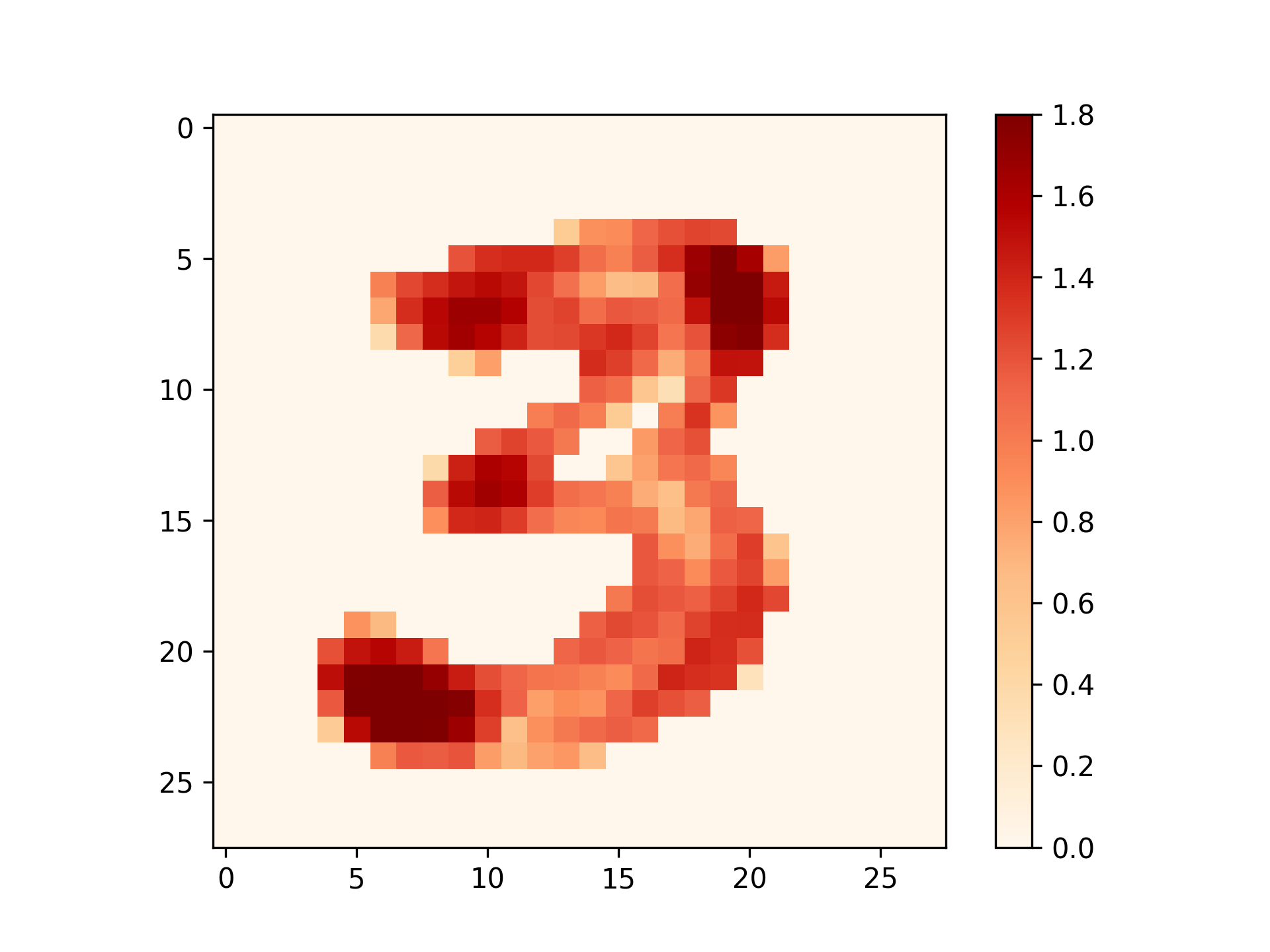}}
   \\
   \subfloat{\includegraphics[trim={1.8cm 0.7cm 2cm 1.2cm}, clip, width=0.5\linewidth]{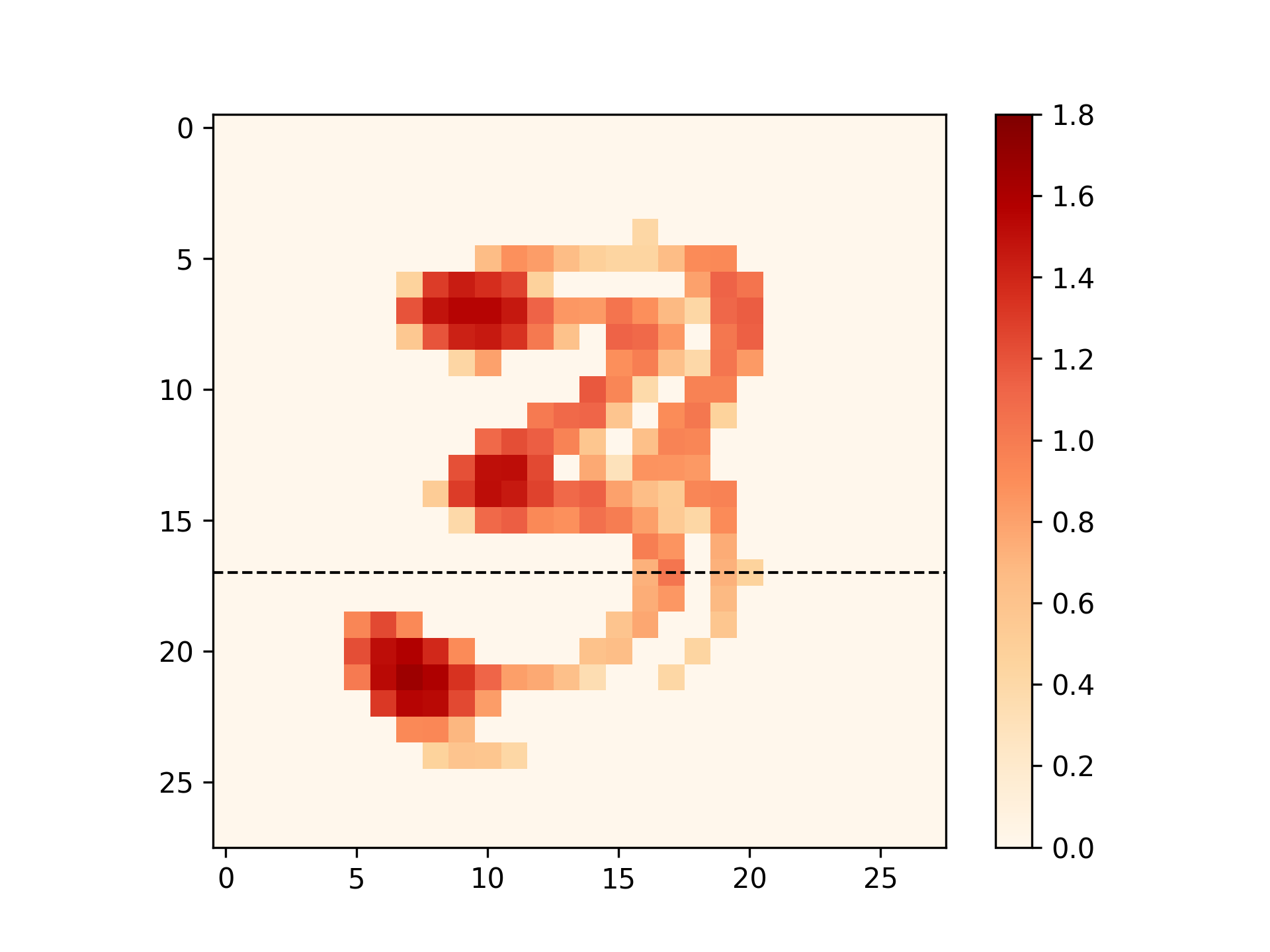}}
   \hfill
   \subfloat{\includegraphics[trim={1.8cm 0.7cm 2cm 1.2cm}, clip, width=0.5\linewidth]{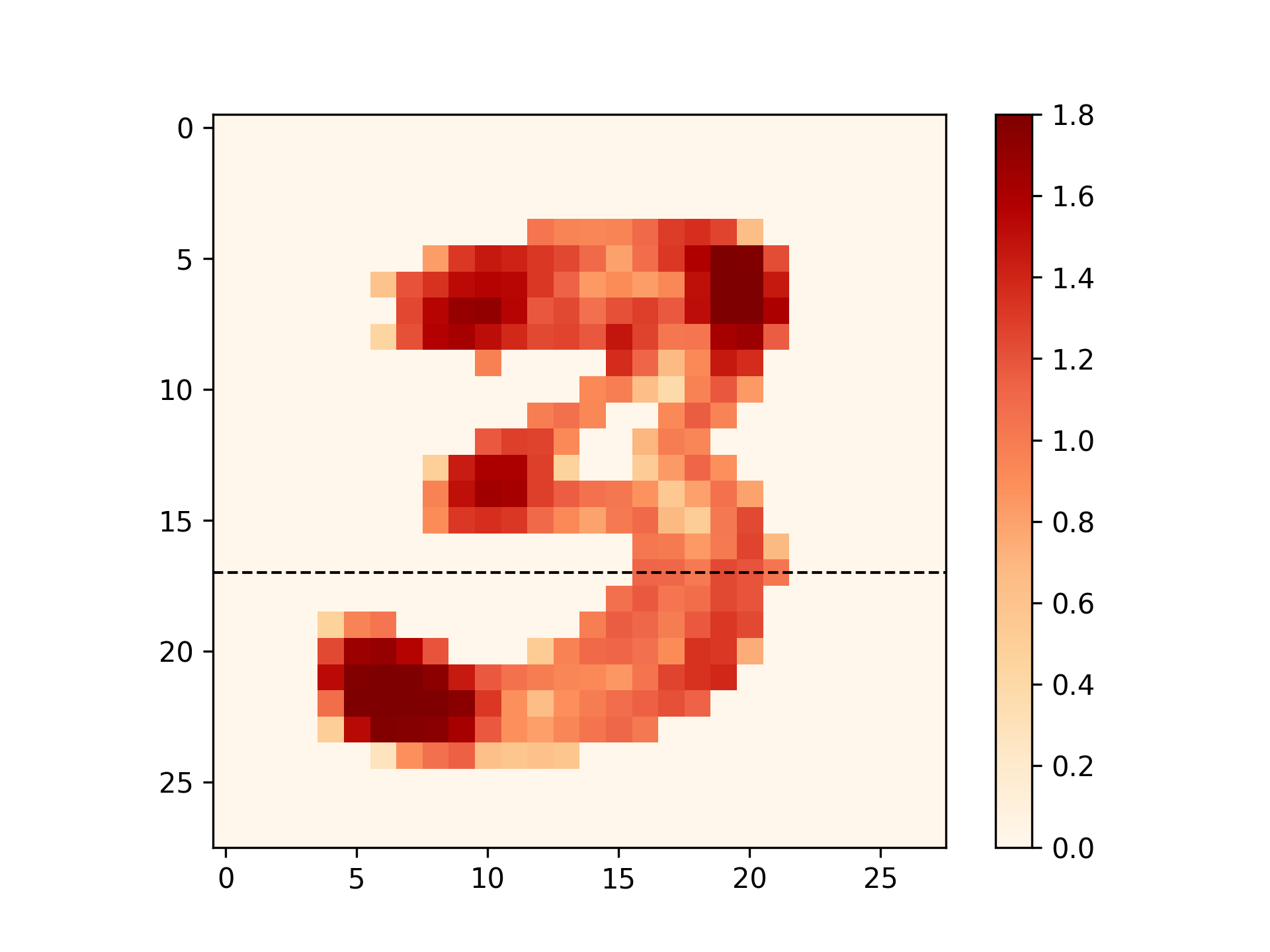}}
   \caption{Template resolution measure $\sigma^*$ based on the $L^2$-norm (top row) and the $L^1$-norm (bottom row) for affine (left column) and rigid (right column) transformations. Figure \ref{fig:mnist_slice} shows 1d slices (dashed lines in the bottom row) for the $L^1$ case.}
   \label{fig:mnist_tmp_res}
\end{figure}

\subsection{2d digits}\label{sec:app_2d}

First, the classic MNIST dataset \citep{mnist_data} is used to demonstrate the method on a relatively simple collection of 2d images.
The image intensities are all normalized to the range $[0, 1]$ and the effective height $\eta \approx 0.6$, i.e. 60\% of the intensity range, is chosen such that the resulting visualization in Figure \ref{fig:mnist_tmp_bars} is not too cluttered.

For the first $100$ samples of each type of digit in the dataset a groupwise registration and template generation was performed.
Here the images of the digit \digit{} are registered with affine and rigid transformations and with the $L^2$- and $L^1$-norm as similarity metrics. The resulting templates are shown in Figure \ref{fig:mnist_tmp}, where one can see that the $L^1$-norm templates appear sharper than the corresponding $L^2$-norm versions, but the template resolution measure given by Algorithm \ref{alg} shows that the horizontal variability of the registered digits is actually similar in both cases. The TRM $\sigma^*$ is shown in Figure \ref{fig:mnist_tmp_res} and the visualization proposed in Section \ref{sec:vis} is shown on top of the templates in Figure \ref{fig:mnist_tmp_bars}. As expected, the affine transformations lead to a better registration than the rigid ones. Here we can see that the horizontal variation quantified by $\sigma^*$ is in accordance with what one would expect in the case of handwritten digits, in particular the position of the end pieces (where the pen is first put down or lifted) and the corner where the arcs meet are highly uncertain in contrast to the top of the upper arc and the curving part of the bottom arc (which are the features the digits are aligned on during the registration).

% 1d slices (l1 rigid vs affine)
\begin{figure}[t!]
   \centering
   \includegraphics[width=\linewidth]{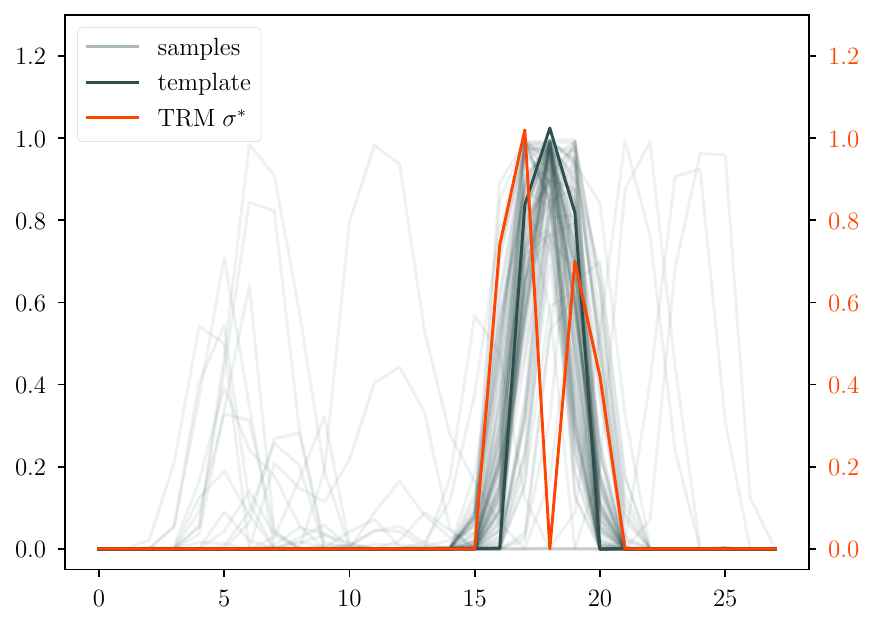}
   \\
   \includegraphics[width=\linewidth]{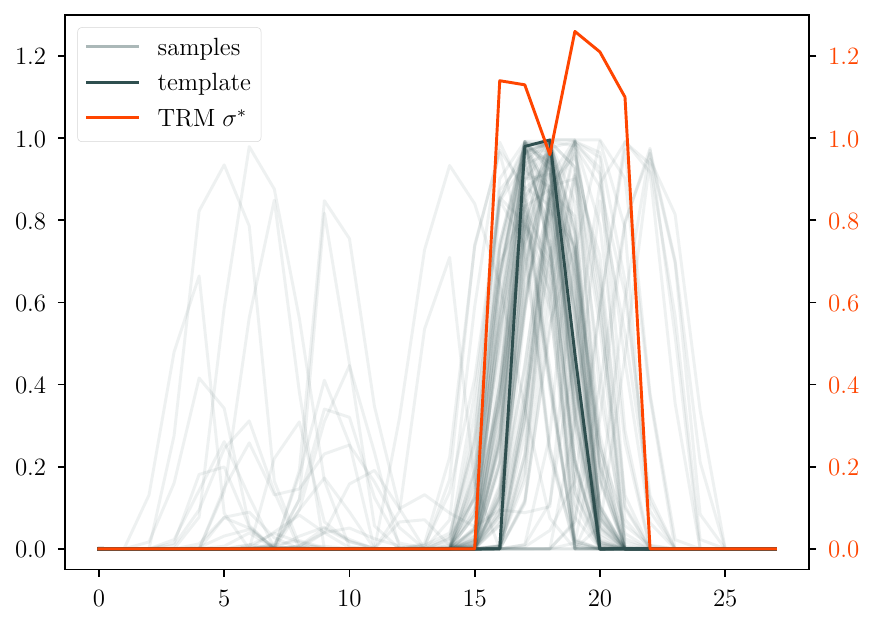}
   \caption{One dimensional slices of $\sigma^*$ (along the dashed lines in Figure \ref{fig:mnist_tmp_res}) for the template based on the $L^1$-norm and affine (top) or rigid (bottom) transformations. The corresponding templates and registered samples are also shown.}
   \label{fig:mnist_slice}
\end{figure}

In Figure \ref{fig:mnist_slice} one dimensional slices (along the dashed lines in Figure \ref{fig:mnist_tmp_res}) through the resolution measures (together with the templates and the registered sample images) are shown for the $L^1$-norm (slices for the $L^2$-norm look similar). Here one can see that the misalignments quantified by $\sigma^*$ are in good correspondence with the visible misalignments of the sample images, e.g. the samples are less aligned with the right slope of the rigid template (bottom) than in the affine version (top), and the size of the misalignment is consistent with the value of $\sigma^*$ at that location.

The digit dataset was also used to examine the robustness of the resolution measure to changes in the probability thresholds $p_0$ and $p_1$, which showed that a quantile range ($p_1 - p_0$) between 0.7 and 0.9 gives qualitatively similar results as the choice $p_0 = 0.1$ and $p_1 = 0.9$ (data not shown).

\begin{figure}[htb!]
   \centering
   \includegraphics[trim={1.8cm 0.7cm 2cm 1.2cm}, clip, width=\linewidth]{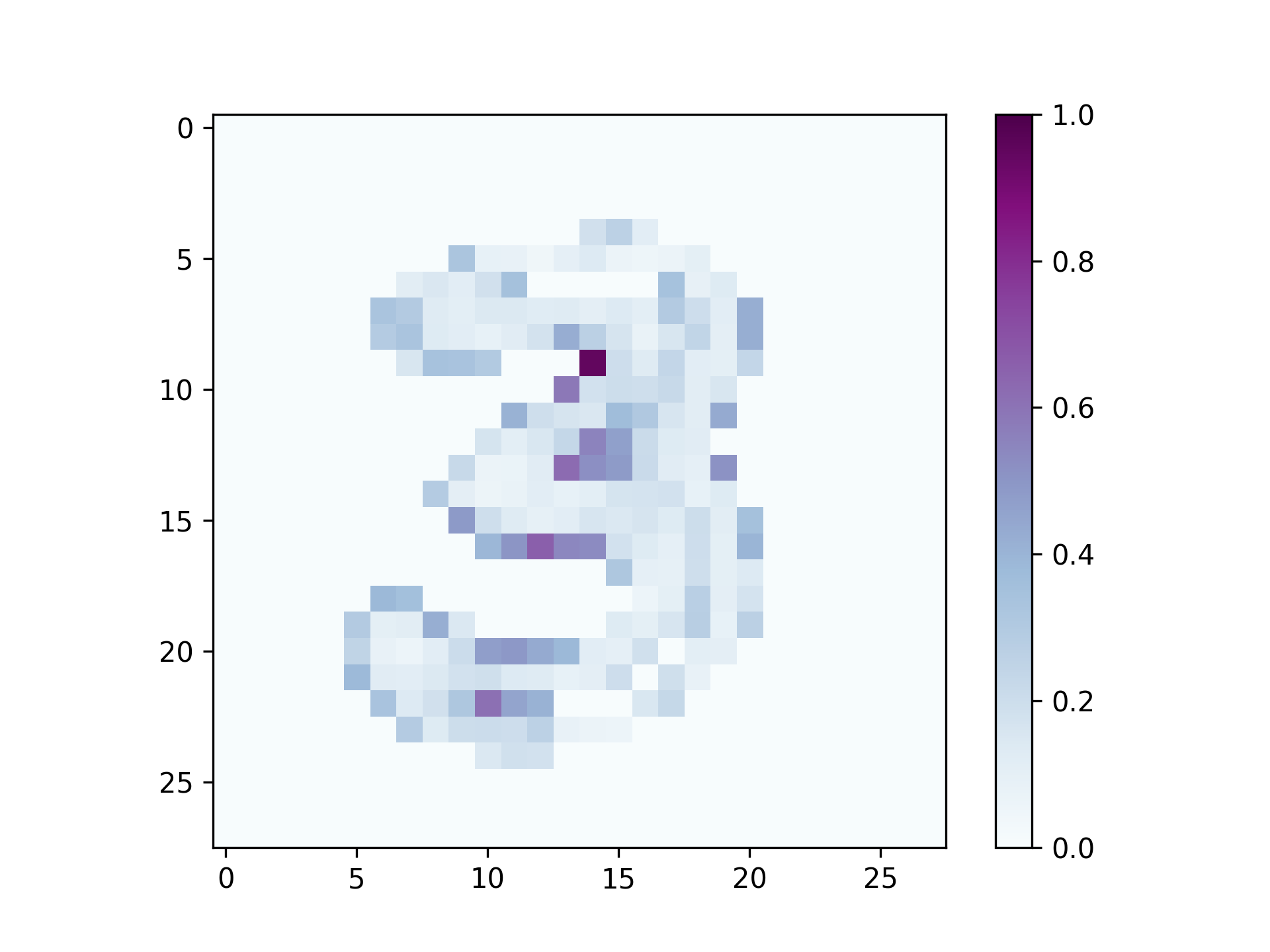}
   \caption{Estimated standard deviation of the TRM for the digit 3 based on the $L^2$-norm and affine transformations.}
   \label{fig:mnist_std}
\end{figure}

\begin{figure*}[hbt!]
   \centering
   \subfloat{\includegraphics[trim={1.8cm 0.7cm 2cm 1.2cm}, clip, width=0.2\linewidth]{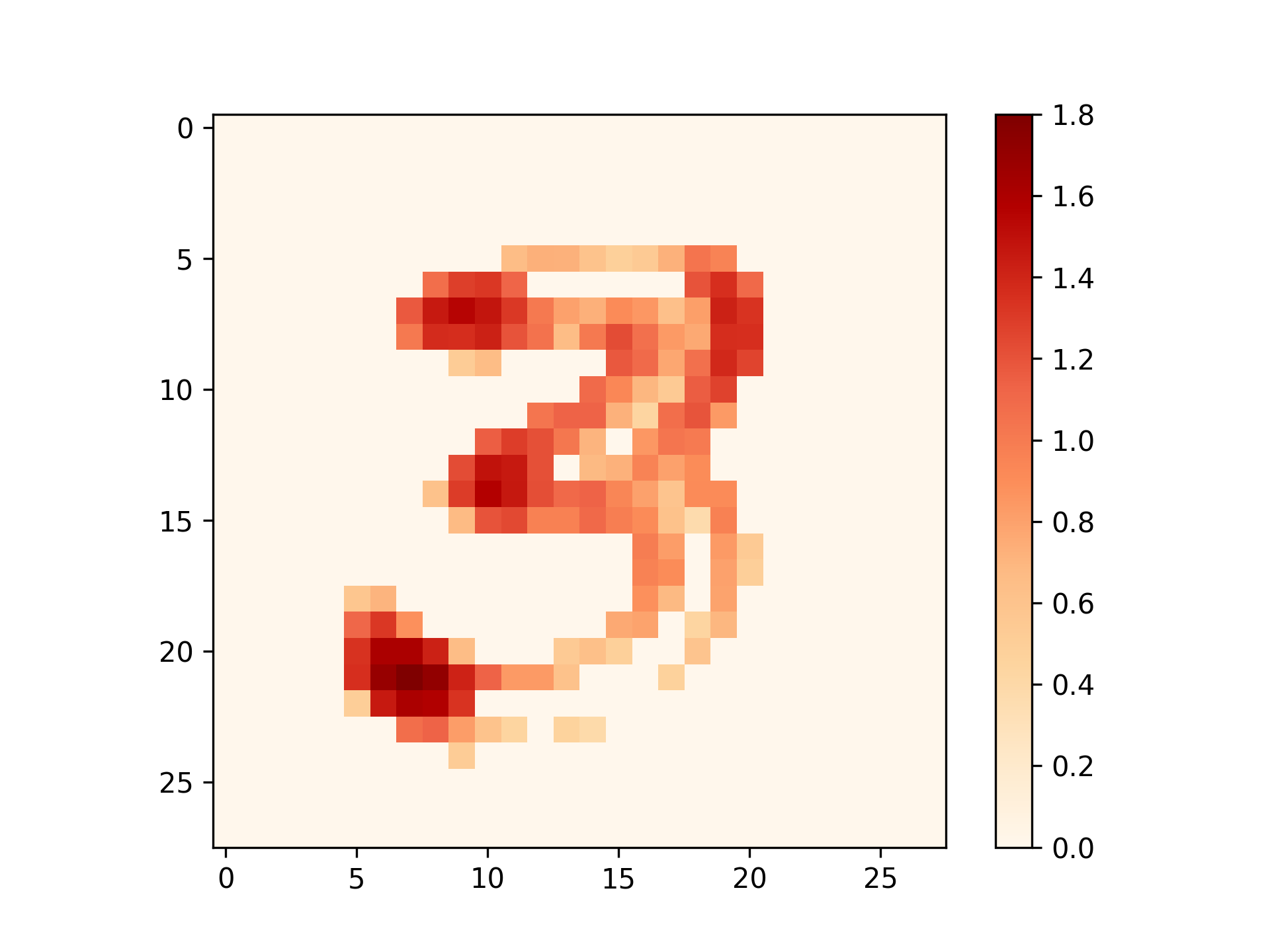}}
   \hfill
   \subfloat{\includegraphics[trim={1.8cm 0.7cm 2cm 1.2cm}, clip, width=0.2\linewidth]{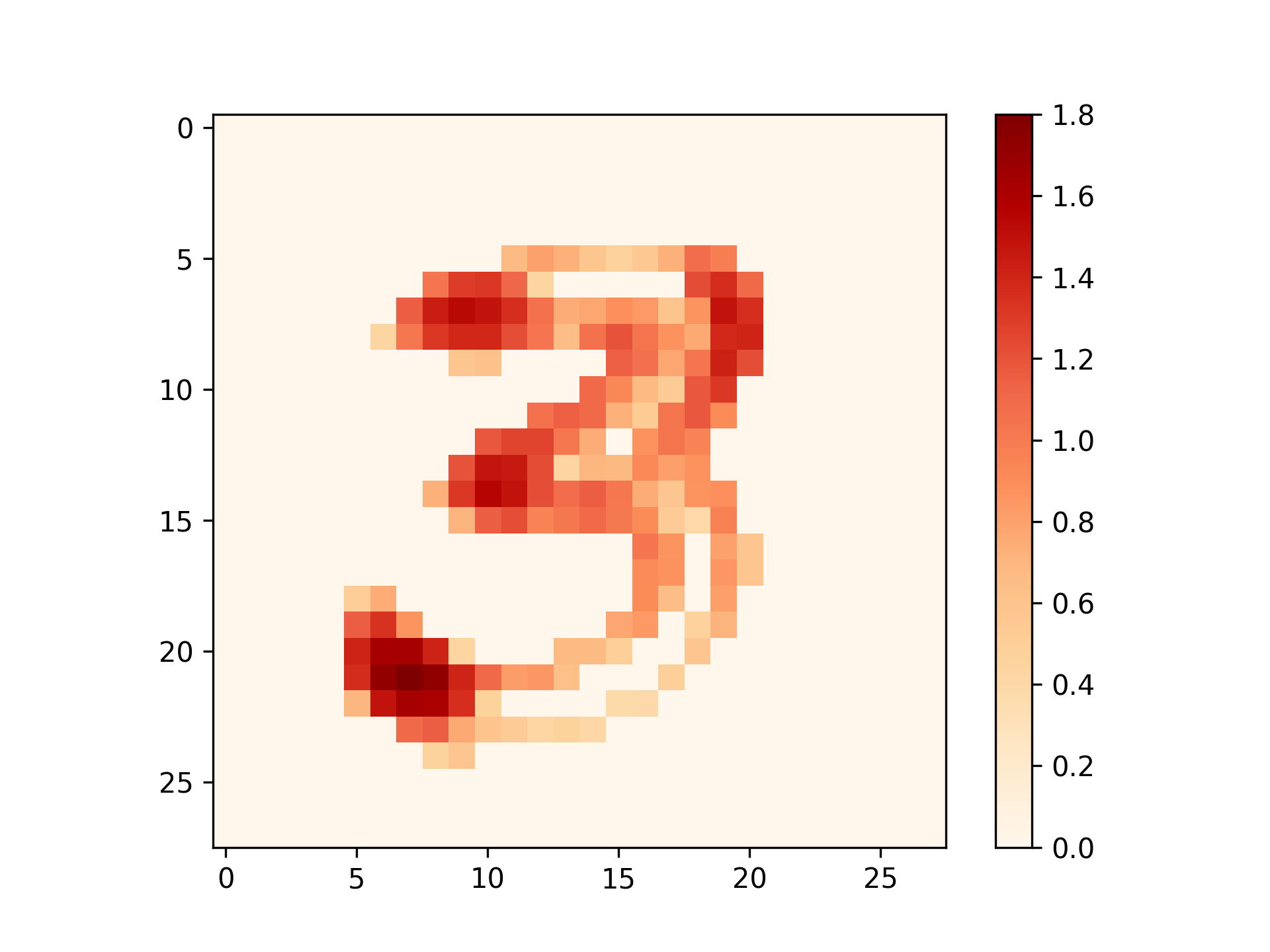}}
   \hfill
   \subfloat{\includegraphics[trim={1.8cm 0.7cm 2cm 1.2cm}, clip, width=0.2\linewidth]{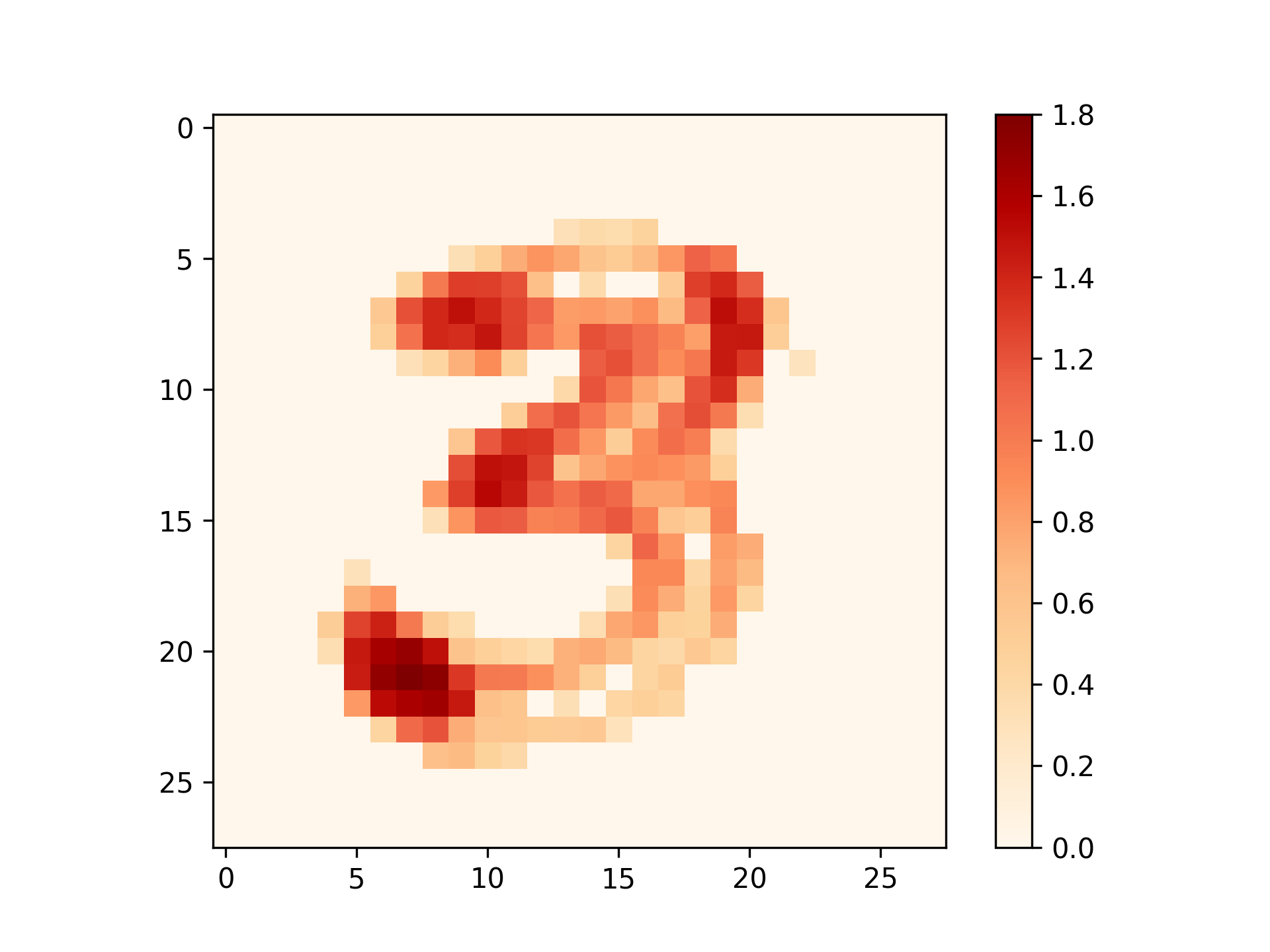}}
   \hfill
   \subfloat{\includegraphics[trim={1.8cm 0.7cm 2cm 1.2cm}, clip, width=0.2\linewidth]{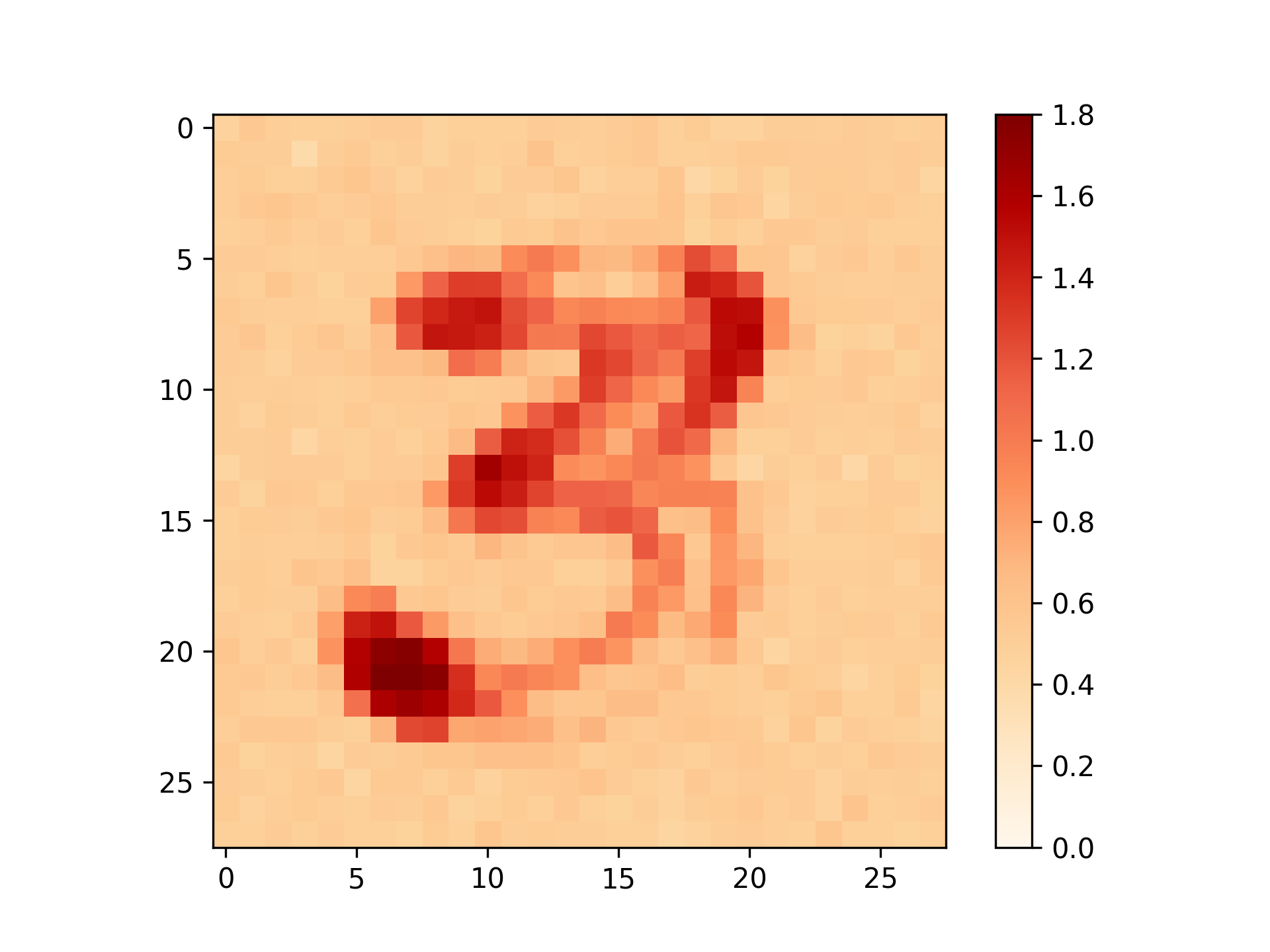}}
   \hfill
   \subfloat{\includegraphics[trim={1.8cm 0.7cm 2cm 1.2cm}, clip, width=0.2\linewidth]{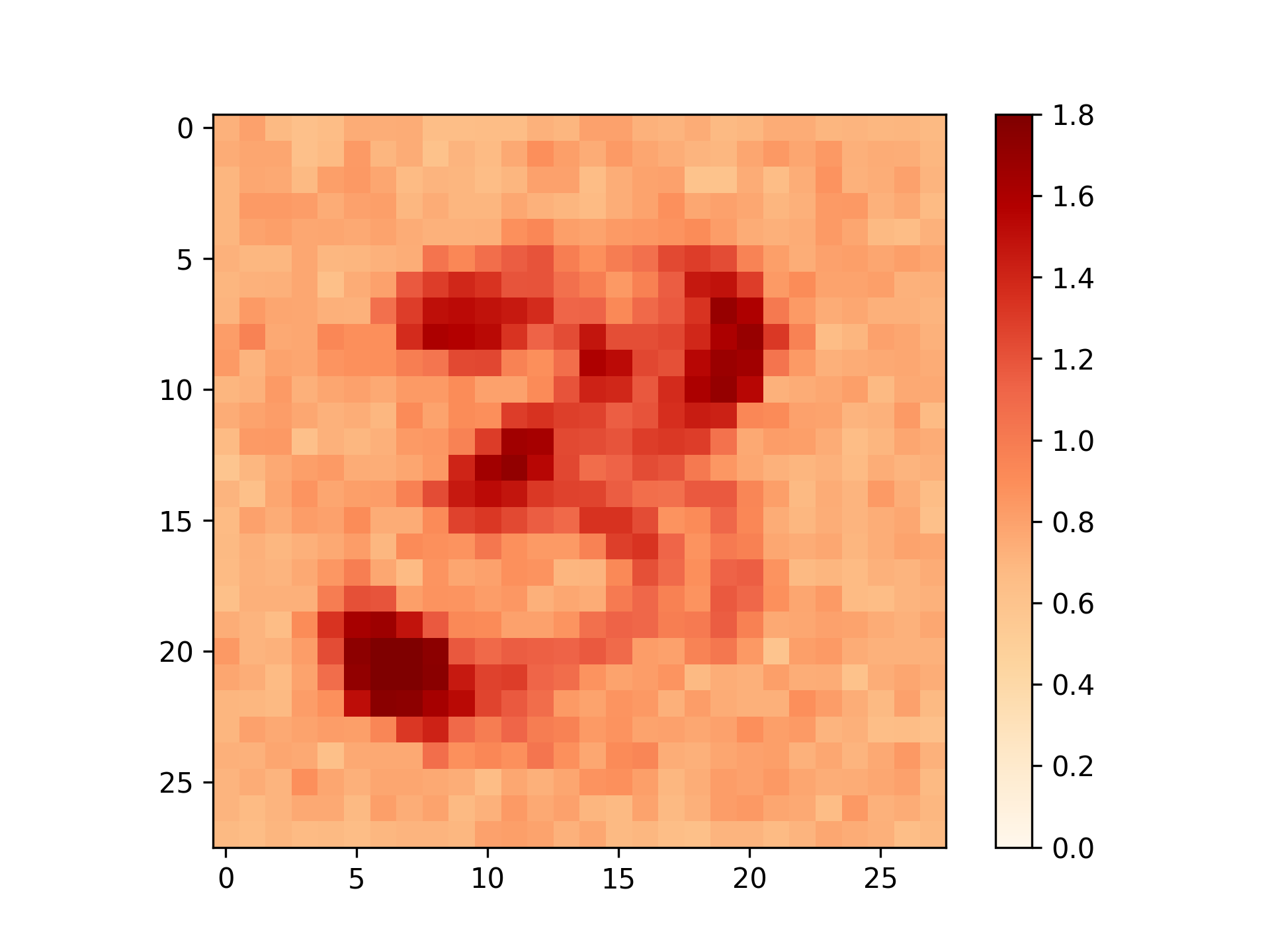}}
   \caption{TRM for noisy images of the digit \digit{} from the MNIST dataset based on the $L^2$-norm and affine transformations. The images are perturbed by additive Gaussian white noise with standard deviation $\sigma_\text{noise} = 0.0$, $0.05$, $0.1$, $0.15$, $0.2$.}
   \label{fig:mnist_noise}
\end{figure*}

%\subsubsection*{Robustness and stability / standard deviation estimation}
\subsubsection*{Robustness and uncertainty quantification}

The TRM is robust to mild noise in the images, since the quantile range is not affected by small perturbations of the image intensities. In Figure \ref{fig:mnist_noise} we show how the TRM changes when the registered 2d digits are perturbed by additive Gaussian white noise with standard deviation $\sigma_\text{noise} = 0.0$, $0.05$, $0.1$, $0.15$, $0.2$. Here we see a gradual decline of the TRM values, though the most prominent features of the TRM remain visible even for strong noise where the images themselves are hardly recognizable anymore.

We can also estimate the standard deviation of the TRM values at every pixel by a subsampling approach. Starting with $n = 1000$ images of the digit 3 of the MNIST dataset we can perform a registration and template estimation once for the full dataset. We then compute the TRM for the full dataset resulting in $\sigma^*$ and for $10$ subsamples of $100$ images each giving $\sigma^*_i$.
%The resulting TRM values $\sigma^*$ for the full sample and $\sigma^*_i$ of the subsamples are then used
These values can then be used to estimate the standard deviation of the TRM at every pixel as
$$\textstyle \text{sd}\,\sigma^* \approx \sqrt{\sum_{i=1}^{10} (\sigma^*_i - \sigma^*)^2}\,.$$
Figure \ref{fig:mnist_std} shows the resulting standard deviation of the TRM for the digit 3 based on the $L^2$-norm and affine transformations. Here we see that the TRM is quite stable, with the standard deviation being below half a pixel for most of the image with only a few locations varying up to one pixel.
In practice, one can always resort to subsampling methods (e.g. bootstrap) to obtain such an estimate of the standard deviation of the TRM in order to get a quantification of the uncertainty of the TRM values themselves.

\begin{figure}[ht!]
   \centering
   \subfloat{\includegraphics[trim={4.1cm 0.7cm 2.1cm 1.2cm}, clip, width=0.495\linewidth]{NFBS_brain_affine_l2_template.png}}
   \hfill
   \subfloat{\includegraphics[trim={4.1cm 0.7cm 2.1cm 1.2cm}, clip, width=0.495\linewidth]{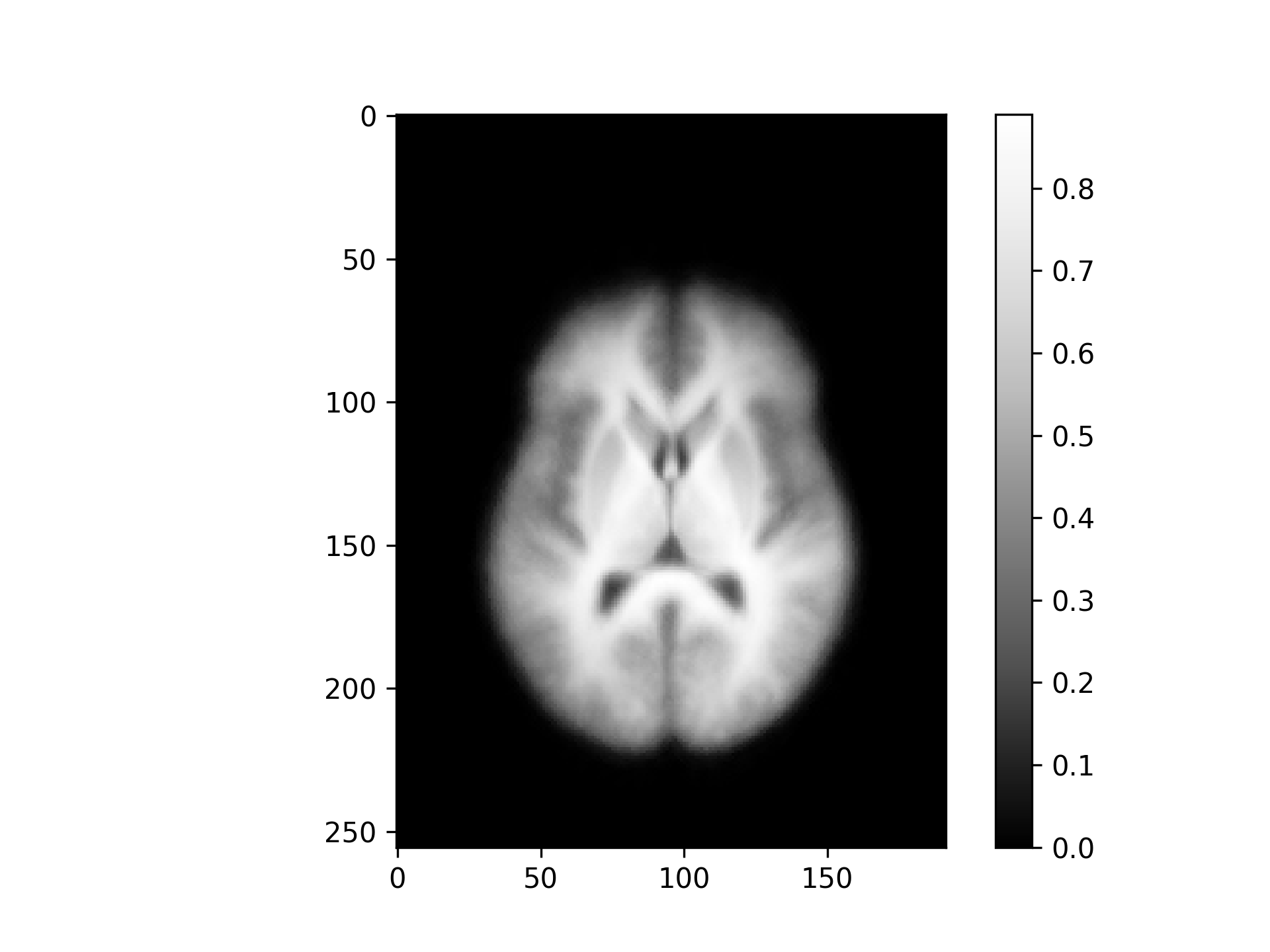}}
   \\
   \subfloat{\includegraphics[trim={4.1cm 0.7cm 2.1cm 1.2cm}, clip, width=0.495\linewidth]{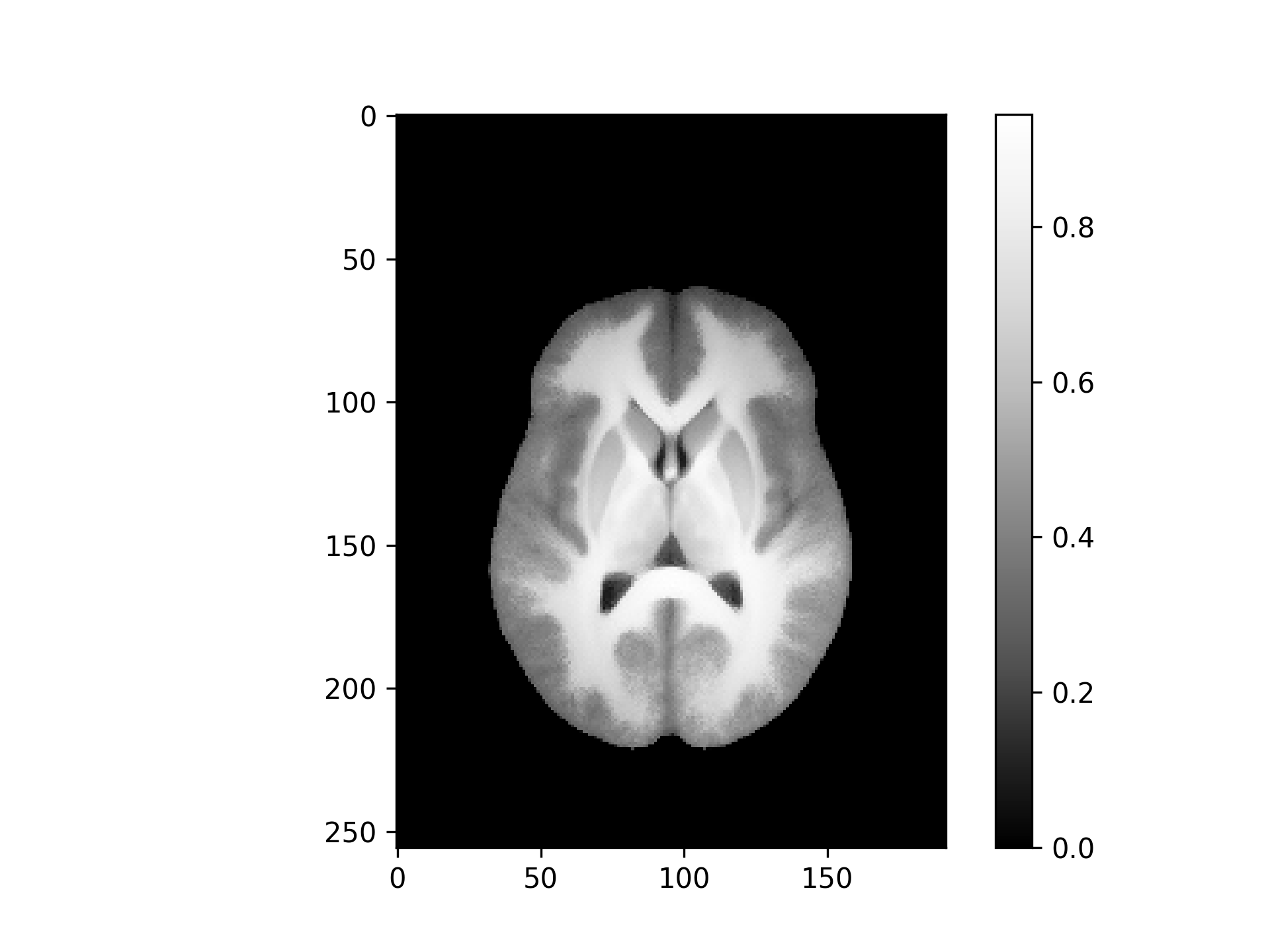}}
   \hfill
   \subfloat{\includegraphics[trim={4.1cm 0.7cm 2.1cm 1.2cm}, clip, width=0.495\linewidth]{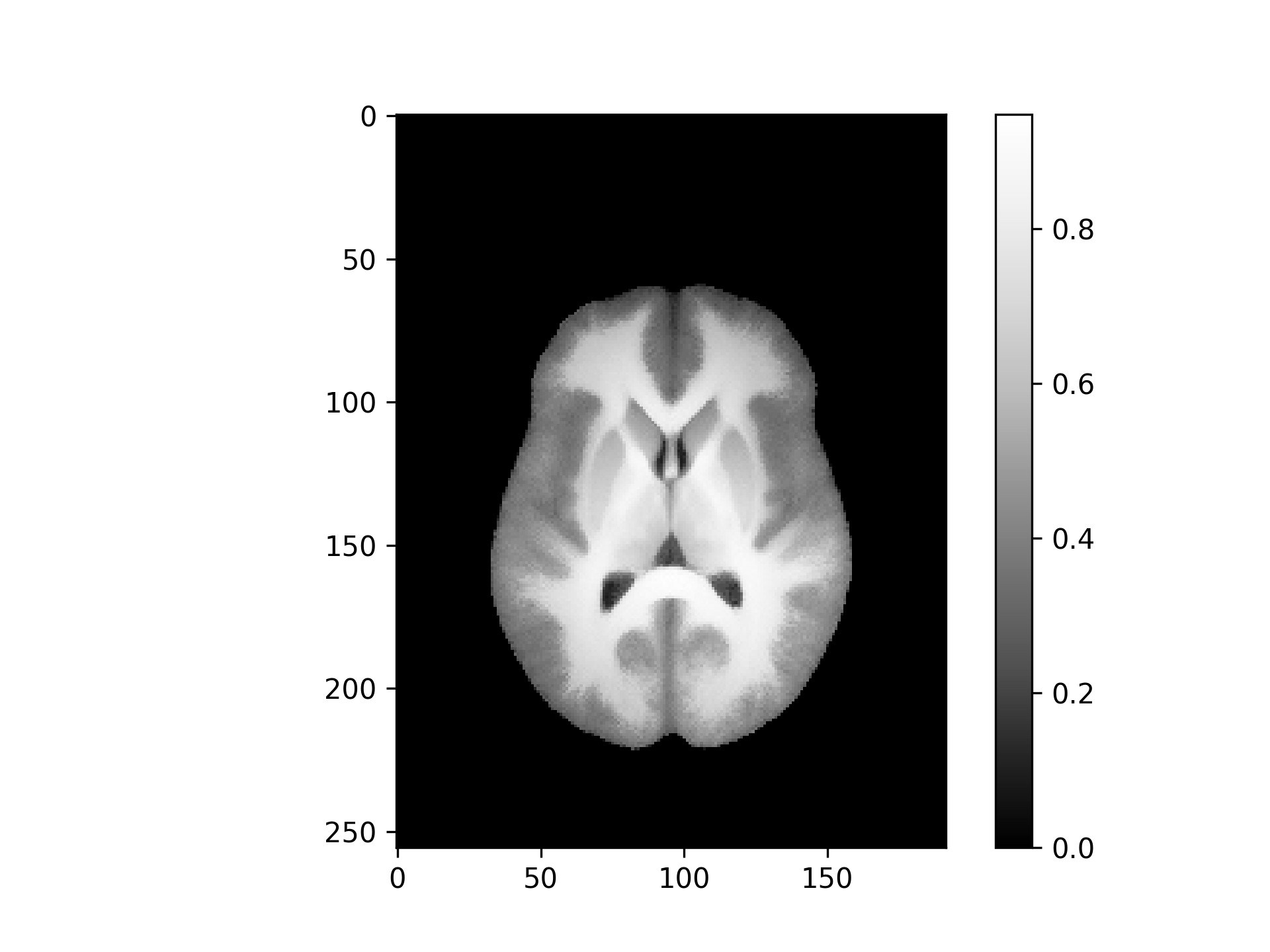}}
   \caption{Horizontal slices of the templates for the 3d NFBS dataset \citep{nfbs_data} based on the $L^2$-norm (top row) and the $L^1$-norm (bottom row) for affine (left column) and rigid (right column) transformations.}
   \label{fig:brain_tmp}
\end{figure}

\begin{figure}[ht!]
   \centering
   \subfloat{\includegraphics[trim={4.1cm 0.7cm 2.1cm 1.2cm}, clip, width=0.495\linewidth]{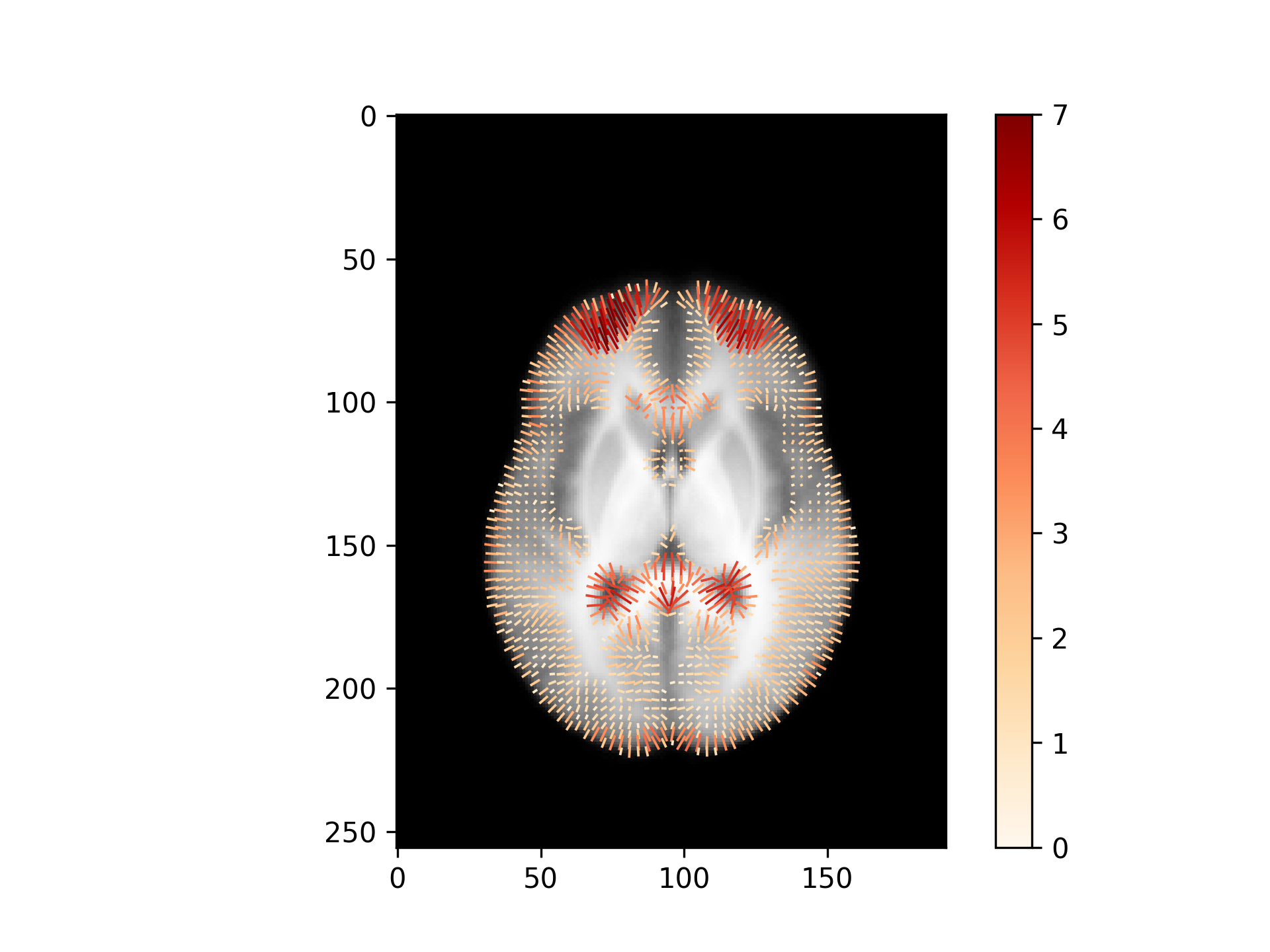}}
   \hfill
   \subfloat{\includegraphics[trim={4.1cm 0.7cm 2.1cm 1.2cm}, clip, width=0.495\linewidth]{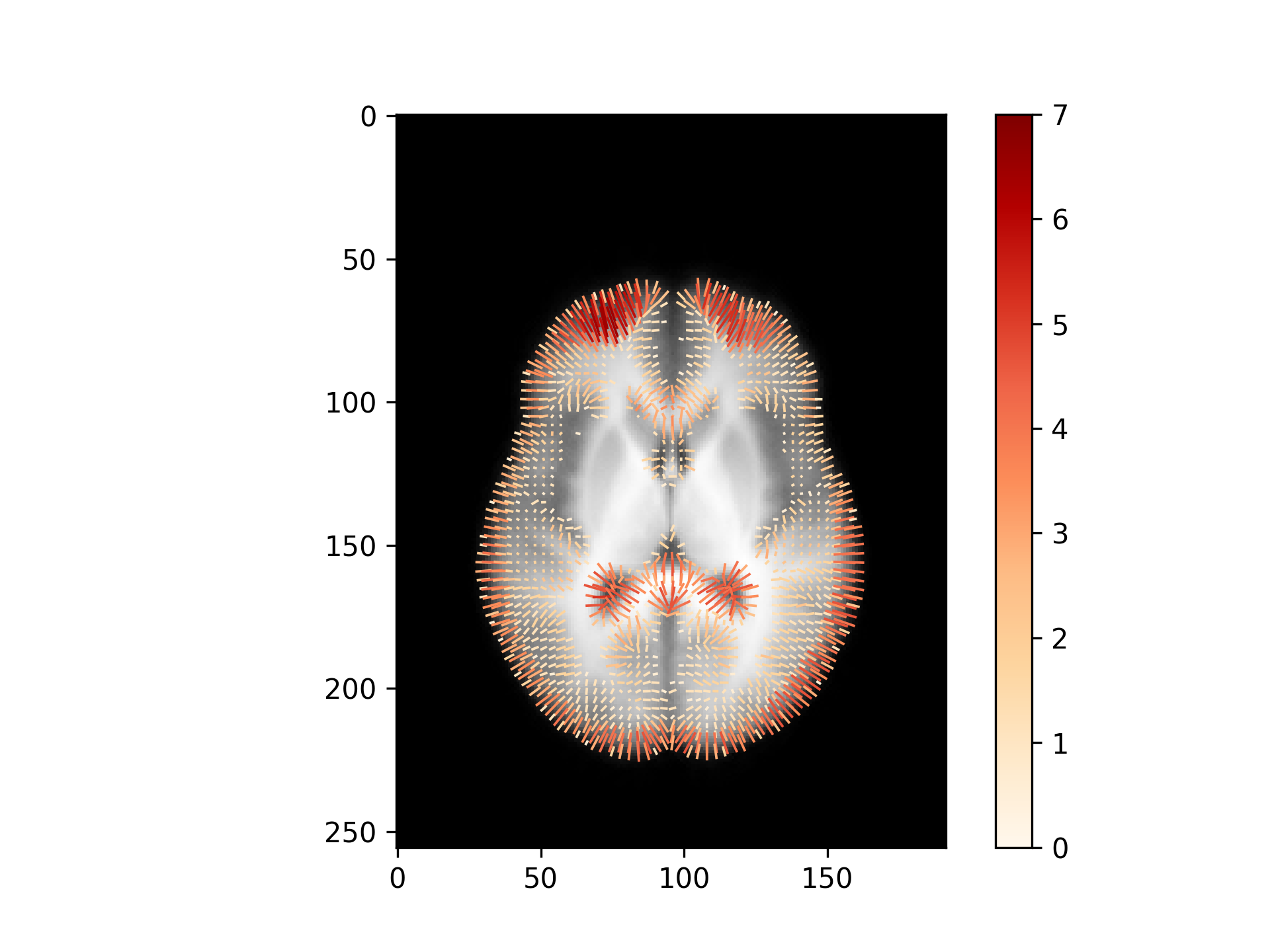}}
   \\
   \subfloat{\includegraphics[trim={4.1cm 0.7cm 2.1cm 1.2cm}, clip, width=0.495\linewidth]{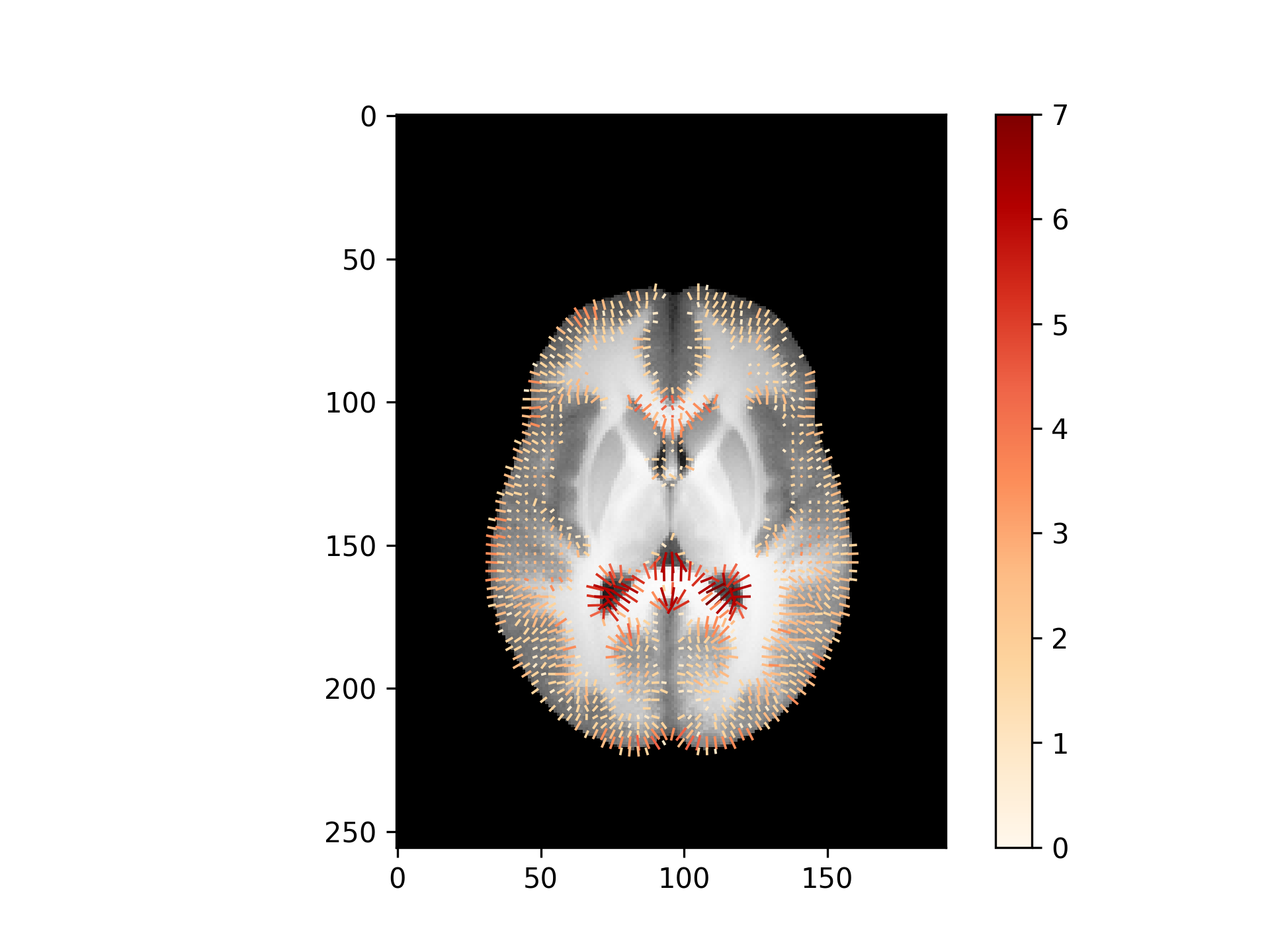}}
   \hfill
   \subfloat{\includegraphics[trim={4.1cm 0.7cm 2.1cm 1.2cm}, clip, width=0.495\linewidth]{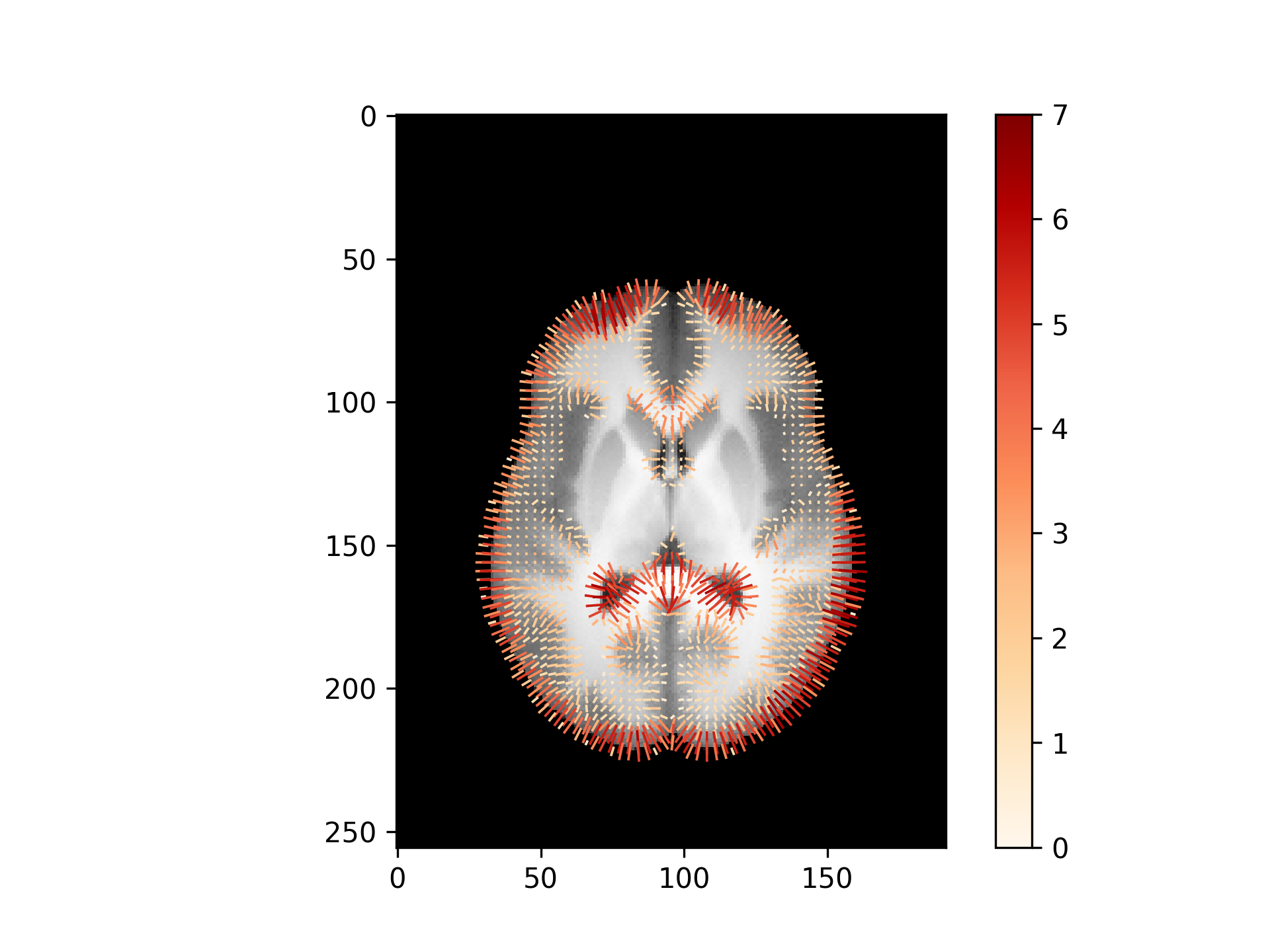}}
   \caption{Visualization of $\sigma^*$ on top of the templates based on the $L^2$-norm (top row) and the $L^1$-norm (bottom row) for affine (left column) and rigid (right column) transformations.}
   \label{fig:brain_tmp_bars}
\end{figure}

\begin{figure}[ht!]
   \centering
   \subfloat{\includegraphics[trim={4.1cm 0.7cm 2.1cm 1.2cm}, clip, width=0.495\linewidth]{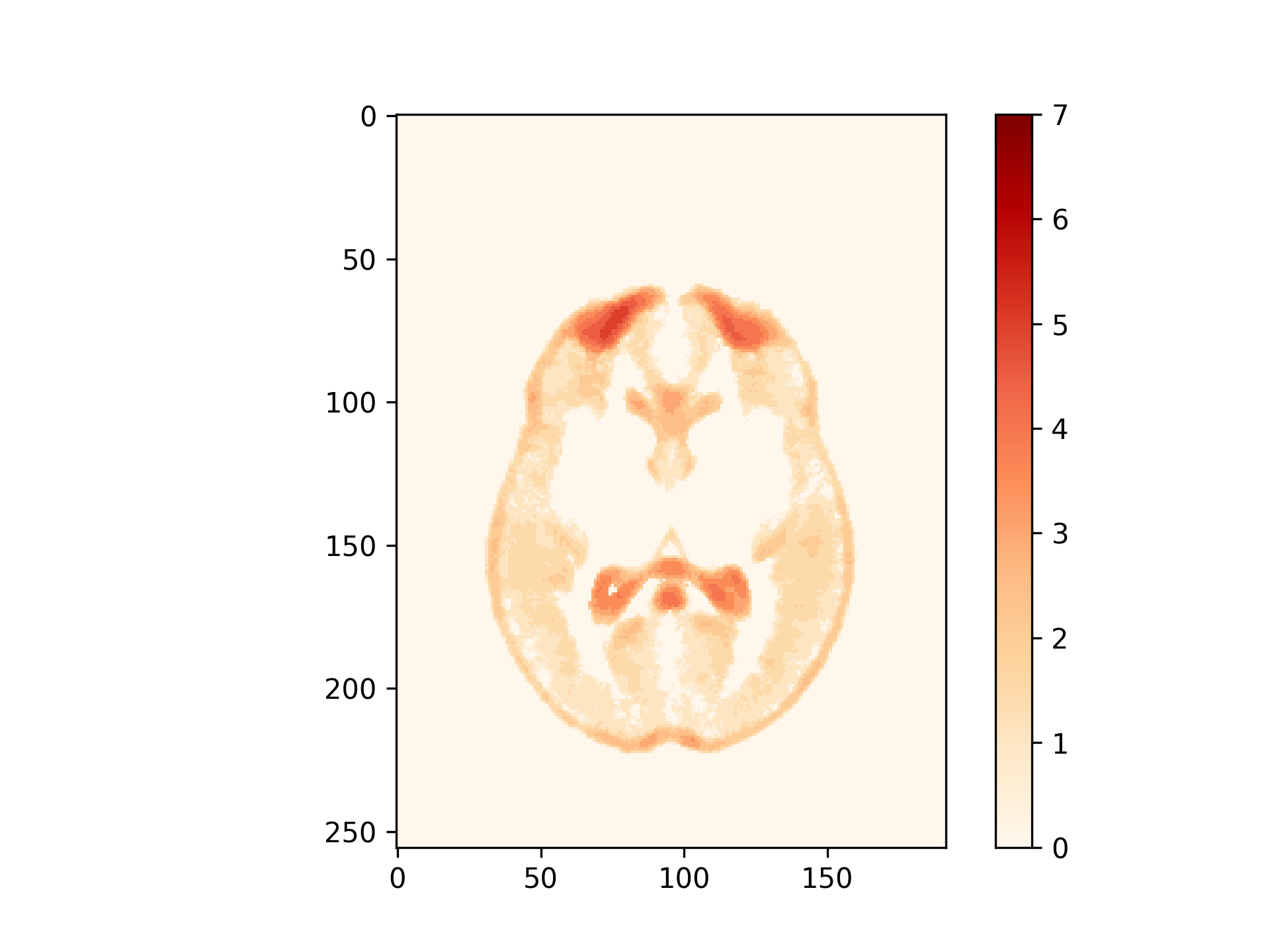}}
   \hfill
   \subfloat{\includegraphics[trim={4.1cm 0.7cm 2.1cm 1.2cm}, clip, width=0.495\linewidth]{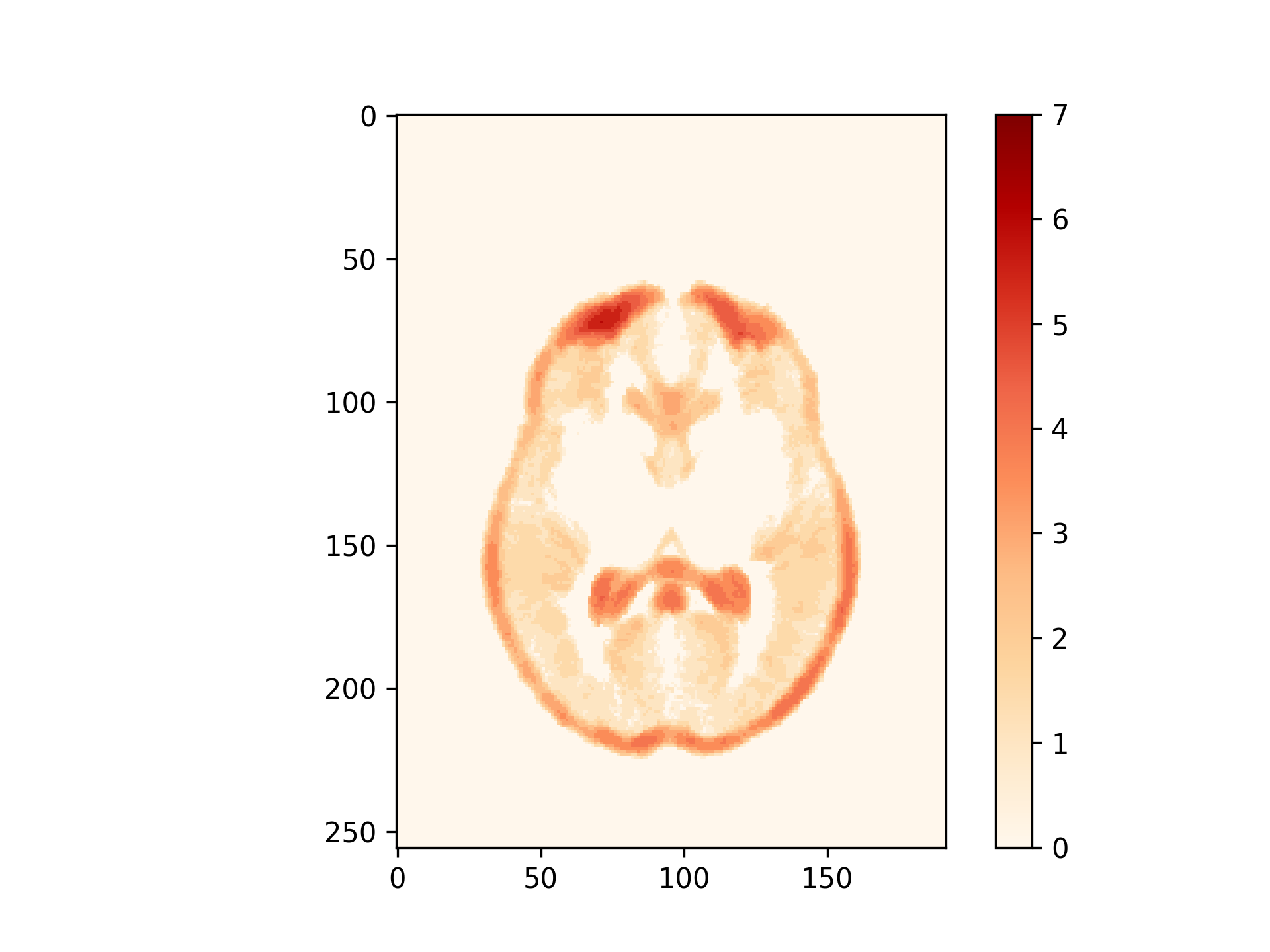}}
   \\
   \subfloat{\includegraphics[trim={4.1cm 0.7cm 2.1cm 1.2cm}, clip, width=0.495\linewidth]{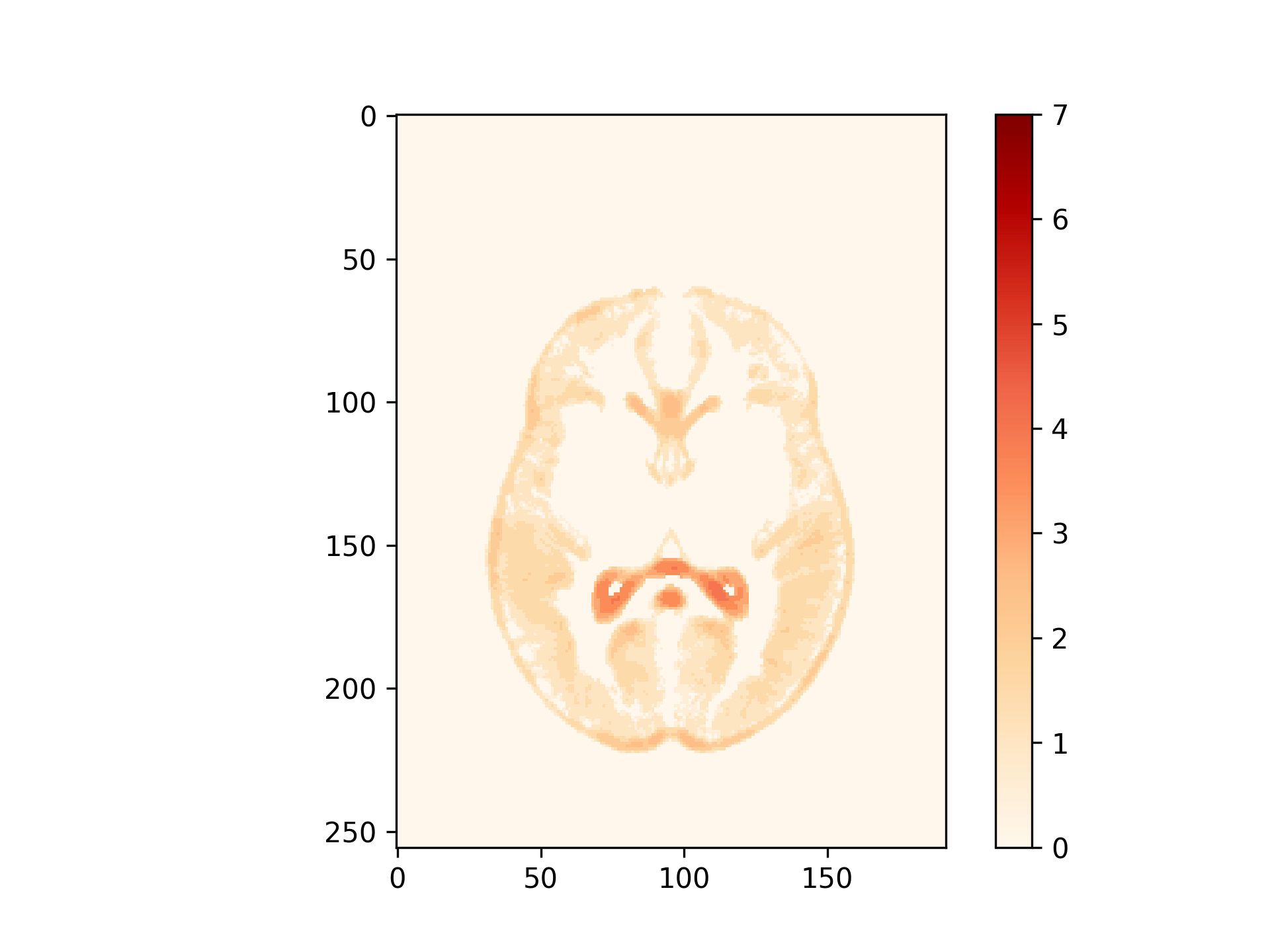}}
   \hfill
   \subfloat{\includegraphics[trim={4.1cm 0.7cm 2.1cm 1.2cm}, clip, width=0.495\linewidth]{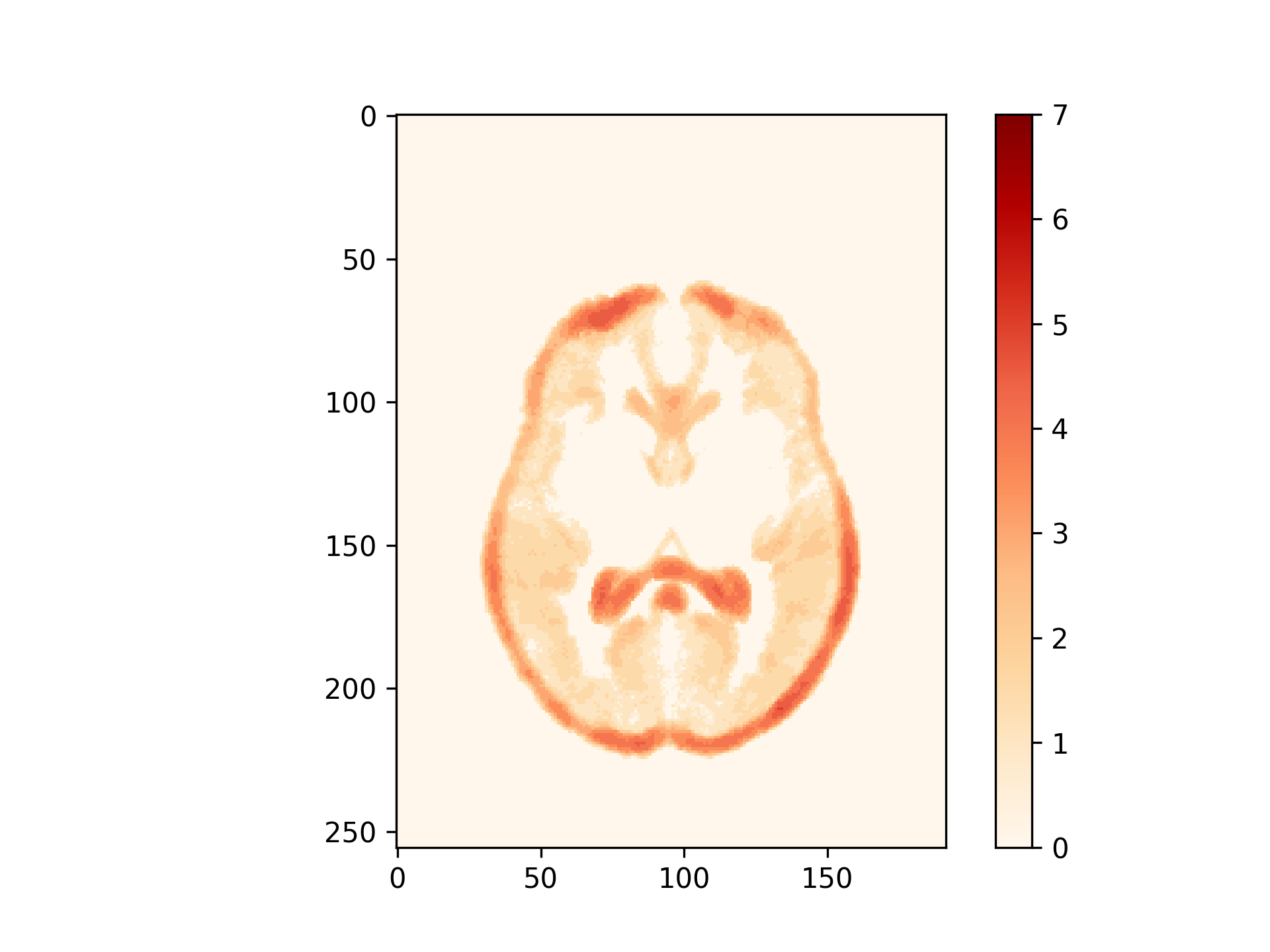}}
   \caption{Template resolution measure $\sigma^*$ based on the $L^2$-norm (top row) and the $L^1$-norm (bottom row) for affine (left column) and rigid (right column) transformations.}
   \label{fig:brain_tmp_res}
\end{figure}

\begin{figure}[htb!]
   \centering
   \subfloat{\includegraphics[trim={1.1cm 0.7cm 2.1cm 1.2cm}, clip, width=0.7\linewidth]{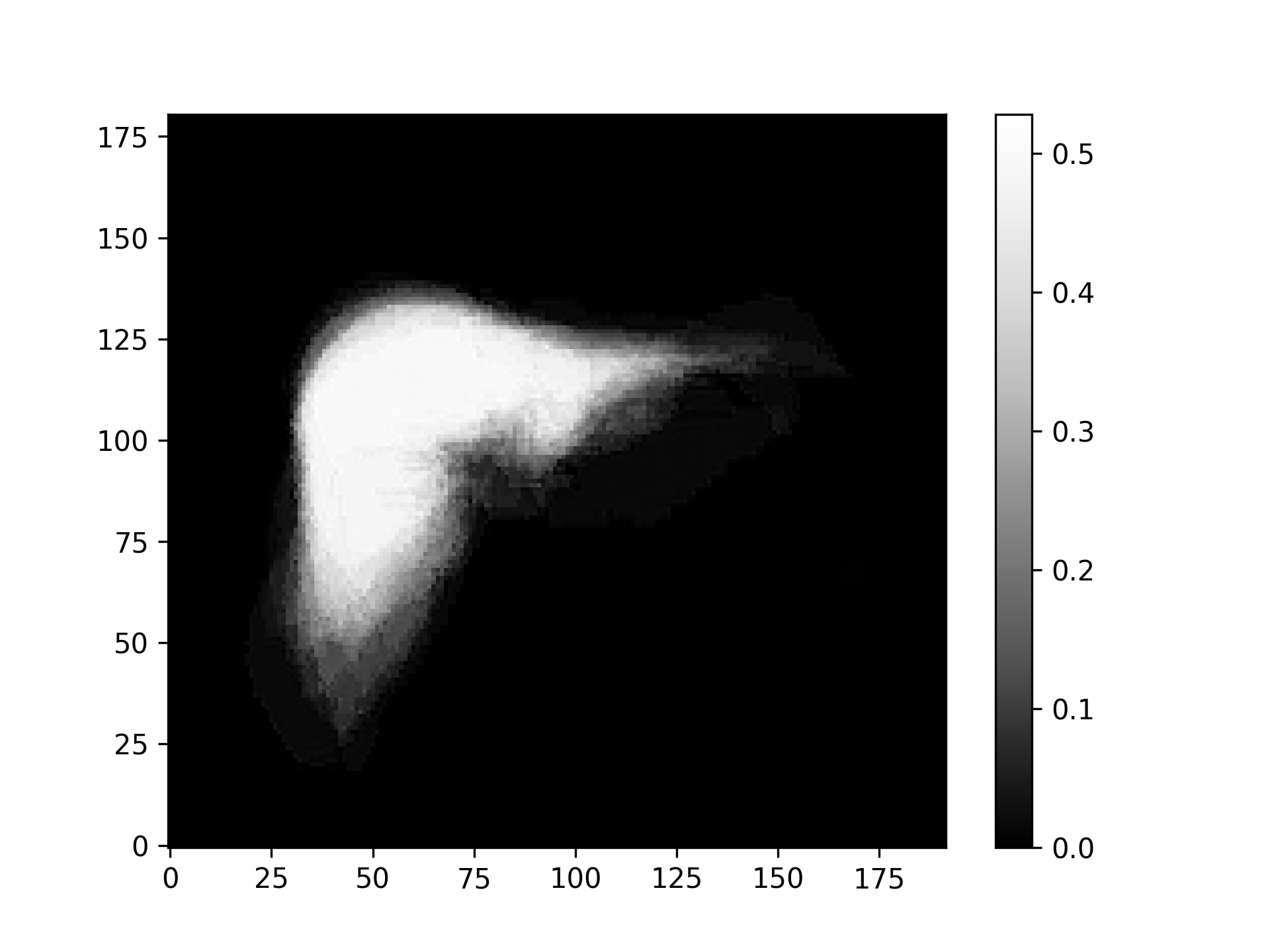}}
   \\
   \subfloat{\includegraphics[trim={1.1cm 0.7cm 2.1cm 1.2cm}, clip, width=0.7\linewidth]{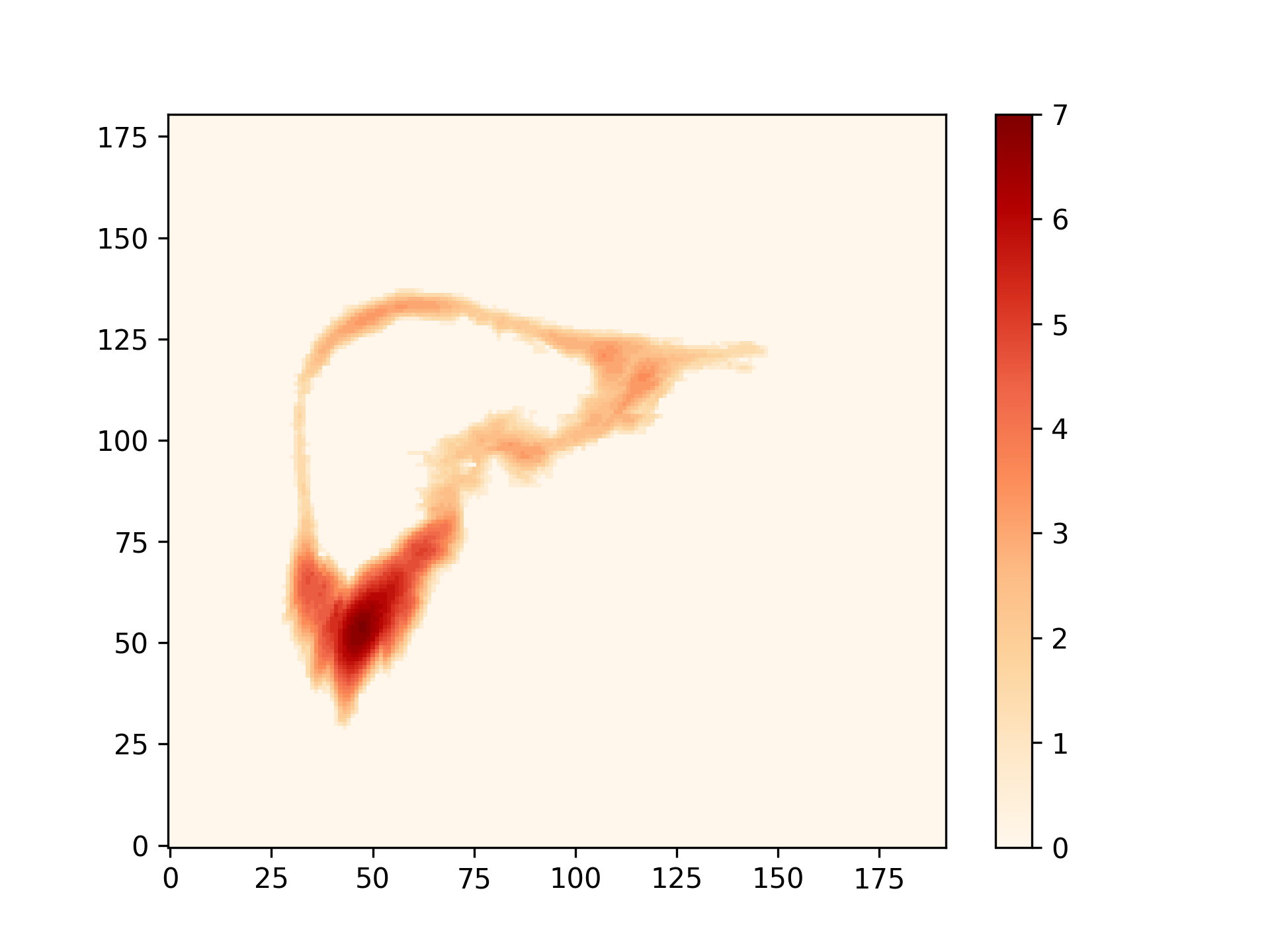}}
   \\
   \subfloat{\includegraphics[trim={1.1cm 0.7cm 2.1cm 1.2cm}, clip, width=0.7\linewidth]{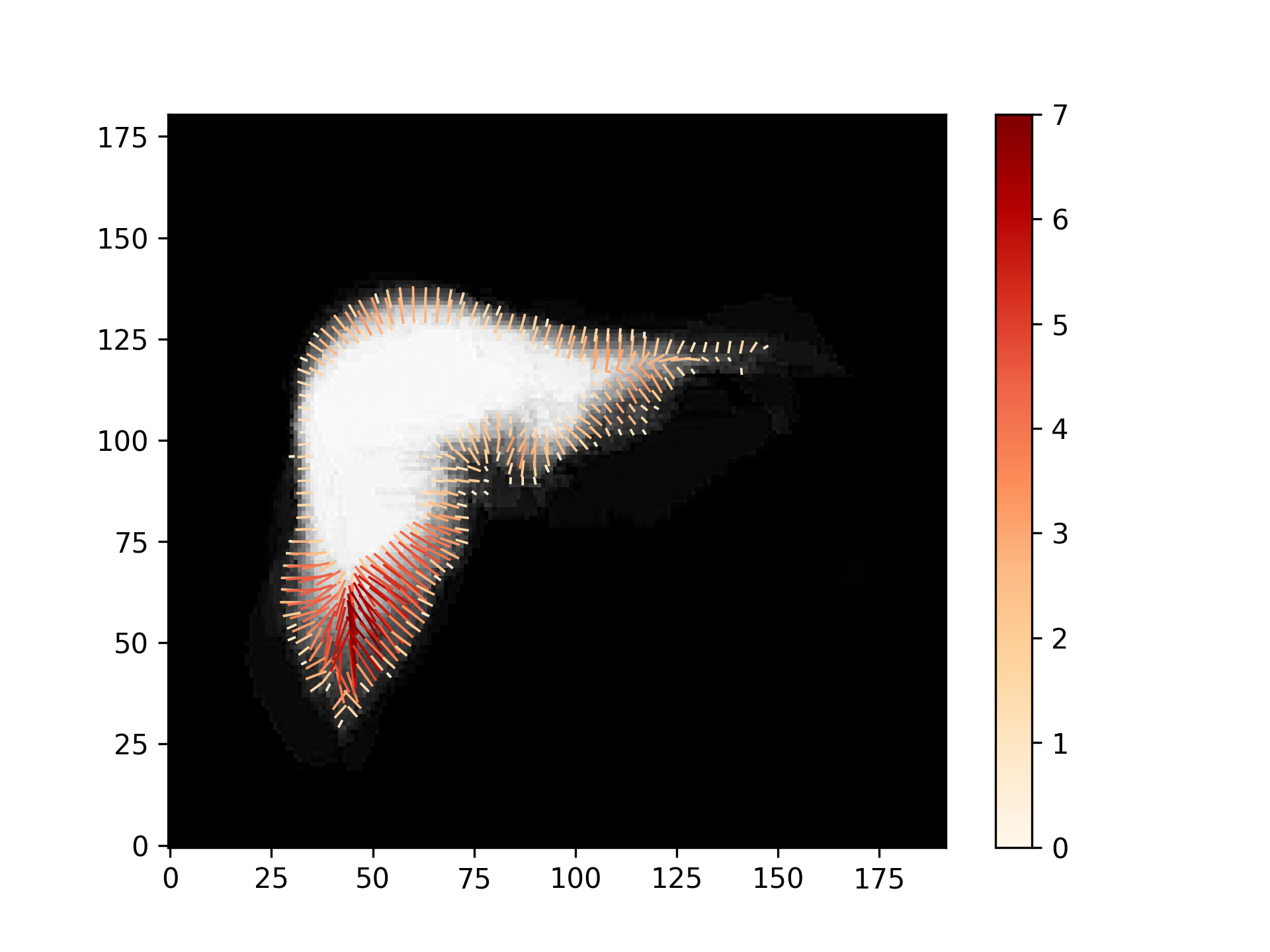}}
   \caption{Template image, TRM and visualization of the TRM on top of the template for the liver using the $L^2$-norm and affine transformations.}
   \label{fig:liver_hippo_tmp_res}
\end{figure}

\subsection{3d brains}\label{sec:app_3d}

An important application of image registration is the preprocessing of 3d MRI brain images. Here a collection of brain images (possibly from different subjects) are registered with a template, here also called an atlas, which enables more direct comparisons between different brains despite differences in the respective brain geometry before registration. The dataset used here for demonstration purposes is the Neurofeedback Skull-stripped (NFBS) dataset \citep{nfbs_data}, which contains 125 raw MRIs and their skull stripped versions.
The intensities of the sample images are normalized such that the median of each image (without the background zeros) is mapped to $0.5$. Additionally, we are optimizing an intensity scaling factor for each image during the registration such that the scaled template matches the registered image under the similarity metric. This minimizes the vertical variability and brings us a bit closer to the model situation of the shifted edges. The effective height for the resolution measure is chosen as $\eta \approx 0.4$.

For the skull stripped brains a template is generated for affine and rigid transformations and for $L^2$- and $L^1$-norm similarity measures, as shown in Figure \ref{fig:brain_tmp}.
In each of the four cases the template resolution is computed with Algorithm \ref{alg} and horizontal slices of the resulting 3d images of the TRM $\sigma^*$ are shown in Figure \ref{fig:brain_tmp_res} and visualized on top of the template, as described in Subsection \ref{sec:vis}, in Figure \ref{fig:brain_tmp_bars}.

Again, the affine registration is more accurate, especially at the boundary of the brain, and the $L^1$-norm templates appear sharper than the $L^2$-norm versions but -- as revealed by the template resolution measure -- are only slightly better in terms of registration accuracy. The rigid registration results are almost identical. In the affine case the $L^1$-norm leads to better registration at the upper part of the brain (frontal lobe; see Figure \ref{fig:brain_tmp_sample} for locations), while being similarly inaccurate around the lateral ventricles (twin holes in the lower middle). In all cases the regions in the upper middle of the brain (around the putamen and thalamus) are quite well registered and are also the sharpest regions of the templates across all registration variants.
This is in agreement with measurements of the volumes of these brain regions (see e.g. Table 4 in \citep{brain_vol_var} for healthy controls, or Table 2 in \citep{brain_vol_var_2}) where the coefficients of variation (standard deviation divided by the mean) of the volumes are relatively low for the putamen and thalamus compared with those for the lateral ventricles, the latters' variability ranging among the highest across all measured regions.

% text for later section:
When comparing the TRM in the bottom rows of Figures \ref{fig:brain_tmp_bars} and \ref{fig:brain_tmp_res} with the pixelwise standard deviation in Figure \ref{fig:nfbs_l1_vert_var} we can see that the TRM is able to quantify an improvement in the horizontal misalignment of the frontal lobes when switching from rigid to affine transformations, that is not visible in the purely vertical variation measure given by the pixelwise standard deviation.

\subsection{3d livers}\label{sec:app_3d_2}

In order to demonstrate the method on a 3d dataset with a different modality, we use the publicly available 3d CT dataset Abdomen CT-CT \citep{abdomenCT} from the Learn2Reg challenge.
This dataset contains 30 CT scans of the abdomen together with their corresponding organ segmentations.
We use the organ segmentations to generate a template for the liver in the case of the $L^2$-norm and affine transformations, as shown in Figure \ref{fig:liver_hippo_tmp_res}.% (top row).
For the dataset the TRM is computed using an effective height of $\eta \approx 0.5$.

As the CT scans have a very low intensity contrast the images are approximately binary with the liver as the foreground, which means that we are closer to the model situation of the (sharp) shifted edges.
Thus the TRM mostly measures the horizontal misalignment of the edges of the livers.
One can see in the Figure that the variation is much larger at the bottom corner of the liver (mostly segments V and VI) than the top.

\subsection{Comparing two groups of 2d brain slices}\label{sec:app_2d_2}

\begin{figure}[t!]
   \centering
   \subfloat{\includegraphics[trim={1.8cm 0.7cm 2.1cm 1.2cm}, clip, width=0.495\linewidth]{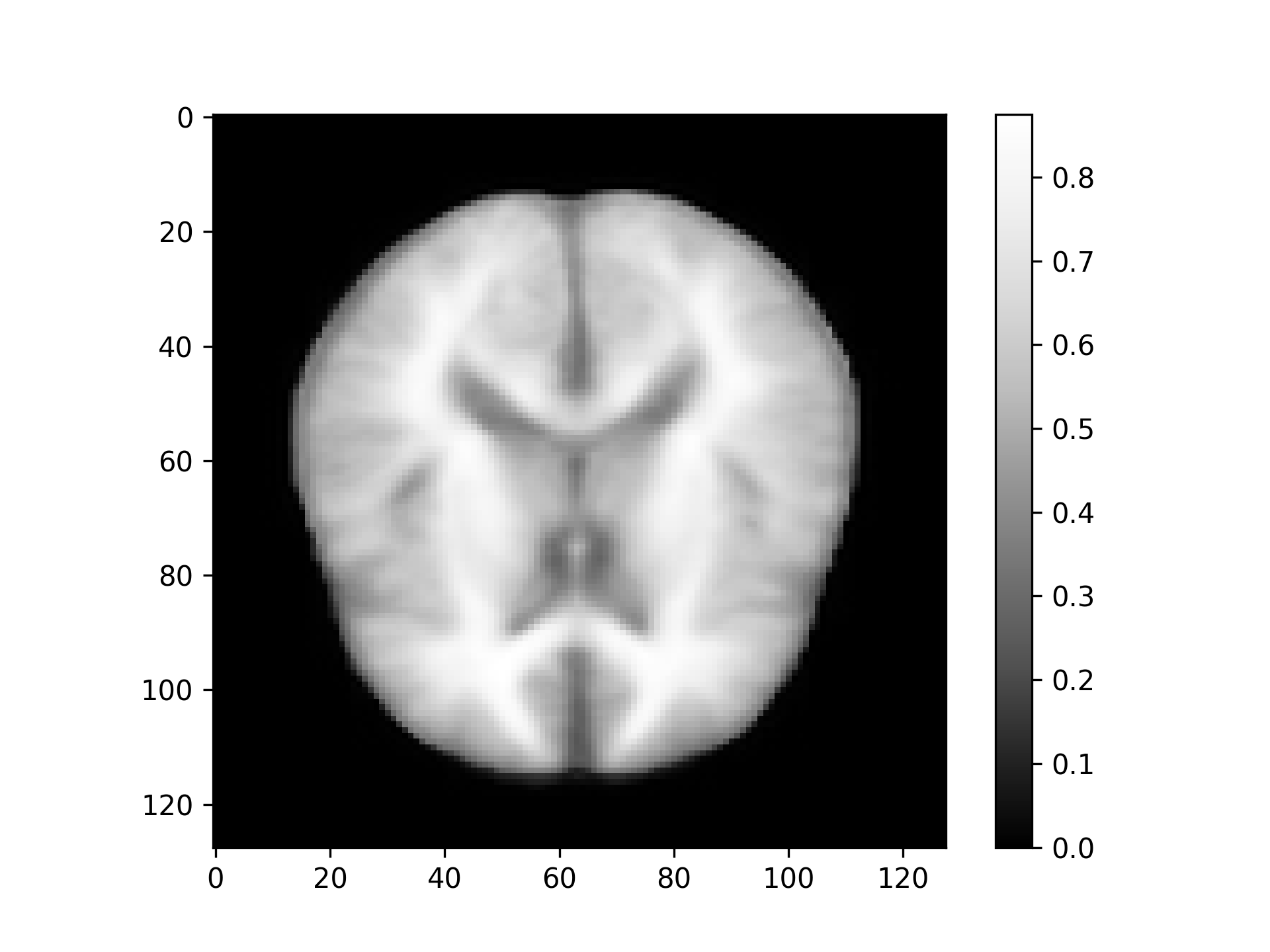}}
   \hfill
   %\subfloat{\includegraphics[trim={1.8cm 0.7cm 2.1cm 1.2cm}, clip, width=0.3\linewidth]{AMR_label_3_affine_l2_template.png}}
   %\hfill
   \subfloat{\includegraphics[trim={1.8cm 0.7cm 2.1cm 1.2cm}, clip, width=0.495\linewidth]{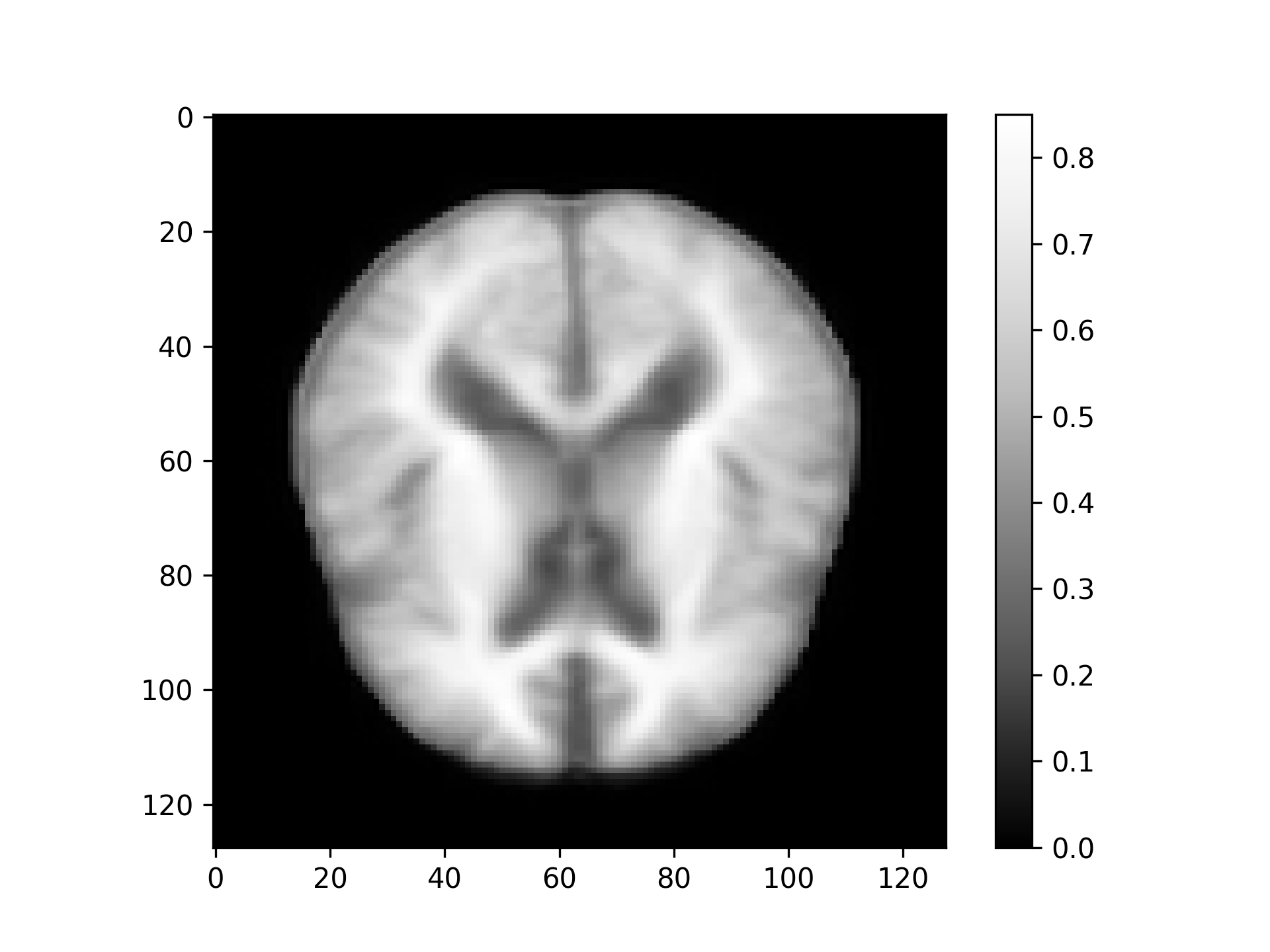}}
   \\
   \subfloat{\includegraphics[trim={1.8cm 0.7cm 2.1cm 1.2cm}, clip, width=0.495\linewidth]{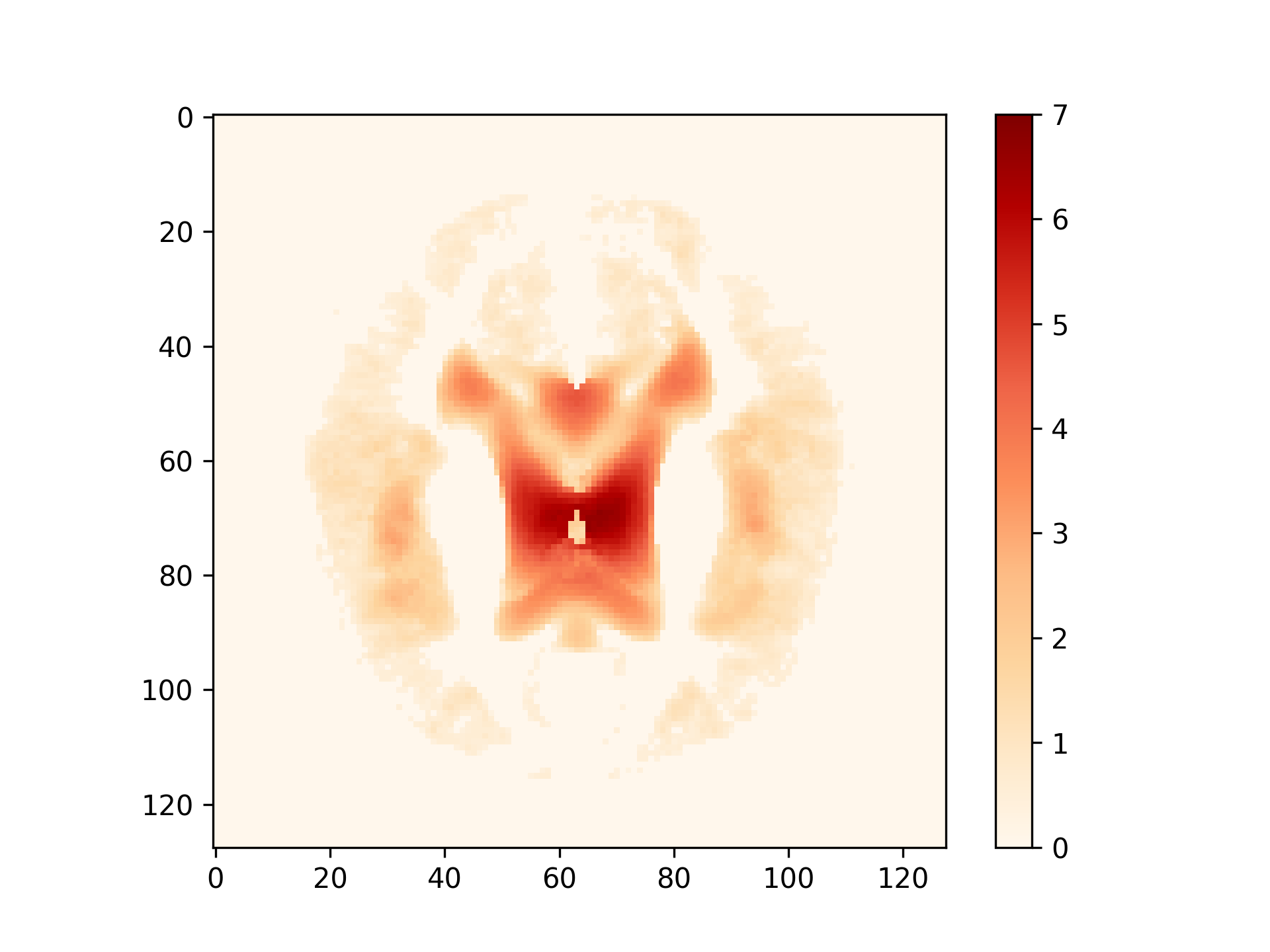}}
   \hfill
   %\subfloat{\includegraphics[trim={1.8cm 0.7cm 2.1cm 1.2cm}, clip, width=0.3\linewidth]{AMR_label_3_affine_l2_template_resolution_eh_0.5.png}}
   %\hfill
   \subfloat{\includegraphics[trim={1.8cm 0.7cm 2.1cm 1.2cm}, clip, width=0.495\linewidth]{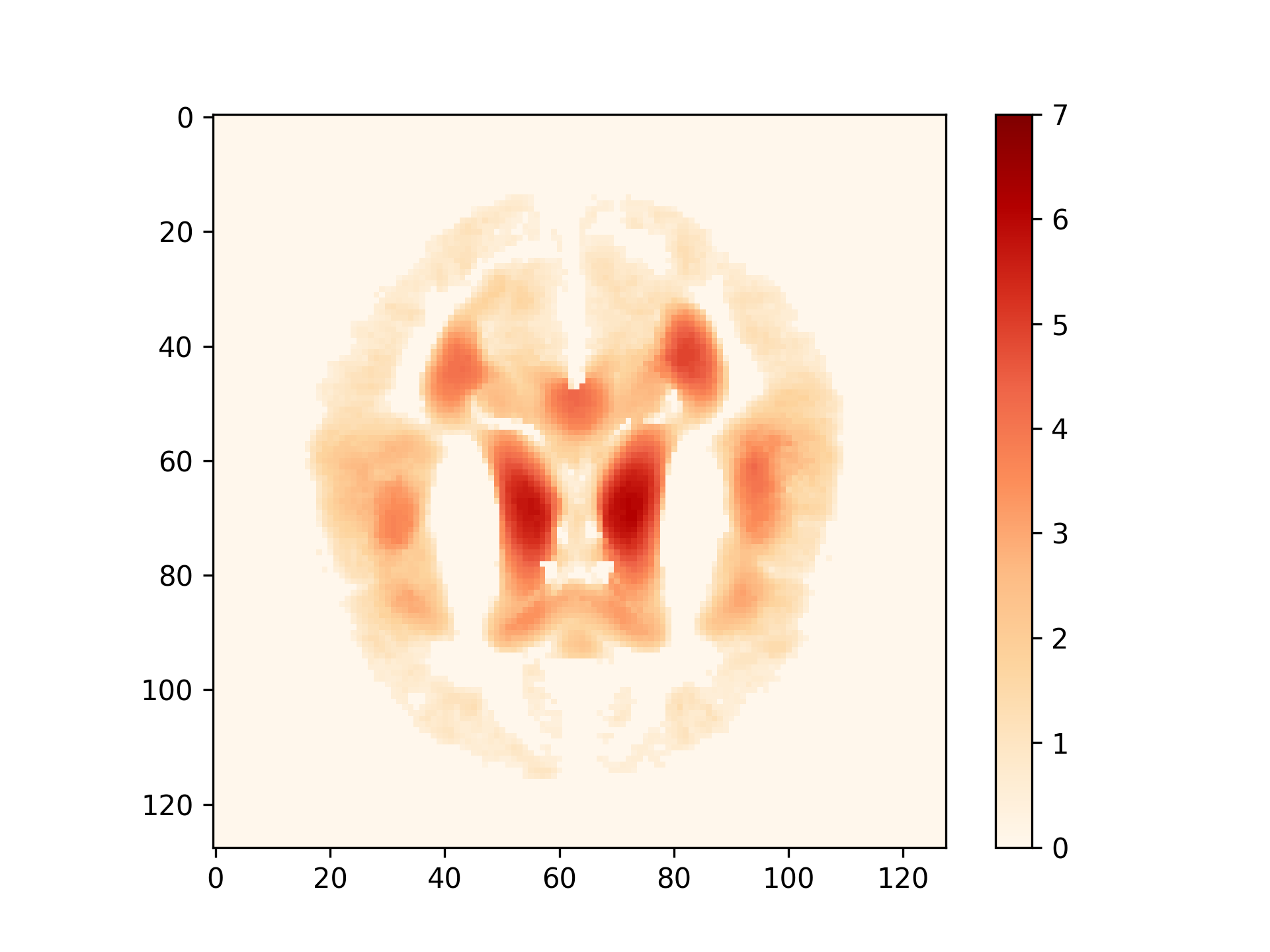}}
   \caption{Templates (top row) and TRM (bottom row) for the Alzheimer's MRI Disease Classification dataset based on the $L^2$-norm and affine transformations. The left column corresponds to the non-demented class and the right column to the mild dementia.}
   \label{fig:alzheimer_tmp_res}
\end{figure}

In order to demonstrate the applicability of our method we also computed the TRM on a sample of 2d slices of brain MRI images from the publicly available Alzheimer MRI Disease Classification dataset \citep{alzheimer_mri}.
This contains 2d images of brains labelled by their Alzheimer's disease status (non demented and mild dementia), where we performed an image registration for these classes with $200$ samples each using the $L^2$-norm and affine transformations. We then computed the TRM for the registered images of each class as shown in Figure \ref{fig:alzheimer_tmp_res} using an effective height of $\eta \approx 0.5$.

Here we can see in the templates that the mildly demented brains have enlarged ventricles. These are also visible in the TRM, where they constitute regions of higher horizontal variability.

The TRM of the mildly demented brains also seem to be higher in the left and right boundary regions showing that the disease not only corresponds to obvious changes in the anatomy of the brain but also increases the variation between the brains in more subtle ways.

\begin{figure}[ht!]
   \centering
   \includegraphics[trim={1.8cm 0.7cm 2.1cm 1.2cm}, clip, width=0.95\linewidth]{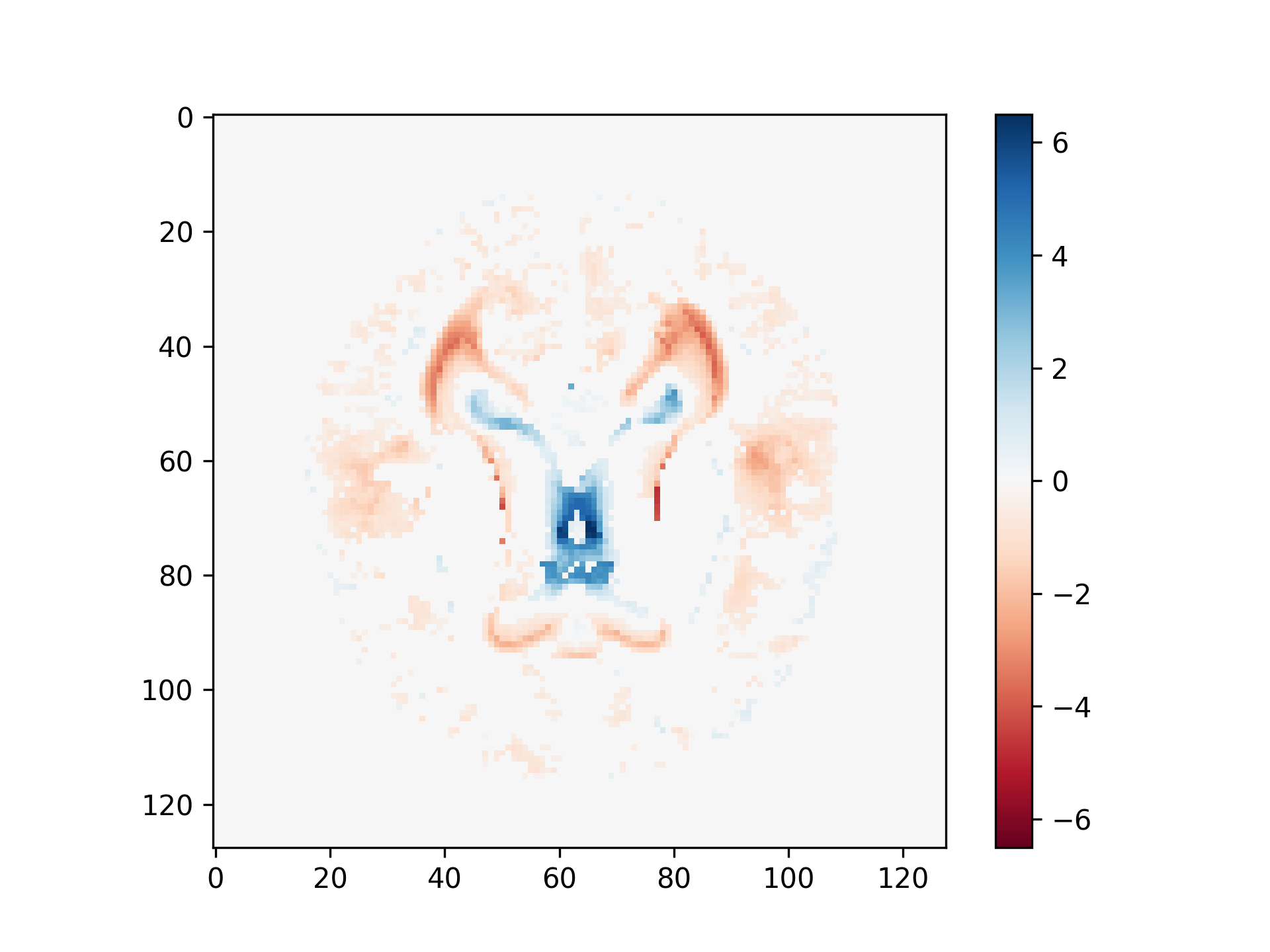}
   \caption{Difference between the TRMs of the non-demented and mildly demented cases. Non significant differences where determined by a permutation test (1000 repetitions) and manually set to zero.}
   \label{fig:alzheimer_tmp_res_diff}
\end{figure}

That these differences between the TRMs are actually significant is quantitatively verified with a permutation test: we repeat the registration and TRM computation for the two groups $1000$ times, where in each repetition the labels of the samples are randomly permuted. The empirical distribution of the resulting TRM differences is then compared to the original difference between the TRMs of the two groups, where for each pixel the $\tfrac\alpha2$ and $1 - \tfrac\alpha2$ quantiles (here for $\alpha = 0.05$) of the empirical distribution are computed and compared to the original difference. If the original difference is outside of this range, it is considered significant. Figure \ref{fig:alzheimer_tmp_res_diff} shows the resulting difference between the TRMs of the non-demented and mildly demented cases, where non-significant differences are manually set to zero.
Here the red regions indicate significant differences where the TRM of the mildly demented group is larger than that of the non-demented group.

\section{Discussion and Conclusion}

From the theoretical analyses as well as the experiments presented above, we conclude that the proposed template resolution measure (TRM) allows to quantify and visualize the horizontal uncertainty remaining after registration at each location of the template.
In fact, in both applications the horizontal variability as measured by the TRM in the template's units of length is in agreement with otherwise obtained measures for it.

A major advantage of the proposed method is that it uses only the registered images as input, see Algorithm~\ref{alg}; in particular, no labelling or segmentation is required (cf. the last paragraph of Section~\ref{sec:intro} on the state of the art).
The computation and visualization of the TRM can thus be fully automatized given three parameters, namely the effective height $\eta$, as well as the threshold probabilities $p_0$ and $p_1$.
However, further experiments implied that changing the threshold probabilities barely affects the results; hence, their selection is more a matter of the corresponding quantiles' statistical robustness given the sample size.
Choosing the effective height is somewhat more important, but fortunately, it has a well specified meaning as the height of a sharp edge which is to be detected as such.
Therefore, about half the difference between the intensity of foreground and background will often be a reasonable choice for $\eta$.

%Of course, there are situations where the proposed resolution measure's behaviour differs from the intended one.
%Indeed, large values may not only correspond to horizontally misaligned features, but also to a feature mismatch, where some sample images contain a certain feature which the other samples are lacking.
%But even in this case the TRM is useful as low values still correctly identify regions where the samples agree.
Of course, the proposed TRM comes with some limitations, in particular the following:
\begin{itemize}
   \item There are situations where the proposed resolution measure's behaviour differs from the intended one.
   Indeed, large values may not only correspond to horizontally misaligned features, but also to a feature mismatch, where some sample images contain a certain feature which the other samples are lacking.
   But even in this case the TRM is useful as low values still correctly identify regions where the samples agree.
   \item %only groupwise
   Our entire analysis is based on the groupwise setting of image registration. In the pairwise setting the proposed algorithm can not be used to evaluate the misalignment of the images.
   %does not make sense, since ... only evaluates the registered images not the transformations themselves.
   \item %only affine/rigid (only if still variation left)
   As we assumed that the images still have horizontal variation left after registration, the method is restricted to simpler types of transformations, such as affine or rigid transformations.
   Additionally the method only looks at the registered images and does not evaluate the transformations themselves.
\end{itemize}
On the bright side we showed that the TRM is robust to noise in the images and can be used to learn about the horizontal variation in a dataset especially when comparing different groups of images.

In summary, we developed a new tool to help with the evaluation of images registered groupwise to a template, providing insight into the reliability of the structures visible in the template, and quantifying as well as visualizing this in terms of a local resolution measure.

\backmatter

\bmhead{Data availability}

All the datasets used in this paper are available online. MNIST (\url{http://yann.lecun.com/exdb/mnist/}) and the NFBS dataset (\url{http://preprocessed-connectomes-project.org/NFB_skullstripped/}) can be downloaded from their official website accordingly.
The Abdomen CT-CT dataset
%and the Hippocampus MR dataset are
is part of the Learn2Reg challenge (\url{https://learn2reg.grand-challenge.org/Datasets}). The Alzheimer's MRI Disease Classification dataset can be downloaded from huggingface (\url{https://huggingface.co/datasets/Falah/Alzheimer_MRI}).

\bmhead{Code availability}

The source code is made publicly available on GitHub (\url{https://github.com/Stochastik-TU-Ilmenau/image-template-resolution}).

\bibliography{lit.bib}% common bib file
%\input{TemplateResolution_input.bbl}

\end{document}